\documentclass[a4paper,11pt]{article}
\pdfoutput=1 

\usepackage{jheppub} 

\usepackage[OT1]{fontenc} 
\graphicspath{{Figures/}} 	     
\usepackage{booktabs} 
\usepackage{bbold} 
\usepackage{mathtools}
\usepackage{multirow}
\usepackage[version=4]{mhchem} 
\usepackage{physics}
\usepackage{pdflscape}
\usepackage{slashed}
\usepackage{xcolor}
\usepackage{caption}
\usepackage{ragged2e}
\usepackage{orcidlink}

\newcommand{\bea} {\begin{eqnarray}}
\newcommand{\eea} {\end{eqnarray}}

\def\tanb {\tan\beta}
\def\tb {t_\beta}
\def\sb  {s_{\beta}}
\def\cb  {c_{\beta}}
\def\stwob  {s_{2\beta}}

\newcommand{\beq}{\begin{equation}}
\newcommand{\eeq}{\end{equation}}
\newcommand{\Eq}[1]{Eq.~(\ref{#1})}

\DeclareMathOperator{\GeV}{\text{GeV}}

\renewcommand{\sb}{s_{\beta}}
\renewcommand{\cb}{c_{\beta}}

\def\vev{{\it vev}}
\def\vevs{{\it vevs}}

\title{ Self Consistent Thermal Resummation: \\  \vskip 4pt  \large{A Case Study of the Phase Transition in 2HDM}}  

 \author[a]{Pedro Bittar \orcidlink{0000-0002-3684-5692},}
  \author[b]{Subhojit Roy \orcidlink{0000-0001-6434-5268},}
  \author[b,c,d,e]{and Carlos E.M. Wagner \orcidlink{0000-0001-6407-623X}}
  \affiliation[a]{Department of Mathematical Physics,
Institute of Physics, 
University of São Paulo,
R. do Matão 1371, São Paulo, 
SP 05508-090, Brazil}
\affiliation[b]{HEP Division, Argonne National Laboratory, 9700 Cass Ave., Argonne, IL 60439, USA}
\affiliation[c]{Enrico Fermi Institute, Physics Department, University of Chicago, Chicago, IL 60637, USA}
\affiliation[d]{Kavli Institute for Cosmological Physics, University of Chicago, Chicago, IL 60637, USA}
\affiliation[e]{Perimeter Institute for Theoretical Physics, Waterloo, Ontario N2L 2Y5, Canada}
\emailAdd{bittar.hep@gmail.com}
    \emailAdd{sroy@anl.gov}
    \emailAdd{cwagner@uchicago.edu}
    \preprint{EFI-25-3} 
\abstract{
An accurate description of the scalar potential at finite temperature is crucial for studying cosmological first-order phase transitions (FOPT) in the early Universe. 
At finite temperatures, a precise treatment of thermal resummations is essential, as bosonic fields encounter significant infrared issues that can compromise standard perturbative approaches. The Partial Dressing (or the tadpole resummation) method provides a self consistent resummation of higher order corrections,  allowing the computation of thermal masses and the effective potential including the proper Boltzmann suppression factors and without relying on any high-temperature approximation. 
We systematically compare the Partial dressing resummation scheme results with the Parwani and Arnold Espinosa (AE) ones to investigate the thermal phase transition dynamics in the Two-Higgs-Doublet Model (2HDM).
Our findings reveal that different resummation prescriptions can significantly alter the nature of the phase transition within the same region of parameter space, 
confirming the differences that have already been noticed between the Parwani and AE schemes.
Notably, the more refined resummation prescription, the Partial Dressing scheme, does not support symmetry non-restoration in 2HDM at high temperatures observed using the AE prescription. Furthermore, we quantify the uncertainties in the stochastic gravitational wave (GW) spectrum from an FOPT due to variations in resummation methods, illustrating their role in shaping theoretical predictions for upcoming GW experiments. Finally, we discuss the capability of the High-Luminosity LHC and proposed GW experiments to probe the FOEWPT-favored region of the parameter space.  }

\keywords{Thermal resummation, Electroweak phase transition, 2HDM, tadpole resummation, partial dressing, gravitational waves, LHC, LISA}
\begin{document}
\maketitle
\flushbottom 
%
\section{Introduction}
The discovery of the Higgs boson in 2012 at the Large Hadron Collider (LHC)~\cite{ATLAS:2012yve, CMS:2012qbp} was the last step in the observation of all particles expected in the Standard Model (SM). The test of its properties demonstrated that the SM is a valid low-energy effective theory at the electroweak (EW) scale.
The LHC continues to investigate the properties of this scalar particle while also conducting searches for physics beyond the Standard Model (BSM).
One of the major goals of the LHC and future proposed colliders~\cite{Curtin:2014jma, Papaefstathiou:2020iag, Ramsey-Musolf:2019lsf} is to understand the dynamics of EW symmetry breaking (EWSB).
Although EWSB develops through a cross-over transition in the SM, in many BSM scenarios, it can be a first-order phase transition (FOPT).

An interesting aspect of FOPT is that it produces a spectrum of stochastic gravitational waves (GW) during the phase transition.
Detecting such GW signals may become feasible at various proposed future detectors, both space- and ground-based, such as LISA~\cite{LISA:2017pwj}, ALIA~\cite{Gong:2014mca}, TAIJI~\cite{Hu:2017mde}, the Big Bang Observer (BBO)~\cite{Corbin:2005ny}, and Ultimate (U)-DECIGO~\cite{Kudoh:2005as}, within the next few decades. 
These upcoming GW experiments open up a new window into the early Universe, shedding light on the electroweak scale physics and the thermal history of the early Universe~\cite{Sesana:2019vho,Caprini:2019egz,Yagi:2011wg,Punturo:2010zz,Hild:2010id}. 
This potential for investigation is especially relevant to the phenomenology of various scenarios with extended Higgs sectors. Such sectors may lead to phenomena like a first-order electroweak phase transition (FOEWPT), which can enable electroweak baryogenesis (EWBG) to explain the Universe's baryon asymmetry~\cite{Sakharov:1967dj, Cohen:1993nk, Rubakov:1996vz, Trodden:1998ym, Riotto:1998bt, Riotto:1999yt, Cline:2006ts, Morrissey:2012db, Davoudiasl:2004gf, Hooper:2025fda, White:2016nbo, Chatterjee:2022pxf, Wagner:2023vqw}. Additionally, it opens possibilities to study an FOPT in hidden (dark) sectors~\cite{Schwaller:2015tja,Baldes:2018emh,Breitbach:2018ddu,Croon:2018erz,Hall:2019ank,Baldes:2017rcu,Geller:2018mwu,Croon:2019rqu,Hall:2019rld,Chao:2020adk, Ghosh:2022fzp, Roy:2022gop, Dent:2022bcd, Borah:2024emz}, vacuum trapping, electroweak symmetry non-restoration (EWSNR)~\cite{Weinberg:1974hy,Meade:2018saz,Baldes:2018nel,Cline:1999wi,Baum:2020vfl,Biekotter:2021ysx,Biekotter:2022kgf,Chatterjee:2022pxf,Chang:2022psj,Aoki:2023lbz}, and the formation of topological defects (e.g., domain walls, cosmic strings)~\cite{Huang:2017laj,Croon:2018kqn,Hashino:2018zsi,Brdar:2019fur, Dunsky:2021tih, Blasi:2022woz}.

One needs to accurately estimate the finite temperature effects to understand the behavior of the scalar potential and its predicted stochastic GW spectrum. This is crucial to fully exploit and recast the available and incoming experimental data to the various BSM scenarios. However, a major challenge arises due to the breakdown of the perturbative expansion at high temperatures \cite{Weinberg:1974hy,Dolan:1973qd,Kirzhnits:1974as,Linde:1978px,Linde:1980ts}. At finite temperatures, the quadratically divergent contributions from the non-zero Matsubara modes need to be re-summed to accurately capture thermal corrections, ensuring consistency in the perturbative expansion and preventing infrared divergences.
The most commonly used methods for implementing these resummations are the Parwani \cite{Parwani:1991gq} and Arnold-Espinosa (AE) \cite{Arnold:1992rz} schemes. 
Both AE and Parwani schemes are examples of high temperature \textit{Truncated Full Dressing} (TFD) methods.
In general, \textit{Full Dressing} (FD) refers to a strategy for including the thermal corrections in which the thermal mass obtained from the self-consistent gap equation is directly inserted back into the effective potential. These methods use the high-temperature approximation, $T^2 \gg m^2$, and truncate the gap equation to the first order to obtain a simple expression.
Parwani's method inserts the thermal masses throughout the one-loop Coleman-Weinberg and finite-temperature potentials. In contrast, the AE scheme resums only the so-called daisy diagram contributions by including thermal masses in cubic terms that are the leading contributions to infrared divergences \cite{Dolan:1973qd,Weinberg:1974hy, Laine:2016hma}.
Both prescriptions effectively mitigate these divergences by incorporating thermally improved masses. These approaches are relatively straightforward to implement at the one-loop level and do not demand high computational power. Another approach that also relies on the high-temperature behavior is dimensional reduction (DR). In this case, the compactified $4d$ thermal theory is reduced to a $3d$ effective field theory~\cite{Farakos:1994kx, Kajantie:1995dw, Braaten:1995cm, Ekstedt:2022bff}. This approach leverages the fact that, at high temperature, only the so-called Matsubara zero modes significantly contribute to the low-energy physics. In contrast, non-zero modes become massive and can be integrated out from the theory. DR effectively incorporates higher order corrections beyond leading daisy diagrams but is limited to specific scale hierarchies, where the high-temperature effects can be effectively separated from the low-temperature ones.
This restriction complicates DR implementation and parameter scans of various BSM scenarios, since different effective theories must be used depending on the mass hierarchies. In summary, the AE, Parwani, and DR methods rely on the high-temperature approximation to consistently include thermal corrections.

There are cases, however, where the high-temperature approximation breaks down, or a clear scale separation is not available. In these cases, a self-consistent resummation prescription is required. 
A notable example occurs for FOEWPTs at the early Universe and their role in EWBG~\cite{Sakharov:1967dj, Cohen:1993nk, Rubakov:1996vz, Trodden:1998ym, Riotto:1998bt, Riotto:1999yt, Cline:2006ts, Morrissey:2012db, Davoudiasl:2004gf,White:2016nbo,Wagner:2023vqw}. In such a scenario, the transition from the false vacuum to the true electroweak vacuum occurs through a bubble nucleation mechanism, which serves as a source for the out-of-equilibrium processes necessary for successful baryogenesis \cite{Sakharov:1967dj}. 
%
To avoid washing out the generated baryon asymmetry in the true vacuum, it is necessary to sufficiently suppress the sphaleron rate in the broken phase~\cite{Cline:2006ts}. A commonly used approximate criterion for this suppression is $v_n/T_n \gtrsim 1$, where $v_n$ is the $\vev$ of the field at the nucleation temperature $T_n$. However, this condition is subject to theoretical uncertainties in the determination of the sphaleron rate (see Refs.~\cite{Zhou:2020irf, Morrissey:2012db, Patel:2011th, Garny:2012cg, Blinov:2015sna} and references therein) and should be viewed as an indicative rather than an exact bound.
This observation implies that it is crucially important to properly account the degrees of freedom (dof) that participate in the phase transition near the nucleation temperature.
The breakdown of the high-temperature approximation  suggests that resummation methods like AE and Parwani- which include effects from dof that should be Boltzmann-suppressed - may not reliably assess the nature of the phase transition. Moreover, these schemes fail to properly incorporate higher-order corrections, as neither AE nor Parwani consistently includes non-Daisy diagrams from higher orders. Resummations are meant to improve the perturbative convergence of the observables, but an inconsistent inclusion of these corrections might lead to spurious effects \cite{Bahl:2022lio}. Thus, it is useful to have an alternative resummation scheme that is self-consistent, incorporates the thermal effects of various dof, and systematically includes higher-order corrections.

To overcome these issues, one might consider solving the exact gap equations without relying on the field-independent thermal mass obtained from the truncated high-temperature prescriptions. Then, adopting the FD perspective, one can resum the relevant contributions by inserting the full thermal mass back into the effective potential. While in principle this approach to resummation might include higher-order terms beyond the one-loop expansion, FD without truncation actually miscounts the two-loop daisy diagrams and leads to unphysical linear terms \cite{Dine:1992vs,Dine:1992wr,Arnold:1992rz,Boyd:1992xn,Laine:2017hdk}. A good alternative that is more consistent in higher-order terms is the \textit{Partial Dressing} (PD) resummation scheme, which obtains the thermal masses from the gap equations and inserts them back into the first derivative of the effective potential, which can then be integrated to obtain the effective potential~\cite{Boyd:1993tz,Curtin:2016urg,Curtin:2022ovx,Bahl:2024ykv}. This approach focuses on dressing the propagator alone and has been explicitly demonstrated, through calculations up to four loops, to accurately account for the so-called daisy and superdaisy diagrams~\cite{Boyd:1993tz}.

In this work we adopt the PD resummation scheme, since it provides a more robust and self-consistent approach to treat cosmological phase transition dynamics. It consistently incorporates effects at any temperature while resolving the problems of the full dressing, i.e., miscounting diagrams beyond one-loop order.
Thus, the PD effective potential can be consistently evaluated for the interest region of SFOEWPT, i.e., beyond the high temperature approximation region.
Because of this feature, we are motivated to consider the effects of a PD calculation in models with extended scalar sectors. Until recently, the implementation of PD with mixing scalars had not been explored in the literature. In Ref.~\cite{Bahl:2024ykv}, a consistent prescription was presented for the case of two mixing scalar singlets. For the first time, in this work, we study the PD resummation prescription in a realistic BSM scenario such as 2HDM.

The FOEWPT in  2HDM has been extensively studied in the literature considering Parwani and AE resummation schemes~\cite{Turok:1990zg,Cline:1996mga,Fromme:2006cm,Cline:2011mm,Dorsch:2013wja,Basler:2016obg,Dorsch:2016tab,Bernon:2017jgv,Dorsch:2017nza,Andersen:2017ika,Kainulainen:2019kyp,Su:2020pjw,Davoudiasl:2021syn,Biekotter:2021ysx,Biekotter:2022kgf,Aoki:2021oez,Goncalves:2021egx,Goncalves:2023svb}. Recently, some intriguing features have been highlighted that arise in part of the parameter space of the 2HDM considering the AE scheme when evaluating the effective potential at finite temperature~\cite{Biekotter:2021ysx,Biekotter:2022kgf,Aoki:2023lbz}. In regions with large quartic couplings, the AE daisy contributions lead to the EW symmetry not being restored, even at high temperatures. This is a particularly interesting case of EWSNR since the truncated high-temperature thermal mass is not negative, and symmetry non-restoration is induced by resummation.
Meanwhile, in some of these parameter regions, the one-loop contributions generate a zero-temperature barrier, enhancing the strength of the FOPT.
In some cases, the presence of the zero-temperature barrier leads to vacuum trapping, where the Universe becomes stuck in a metastable vacuum rather than transitioning to the true electroweak symmetry-breaking minimum. 
While all of these intriguing features were observed using the AE method, some of them are notably altered when the Parwani resummation scheme is employed~\cite{Basler:2016obg, Biekotter:2021ysx, Biekotter:2022kgf, Aoki:2023lbz}. 
Specifically, EWSNR behaviour at high temperatures is present in AE, while it is absent in the Parwani scheme. Additionally, the critical temperature and the $\vev$ at that temperature differ significantly between the two approaches, with the Parwani method generally predicting a lower critical temperature. The height of the barrier separating the vacua is also considerably altered in Parwani, affecting the phase transition dynamics and strengthening the first-order transition in these parameter regions~\cite{Basler:2016obg}.

In this work, we implement the PD resummation scheme in the 2HDM scenario to address the disagreement in the results obtained by the AE and Parwani resummation methods. We examine the thermal masses of particles in the plasma at the finite temperature within the 2HDM using various methods. We estimate these masses through the high-temperature approximation, truncated gap equations, and full gap equation solutions and compare the results from these approaches. 
As mentioned before, we further investigate the behavior of the effective scalar potential at high temperatures in the context of the EWSNR, considering various resummation schemes. 
Our findings reveal that the occurrence of EWSNR is sensitive to the choice of the resummation scheme, and we explore the underlying reasons for this dependency. Furthermore, we explore the impact of different thermal resummation schemes on the prediction of stochastic GW production from an FOEWPT. Finally, we discuss the experimental probes of FOEWPT-favored regions at the High-Luminosity LHC (HL-LHC) and various proposed GW experiments.

The paper is organized in the following way. In Sec.~\ref{sec:schemes}, we review the different resummation schemes we use in the paper. In particular, we focus on the case of multiple mixing scalar fields. The 2HDM is discussed in sec.~\ref{2hdm}. In Sec.~\ref{results}, we discuss the results of the work.
The estimation of the thermal masses of the fields in the plasma after solving full gap equations is discussed in Sec.~\ref{TMGAP}.
 FOEWPT-favored region of parameter space
 considering PD, Parwani, and AE resummation schemes is discussed in Sec.~\ref{EWPTPD}.
The occurrence of EWSNR at high temperatures with different resummation schemes is further discussed in Sec.~\ref{SNRsec}. In Sec.~\ref{GWimpact}, the uncertainty caused by choosing different resummation schemes in predicting GW amplitude from an FOEWPT is discussed. Prospects of probing the FOEWPT-favored parameter space through HL-LHC and proposed GW experiments are then discussed in Sec.~\ref{sec:LHCandGWs}. Finally, we conclude in Sec.~\ref{conclusion}. Various calculation details of this work are presented in Appendix.
\section{Effective potential at finite temperature}
\label{Pot}
The following sections review the zero-temperature and finite-temperature radiative corrections, emphasizing resummation techniques and their significance for perturbative effective potential calculations.
\subsection{One-loop Potential at zero temperature}
We consider a tree-level theory defined by the interacting Lagrangian $\mathcal{L}$, with a potential $V_0(\phi)$, where $\phi$ collectively represents the scalar dof of the model. The scalar particles generically have interactions among themselves and with the other fermionic and vectorial dof of the theory. Then, the one-loop quantum corrections at zero temperature can be obtained from the well-known Coleman-Weinberg (CW) potential~\cite{Coleman:1973jx}. Using the $\overline{\rm MS}$ renormalization scheme and in the Landau gauge the CW potential is
\begin{equation}\label{eq:CW_potential}
    V_{\text{CW}}(m_i^2(\phi))=\frac{1}{64\pi^2}\sum_{i}(-1)^{2s_i} n_i |m_i^2(\phi)|^2\left[\log\left(\frac{m_i^2(\phi)}{\mu^2}\right)-k_i\right].
\end{equation}
The species index $i=S,F,B$ corresponds to the scalar ($S$), fermion ($F$), and vector ($B$) dof, respectively, running in the loops of the effective potential. The constant $k_i$ is $\frac{3}{2}$ for scalars and the longitudinal modes of the gauge bosons and $\frac{1}{2}$ for the fermions and the transverse modes of the gauge bosons. Here, $\mu$ represents the renormalization scale.
To analyze the phase transition in 2HDM, we set $\mu = 246$~GeV. 
$s_i$ and $n_i$ denote the spin and dof of the $i$-th state.
 Finally, the field-dependent masses, $m_i^2(\phi)$, are obtained by diagonalizing the tree-level mass matrix, $\mathbb{M}^2$. The scalar mass matrix is given from the second derivatives of the potential with respect to the scalar fields
\begin{equation}\label{eq:thermalmassmatrix}
    \mathbb{M}_{ij}^2(\phi)=\frac{\partial^2 V_0(\phi)}{\partial \phi_i \partial \phi_j}, \qquad\text{$i=$ (all scalar fields).}
\end{equation}
Therefore, the field-dependent masses can be expressed as,
\begin{equation}
    m_k^2(\phi) = U_{ki}(\theta) \mathbb{M}_{ij}^2(\phi) U_{jk}^{\dagger}(\theta) 
\end{equation}
where, $U$ is the unitary matrix that diagonalizes $\mathbb{M}^2$ and $\theta$ collectively denotes the mixing angles of the scalar sector.

To keep the same tree-level masses and mixing angles at the one-loop level, we modify the $\overline{MS}$ CW potential by adding finite counterterms, $V_{\text{CT}}$ to the potential. Then, the counterterms are fixed by imposing the following on-shell renormalization conditions at zero temperature:
\begin{align}
\label{CTcond.}
    &\left.\frac{\partial (V_\text{CW}+V_{\text{CT}})}{\partial \phi_{i}}\right|_{\langle\phi_k\rangle = v_{k_{\rm EW}}} = 0 \, \, , \\
    &\left.\frac{\partial^2 (V_\text{CW}+V_{\text{CT}})}{\partial \phi_{i}  \partial \phi_{j}}\right|_{\langle\phi_k\rangle = v_{k_{\rm EW}}} = 0 \, \, .
\end{align}
The general form of $V_{\text{CT}}$ for the 2HDM, along with various relations for its coefficients obtained from the derivatives of $V_{\text{CW}}$ are shown in Appendix~\ref{CTcoeff}.
\subsection{One-loop thermal correction}
We include the leading effects of the thermal plasma at equilibrium composed from the dof of the theory in the compactified imaginary time formalism. In the Landau gauge, the one-loop effective potential induced at finite temperature is given by~\cite{Dolan:1973qd, Weinberg:1974hy}
\begin{equation}\label{vt1loop}
    V_T (m^2(\phi), T)= \frac{T^4}{2\pi^2}\left[\sum_{k} n_k J_B\left(\frac{m_k^2(\phi)}{T^2}\right)-\sum_{k=\rm F} n_k J_F\left(\frac{m_k^2(\phi)}{T^2}\right) \right],
\end{equation}
where $n_k$ are the numbers of dof for particles as discussed earlier. The sum includes all the particles as described in the previous section. The thermal functions $J_{B,(F)}$ for Bosonic (Fermionic) dof are defined as 
\begin{equation}\label{JBF}
    J_{B,F}(y)=\int_{0}^\infty dk \,k^2 \log\left[1\mp e^{-\sqrt{k^2+y}}\right],
\end{equation}

At the high temperature (HT) limit, with $m_k^{2}(\phi) \ll T^2$, the thermal functions can be expanded as,
\begin{align}
    &\label{eq:JBapprox}
    J_{B}(y) \big|_{HT} \approx -\frac{\pi^4}{45}+\frac{\pi^2}{12}y-\frac{\pi}{6} y^{3/2}-\frac{1}{32}y^2 \log\left(\frac{y}{a_B}\right) +\mathcal{O}\left(y^3\right) \, ,\\
    &\label{eq:JFapprox}
    J_{F}(y)\big|_{HT} \approx \frac{7\pi^4}{360}-\frac{\pi^2}{24}y -\frac{1}{32}y^2 \log\left(\frac{y}{a_F}\right) +\mathcal{O}\left(y^3\right) \, ,
\end{align}
where
$a_b= 16\pi^2 \exp(3/2 - 2 \gamma_E)$ and $a_f= \pi^2 \exp(3/2 - 2 \gamma_E)$,
$\gamma_E$ being the Euler-Mascheroni constant ($\approx 0.577$). 
The term $- \frac{\pi}{6} y^{3/2}$ appearing in the high-temperature approximation of $J_B$ in~\Eq{eq:JBapprox} contributes a negative cubic term to the finite-temperature effective potential. As noted earlier, the presence of this term can generate an energy barrier between two degenerate vacua, thus facilitating an SFOPT. Such a cubic term appears only for bosonic dof as it comes from the (Matsubara) zero mode propagator, which exists only for them. This term is associated with divergences in the IR limit.
 Conversely, in the low-temperature limit, the thermal functions are given by,
\begin{align}
    &\label{eq:JLTapprox}
    J_{B,F}(y) \big|_{LT} \approx - \left(\frac{\pi}{2}\right)^{1/2} y^{3/4} e^{-\sqrt{y}} \left( 1+ \frac{15}{8} y^{-1/2}\right)
\end{align}
This limit reveals that for $m_k^{2}(\phi) \gg T^2$ , i.e., for large $y$, these thermal functions are exponentially (Boltzmann-) suppressed. Therefore, any massive new physics excitations that can be integrated out from the theory should have only a limited impact at finite temperatures. As we discuss in the next section, the consistent inclusion of the Boltzmann suppression effects should be carefully considered when improving the perturbative convergence by resumming higher-order diagrams.
\subsection{Resummation methods}
\label{sec:schemes}
As discussed in the Introduction, the perturbative expansion at finite temperatures breaks down as the self-energy receives large corrections from higher-order loop diagrams. In the high-temperature limit, the self-energy contributions of the daisy-type diagrams require $\lambda T^2/m^2 <1$ for the perturbative expansion to make sense. However, assuming an order $\lambda T^2 \phi^2$ correction to the mass, the tree-level and thermal masses should balance each other at the critical temperature ($T_c$) and one should have  $T_c \sim m/\sqrt{\lambda}$. This scaling of $T_c$ means that the one-loop effective potential is not enough to describe the transition as the expansion parameter $\lambda T^2/m^2$ is not small. The idea of resummation techniques is to define a modified thermal mass that cuts off the problematic divergences and regulates the infrared behavior of the theory. Effectively, this is done by dressing the theory with the thermal mass in different schemes.

The starting point for all schemes is to obtain the thermal mass through the gap equation. The gap equation can be defined from the exact one-particle-irreducible (1PI) resummed propagator,
\begin{equation}
    G(\omega,\mathbf{p}) = \frac{1}{\omega^2 + |\mathbf{p}|^2 - m_0^2(\phi) - \Pi(\omega,\mathbf{p};T)} \, \, \, \, ,
\end{equation}

\noindent where $\Pi$ is the 1PI self-energies and $m_0^2$ is the background-field dependent tree-level mass. Requiring the thermal mass to be defined by the pole of the propagator at zero spatial momentum, $G^{-1}(\omega,0)\big|_{\omega \rightarrow M_T(\phi,T)} = 0$, leads to the gap equation,
\begin{equation}
     M_k^2(\phi,T) = m_{0,k}^2(\phi) + \Pi_k (M^2(\phi,T),0;T) \, ,
     \label{eq:gap_eq_2}
\end{equation}
where the index $k$ runs over the propagating dof. Evaluating the gap equation requires solving the non-linear  \Eq{eq:gap_eq_2}. For that, the first step is to obtain the mass from the effective potential,
\begin{equation}
    M_k^2 (\omega,0;T) = U_{ki}(\theta_T) \frac{\partial^2 V_{n, \rm eff}}{\partial \phi_i \phi_j} U^\dagger_{jk}(\theta_T),
\end{equation}

\noindent where $V_{n, \rm eff}$ is the effective potential at $n$-th order in the loop expansion, and $U$ are the rotation matrices that diagonalize the second derivative of the potential. Notice that we assume a general scalar potential that allows for mixing between the scalar fields and the mixing angle $\theta_T$ is temperature dependent.

In one-loop order, we can write the gap equation as
\begin{equation}
    M^2_k (\phi,T) = U_{ki}(\theta_T) \left[ \frac{\partial^2 V_0}{\partial \phi_i \partial \phi_j} + \frac{\partial^2 V_{CT}}{\partial \phi_i \phi_j} + \frac{\partial^2 V_{CW}}{\partial \phi_i \phi_j} + \frac{\partial^2 V_{T}}{\partial \phi_i \phi_j}\right] U^\dagger_{jk}(\theta_T).
\end{equation}

\noindent One method for solving the gap equation is to iterate the calculation of the right-hand side of \Eq{eq:gap_eq_2}. The first iteration corresponds to inserting the tree-level field dependent mass $m_0^2(\phi)$ into the one-loop effective potential. Then, we obtain a thermal mass matrix that can be diagonalized again, leading to new temperature-dependent mass eigenvalues and mixing angles. Once these are at hand, one needs to insert again into the right-hand side of \Eq{eq:gap_eq_2} and iterate the process until the thermal mass converges. The convergence of this procedure is a notorious challenge as the non-linearity of the gap equation can lead to diverging and oscillatory behavior (see \cite{Curtin:2016urg} for a detailed discussion). 

Instead of solving the full gap equation, which can be numerically demanding, one can truncate the expansion at the first iteration. Then, it is possible to obtain a closed following form for the thermal mass using the high-temperature approximation of the $J_{B,F}$ functions:
\begin{alignat}{5}
    M^2_{ij}(\phi,T) & \simeq m_{0,ij}^2(\phi) + \Pi_{ij} (m_0^2(\phi),T)  \qquad &&\text{(Truncated gap eq.)} \, \, .
    \label{eq:truncated_gap_eq}
\end{alignat}
At the high-temperature limit, if the thermal potential is evaluated at the leading order, i.e., considering \Eq{eq:JBapprox} and \Eq{eq:JFapprox}, the squared thermal mass takes the well-known field-independent form of 
\begin{equation}\label{eq:PiT2}
    \Pi^2_{ij} \sim c_{ij} T^2    \qquad \qquad  (\text{High Temperature}),
\end{equation}
where the couplings, $c_{ij}$, are determined by various model parameters. The relation, defined in \Eq{eq:PiT2}, is known as the Truncated thermal mass as high-temperature approximation. As we discuss next, it is fairly simple to develop resummation schemes that can be evaluated analytically with the high-temperature approximation. 
\subsubsection{Arnold-Espinosa Method}
\label{AEpores}
Among the various schemes for the diagrammatic approach of resumming higher-order loop thermal contributions and solving the IR problem, one notable method is the AE approach.
This method involves modifying the cubic term of the potential, which originates from the Matsubara zero modes of the bosonic dof in the high-temperature approximation, by incorporating the truncated thermal mass evaluated at this approximation.
This is necessary because only the Matsubara zero mode is associated with the infrared divergence, and this scheme specifically resums these modes to address the IR issue while leaving the hard non-zero modes untouched.

The one-loop finite temperature correction due to the bosonic dof to the potential at the high-temperature  limit can be expressed using equations from \Eq{vt1loop} to \Eq{eq:JBapprox},
\begin{align}
    \nonumber
     V^{\rm HT}_{T}(\phi)= &-\frac{T^4\pi^2}{90}N_B+\frac{T^2}{24}\sum_B m_B^2(\phi) -  \frac{T}{12\pi}\sum_B (m_B^2(\phi))^{3/2} \\
    \label{eq:Vht}&- \frac{1}{64\pi^2}\sum_B m_B^4(\phi)\log\left( \frac{m_B^2(\phi)}{a_B}\right) \, ,
\end{align}
where $N_B$ denotes the total number of bosonic dof. The logarithmic dependence in $V^{\rm HT}_{T}(\phi)$  cancels out when combined with the Coleman-Weinberg correction, defined in \Eq{eq:CW_potential}.
The finite temperature contributions proportional to $m^3$, i.e., nonanalytic in $m^2$, originate only from the Matsubara zero modes. 
The IR problem associated with these terms is cured by performing the resummation  via adding the daisy``ring improvement” term, i.e.  the daisy potential, given by,
\begin{align}
\label{Vdaisy}
    V_{\rm Daisy}^{\rm AE}(\phi, T) = - \frac{T}{12\pi}\sum_i \left( (m_{T,i}^2(\phi, T))^{3/2} - (m_i^2(\phi))^{3/2}\right),
\end{align}
where $m_{T,i}^2(\phi, T) = m_i^2(\phi) + c_i T^2$, is the $i$-th mass-squared eigenvalue of the tree-level mass matrix including the thermal corrections at the high-temperature approximation, as defined in~\Eq{eq:PiT2}.
Thus, the full AE effective potential is
\begin{equation}
    V_{\rm eff}^{\rm AE}(\phi,T)= V_0(\phi) +  V_{\rm CW}(m_i^2(\phi)) +  V_{\rm CT}(\phi) +  V_{T}(m_i^2(\phi), T)  + V_{\rm Daisy}^{\rm AE}(\phi, T).
\end{equation}
Thus, it is essential to understand that the AE resummation scheme is fundamentally based on the high-temperature approximation. Therefore, this resummation method becomes unreliable as it does not take into account the proper Boltzmann-suppression of heavy dof.
\subsubsection{Parwani Method}
\label{Parwanipres}
Another well-known diagrammatic approach to resummation prescription is the Parwani method~\cite{Parwani:1991gq}, where all modes are resummed.
In this scheme, $m_i^2 (\phi)$ is replaced by the truncated thermal mass at high-temperature limit $m_{T,i}^2(\phi, T)$  everywhere in the Coleman-Weinberg correction, defined in \Eq{eq:CW_potential}, and the one-loop thermal correction potential, defined in \Eq{vt1loop}. Thus,
\begin{equation}\label{veffpar}
    V_{\rm eff}^{\rm Par}(m_{T,i}^2,T)=  V_0(\phi) +  V_{\rm CW}(m_{T,i}^2(\phi,T)) +  V_{\rm CT}(\phi) +  V_{T}(m_{T,i}^2(\phi, T))  .
\end{equation}
Similar to the AE method, the Parwani method is also based on the high-temperature approximation and focuses solely on resumming the leading contributions in this limit. 
In the case of Parwani, decoupling the effects of heavy dof is a bit more reliable than AE since the thermal potential $V_T$ in \Eq{veffpar} includes the full $J_B(y)$ function. Therefore, even though the truncated thermal masses inserted in $V_T$ are valid only in the high-temperature approximation, a proper Boltzmann suppression is obtained due to the effect of $J_B(y)$. However, Parwani is still not a fully consistent method at all temperatures since the thermal mass insertion into $V_{\rm CW}$ leads to a contribution to the effective potential that is only valid at high-temperatures. This inconsistency can also lead to heavy modes contributing to the phase transition when they should be effectively Boltzmann-suppressed. Furthermore, since this scheme resums all the Matsubara modes inconsistently, it overcounts higher order corrections which may induce some spurious effects. For a detailed discussion on this, see Refs.~\cite{Parwani:1991gq, Laine:2016hma, Bahl:2024ykv}.
\subsubsection{Full- and Partial-Dressing Methods}
The approaches that rely on substituting the field independent thermal mass $m_i^2 \rightarrow m_{T,i}^2$ are called \textit{Truncated Full Dressing} resumation. Both previous methods of resumming hard thermal loops rely on truncating the thermal mass in the high-temperature approximation to obtain the simple expression, defined in  \Eq{eq:PiT2}. 
Gap resummation provides an alternative to diagrammatic methods for resummation, which can become complex at higher loop orders. Instead of analytically evaluating these diagrams, gap resummation involves calculating the effective potential $V_{1,\rm eff}$
  and solving the ``gap equation'' for the thermal mass. This equation captures the leading contributions from numerous higher-order diagrams, although it does not account for certain sub-leading contributions, such as parts of the two-loop sunset diagram.

One can define the gap equation  for thermal mass as,
\begin{align}\label{eq:gap_2hdm}
    M_k^2 (\phi,T) = U_{ki}(\theta_T) \frac{\partial^2 V_{\rm eff}(M^2 (\phi,T))}{\partial \phi_i \partial \phi_j} U^\dagger_{jk}(\theta_T),
\end{align}
where the effective potential is 
\begin{equation}
    V_{\rm eff}(M^2(\phi,T))=  V_0(\phi) + V_{\rm CT}(\phi) +  V_{\rm CW}(M^2(\phi,T)) +  V_{T}(M^2(\phi,T),T) \, \, \, .
\end{equation}
Note that the thermal mass appears on both the left- and right-hand sides of this equation, requiring a numerical approach for its solution.
This procedure can be truncated at a given order and the  
 leading order, the truncated squared thermal mass leads 
 to the eigenvalues of \Eq{eq:thermalmassmatrix}.  As previously mentioned, in the high-temperature limit, this leads to \Eq{eq:PiT2}.

It is crucial to emphasize that truncating the expansion and employing the high-temperature approximation is not universally applicable. In scenarios involving an SFOPT, finite field excursions can become comparable to the temperature itself, such that $\phi\sim T$, Under these circumstances, the field-dependent masses associated with the field $\phi$, may no longer be small at the tree level in comparison to the thermal effects, rendering the high-temperature approximation invalid. As the tree-level field-dependent masses increase significantly, they should decouple smoothly from the thermal plasma. Therefore, assessing the field-dependent thermal mass beyond the leading order in temperature is essential. We will explore this issue in detail in Sec.~\ref{sec:schemes}.

In the FD resummation prescription, the field-dependent thermal masses $M_i^2$, determined by solving the gap equations, are directly incorporated into the effective potential. This results in the modified effective potential given by $V_\text{eff}^{\text{FD}} = V_\text{eff}\Big( M^2(\phi,T) \Big)$ where $M^2(\phi,T)$ is the solution of the gap equation \eqref{eq:gap_2hdm}.
This becomes identical to the one-loop effective potential in the Parwani scheme, $V_{\rm eff}^{\rm Par}$, as given in Eq.~(\ref{veffpar}) when the truncated thermal mass at the high-temperature approximation are considered.
These two schemes  differ in general, as the FD schemes use the 
thermal mass from the full solutions of the gap equation, which includes various higher-order diagrams.

While the FD prescription avoids the need to analytically evaluate leading-order diagrams, 
it also encounters several challenges. Starting at two-loop order, certain higher-order diagrams, such as the sunset diagram, are not automatically incorporated and must be added manually. More critically, the FD prescription 
has been shown to inaccurately account for daisy and superdaisy diagrams beginning at two loops. An alternative method that effectively resums the dominant contributions at higher orders is the PD prescription, first introduced in \cite{Boyd:1993tz} as \textit{tadpole resummation}. Instead of directly substituting $m_i^2 \rightarrow M_i^2$
 in the effective potential, the PD prescription applies the substitution to the first derivative of the effective potential, $\partial_\phi V_\text{eff}$.
Then, the resummed effective potential is obtained via the integration,
\begin{equation}
    V_\text{eff}^\text{PD} = \int d\phi \left( \frac{\partial V_\text{eff}(m_i^2(\phi),T)}{\partial \phi} \right)_{m_i^2(\phi) \rightarrow M_i^2(\phi,T)} \,,
\end{equation}
where, $M_i^2(\phi,T)$ denotes the thermal mass of the $i$-th dof obtained from the full solution of the gap equation \eqref{veffpar}.
This scheme involves dressing only the propagator and has been explicitly shown through calculations up to four loops to correctly account for daisy and superdaisy diagrams \cite{Boyd:1993tz}.
However, this scheme also misses a class of subleading diagrams starting at the two-loop level, as discussed at the end of Sec.~\ref{EWPTPD} in the context of the present work.
\section{The two Higgs doublet model}
\label{2hdm}
In this section, we review the Higgs sector of the $CP$-conserving 2HDM scenario. The tree-level potential  is given by,
\begin{align}
V_{0}&=m_{11}^{2}\left|\Phi_{1}\right|^{2}+m_{22}^{2}\left|\Phi_{2}\right|^{2}-m_{12}^{2}\left(\Phi_{1}^{\dagger}\Phi_{2}+\text{h.c.}\right)+\frac{\lambda_{1}}{2}\left(\Phi_{1}^{\dagger}\Phi_{1}\right)^{2}+\frac{\lambda_{2}}{2}\left(\Phi_{2}^{\dagger}\Phi_{2}\right)^{2} \notag \\
&+\lambda_{3}\left(\Phi_{1}^{\dagger}\Phi_{1}\right)\left(\Phi_{2}^{\dagger}\Phi_{2}\right)+\lambda_{4}\left(\Phi_{1}^{\dagger}\Phi_{2}\right)\left(\Phi_{2}^{\dagger}\Phi_{1}\right)+\frac{\lambda_{5}}{2}\left[\left(\Phi_{1}^{\dagger}\Phi_{2}\right)^{2}+
\mathrm{h.c.}\right],
\label{Tree-Level-Potential}
\end{align}
where all the parameters are real due to hermiticity and $CP$-conservation.
The term associated with $m_{12}^2$ softly breaks the discrete $Z_2$-symmetry in equation~\ref{Tree-Level-Potential}, $\Phi_1 \rightarrow \Phi_1$, $\Phi_2 \rightarrow - \Phi_2$.
The $\Phi_1$ and $\Phi_1$ doublet-fields can be decomposed around the electroweak vacuum as,
\begin{equation}
\Phi_1 =
\begin{pmatrix}
\phi_1^+ \\
\left(v_1 +  h_1 + \mathrm{i} a_1 \right)
    / \sqrt{2}
\end{pmatrix} \ , \quad
\Phi_2 =
\begin{pmatrix}
\phi_2^+ \\
\left(v_2 +   h_2 + \mathrm{i} a_2 \right)
    / \sqrt{2}
\end{pmatrix}, \quad
\end{equation}
where $v_1$ and $v_2$ are the zero-temperature real $\vevs$ of the $CP$-even neutral parts $h_1$ and $h_2$, respectively, of the two doublets. This also defines the electroweak scale, which is given by $v=\sqrt{v_{1}^{2}+v_{2}^{2}}\approx 246 \GeV$. 
The minimization conditions along the $h_1$ and $h_2$	field directions can be used to trade $m_{11}^2$ and $m_{22}^2$ for $v_1$ and $v_2$. These conditions are expressed as:
\begin{align}
\label{tadpolecond}
    m_{11}^2-m_{12}^2\frac{v_2}{v_1}+\lambda_1 v_1^2 + \lambda_{345}v_2^2=0 \, ,\\
    m_{22}^2-m_{12}^2\frac{v_1}{v_2}+\lambda_2 v_2^2 + \lambda_{345}v_1^2=0 \, ,
\end{align}
where, $\lambda_{345}=\lambda_3+\lambda_4+\lambda_5$.
Since all parameters are real, there are no bilinear mixing terms of the form  $h_i a_j$, ensuring that the neutral mass eigenstates are $CP$-eigenstates.
After electroweak spontaneous symmetry breaking (SSB), the particle spectrum consists of two CP-even neutral scalars ($h$ and $H$), one CP-odd neutral pseudoscalar ($A$), a pair of charged scalars ($H^{\pm}$), and three massless Goldstone bosons: one neutral $G^0$ and two charged ($G^{\pm}$). These Goldstone bosons are subsequently absorbed as the longitudinal polarization modes of the $Z$ and $W^{\pm}$ bosons, respectively.

Orthogonal rotational matrices can be used to estimate the relations of the 
masses and gauge eigenstates. The charged and $CP$-odd sectors can be diagonalized using the same orthogonal matrix with the rotation angle $\beta$, where $\tan\beta\equiv v_2/v_1$. The rotational angle for the $CP$-even sector is $\alpha$. These rotational matrices, mass relations, and eigenstates are discussed in detail in  Appendix~\ref{modelrealtions}.
These mixing angles $\alpha$ and $\beta$ control the coupling strength of the scalar particles to fermions and gauge bosons~\cite{Branco:2011iw}. Therefore,
instead of the eight parameters in the Higgs potential $m_{11}^2$, $m_{22}^2$, $m_{12}^2$, $\lambda_1...\lambda_5$, it is convenient to phenomenologically
study the model in terms of the physical masses of the scalar particles and the mixing angles, 
\begin{equation}
  \tan\beta,~\cos (\beta-\alpha),~m_{12}^2,~v,~m_h,~m_H,~m_A,~m_{H^\pm} \,.
\end{equation}
The conversion relations are given in equations~\ref{tadpolecond} and \ref{conversionrelation}. 
Since the discovered Higgs boson at the LHC around $m_h = 125$~GeV mostly follows the properties of the SM Higgs boson, we remain in the so-called ``alignment limit" $\cos(\beta-\alpha) = 0$ to comply with various experimental constraints.
In this limit, at the leading order, the couplings of 
$h$ to the SM particles match the predictions of the SM precisely. Deviations from the SM values in the couplings of $h$ start to arise when $\cos(\beta-\alpha) \neq 0$.
As discussed earlier, the $Z_2$ discrete symmetry imposed on the potential in Equation~\ref{Tree-Level-Potential} prevents Higgs-mediated tree-level flavor-changing neutral currents (FCNCs). Among the four independent implementations of this symmetry in the fermion (Yukawa) sector, we focus on the specific case: 
the Type-II scenario, where $\Phi_1$ couples to down-type SM fermions while $\Phi_2$ interacts with up-type SM fermions~\cite{Aoki:2009ha}.
Various theoretical and experimental constraints relevant to the present work in the context of the 2HDM are discussed below.
%
%
\subsection{Theoretical constraints}
\label{theoryconst}
The tree-level stability conditions for the 2HDM potential, as defined in Equation~\ref{Tree-Level-Potential}, ensure that the potential remains bounded from below. These conditions are expressed as follows:
\bea
& \lambda_1, \lambda_2 > 0, \quad \lambda_3 + \lambda_4 - |\lambda_5| > -\sqrt{\lambda_1 \lambda_2}, \quad
\lambda_3 > -\sqrt{\lambda_1 \lambda_2}, & \label{stability}\ .
\eea
Additionally, constraints on the quartic couplings $\lambda_i$, or specific combinations of them, can be derived from the requirements of unitarity and perturbativity of the $S$-matrix. These bounds are discussed in detail in Refs.~\cite{Grinstein:2015rtl, Akeroyd:2000wc, Ginzburg:2005dt, Bahl:2022lio}.
Regarding perturbativity bounds, Refs. \cite{Biekotter:2021ysx,Biekotter:2022kgf} performs a full renormalization group (RG) analysis, tracking the running of the quartic couplings under one- and two-loop RGEs. They verify that all couplings remain well within the perturbative regime ($\lambda_i(\mu) < 4\pi $) across the relevant energy range. In our work, we stay within the same parameter space and preserve perturbativity under RG evolution.
Overall, we focus on the region of parameter space where all these constraints are satisfied.
%
\subsection{The experimental constraints}
\label{sec:LHCbounds}
In this section, we review the latest experimental constraints on the 2HDM parameter space, focusing on those arising from the Higgs sector. These considerations guide us in selecting a viable region of parameter space for this study.

Electroweak precision data (EWPD), particularly the $T$ parameter, impose restrictions on the mass differences between the charged Higgs boson and either the pseudoscalar or the heavy CP-even Higgs boson. To preserve custodial symmetry in the Higgs sector, one of the neutral states should approximately match the charged Higgs boson in mass~\cite{Gerard:2007kn,Haber:2010bw}. In this work, we assume mass degeneracy between the heavy charged Higgs boson and the pseudoscalar, i.e., $m_{H^{\pm}} = m_A$~\footnote{This condition also implies $\lambda_3 = \lambda_4$.}, to satisfy the EWPD constraints.

In the flavor sector, measurements of $BR(B \rightarrow X_{s} \gamma)$~\cite{Belle:2016ufb} exclude charged Higgs masses below approximately $m_{H^\pm}\lesssim 580$~GeV~\cite{Misiak:2017bgg} for the type-II 2HDM scenario. To comply with these constraints, we restrict our analysis to the parameter space where $m_{H^\pm}\gtrsim 600$~GeV for the type-II scenario.
Additionally, direct searches for heavy Higgs bosons at the LHC have already excluded lower mass regions. For smaller values of $\tanb$, doublet-like heavy Higgs bosons can still remain relatively light while satisfying current LHC constraints~\cite{Bagnaschi:2018ofa}. The exclusion limits are typically presented in the $m_{H^{\pm}}-\tanb$~\cite{ATLAS:2021upq} and $m_A-\tanb$~\cite{CMS:2018rmh, CMS:2019bnu, ATLAS:2020zms} planes. In this work, we consider $\tanb=3$ and $m_H > 350$~GeV~\footnote{Note that, one of the co-positivity condition $\lambda_3 + \sqrt{\lambda_1 \lambda_2} > 0$, defined in equation~\ref{stability}, implies $m_{H}^2 < m_{A}^2 + m_{h}^2$.} to ensure compliance with these constraints.

The observed Higgs boson around 125~GeV also imposes significant restrictions on the 2HDM parameter space through measurements of its signal rates. These measurements strongly favor the alignment limit, where the couplings of the light Higgs are SM-like. To align with these observations, we adopt the limit $\cos(\beta-\alpha) = 0$, ensuring that the tree-level couplings of the light  Higgs boson, $h$, resemble those of the SM. However, even in the alignment limit, the loop induced Higgs boson decay processes, such as $h \rightarrow \gamma \gamma$ can alter significantly due to the presence of the charged Higgs in the model.
The signal strength parameter of 
$h^0 \rightarrow \gamma \gamma$ channel is defined as,
\begin{equation}
\label{mugammagamma}
\mu_{\gamma \gamma} = \frac{\Gamma^{\text NP}[h \rightarrow \gamma \gamma ]}{\Gamma^{\text SM}[h \rightarrow \gamma \gamma ]} \, \, ,
\end{equation}
where $\Gamma^{\text{NP (SM)}}[h \rightarrow \gamma \gamma ]$ represents the decay width of the Higgs boson in the presence (absence) of new physics contributions. 
Since we assume $\cos(\beta-\alpha) = 0$, the production cross-section of $h$ remains unchanged even when new physics effects are included. For a more detailed discussion of $\Gamma(h \rightarrow \gamma \gamma)$ within the 2HDM scenario, we refer the reader to Refs.~\cite{Posch:2010hx, Djouadi:2005gj, Branco:2011iw}. The most recent experimental constraints on $\mu_{\gamma \gamma}$ from ATLAS and CMS are reported as $1.04^{+0.10}_{-0.09}$~\cite{ATLAS:2022tnm} and $1.12 \pm 0.09$~\cite{CMS:2021kom}, respectively.
By adhering to these constraints, the chosen parameter space remains consistent with current experimental data, allowing us to explore the phenomenology of the 2HDM in a valid and meaningful way.
\section{
Results on the Electroweak phase transition in the 2HDM}
\label{results}
With the setup described above, we can compute the effective potential in different resummation schemes. Our focus is on PD, which has not been previously implemented in the 2HDM. As we will show, a proper implementation of PD resolves many of the inconsistencies found in the AE and Parwani methods. The main challenge in PD arises from the extended scalar sector, where mixing terms leads to a non-linear system of coupled gap equations. To solve these equations, we keep only the CP-even Higgs directions as background fields, taking care to account for the relevant effects of other dof in the process. We assume that the CP odd and charged fields do not acquire a finite temperature $\vev$, ensuring that no new minima appear in other directions. Then, we can express the gap equation entirely in terms of masses, their derivatives, and mixing angles. This form allows for numerically iterating the equation until convergence. We obtain thermal masses that remain valid at any temperature. With these thermal masses, we compute the tadpole potential and scan the relevant parameter space. In the following, we describe this procedure in detail, comparing different resummation schemes regarding their impact on the phase transition and the predicted GW spectra.
\subsection{Thermal mass from the gap equation}
\label{TMGAP}
From the previous section, the gap equation is given by
\begin{equation}
    M^2_k (\phi,T) = U_{ki}(\theta_T) \left[ \frac{\partial^2 V_0}{\partial \phi_i \partial \phi_j} + \frac{\partial^2 V_{CT}}{\partial \phi_i \phi_j} + \frac{\partial^2 V_{CW}}{\partial \phi_i \phi_j} + \frac{\partial^2 V_{T}}{\partial \phi_i \phi_j}\right] U^\dagger_{jk}(\theta_T).
     \label{eq:gap_eq_3}
\end{equation}
The strategy for using the iteration procedure is to fully write the right-hand side of the gap equation as a function of masses, $m_k^2(\phi)$, and mixing angles. To do that, we need to express the second derivatives of the CW and the thermal potential that enter the self-energy $\Pi_k$ as functions of the masses and their derivatives. The CW second derivative is given by
\begin{align}
    \nonumber 
    \frac{\partial^2 V_{CW}}{\partial \phi_a \partial \phi_b} = \frac{1}{64\pi^2}\sum_{k}(-1)^{2s_k} n_k &\left[ m_k^2 \frac{d^2m_k^2}{d\phi_a d\phi_b}\left( 1-2c_k+2\log\frac{m_k^2}{\mu^2}\right)\right.
    \\
    & + \left. \frac{dm_k^2}{d\phi_a}\frac{dm_k^2}{d\phi_b} \left( 3-2c_k +2\log \frac{m_k^2}{\mu^2} \right)\right],
\end{align}

\noindent and the second derivative of the thermal potential is

\begin{align}
    \frac{\partial^2 V_{T}}{\partial \phi_a \partial \phi_b} = \frac{T^2}{2\pi^2} \sum_k (-1)^{s_k} n_k  &\left[ \frac{d^2m_k^2} {d\phi_a d\phi_b} J'_{B,F}\left(\frac{m_k^2}{T^2} \right) + \frac{1}{T^2}\frac{dm_k^2}{d\phi_a}\frac{dm_k^2}{d\phi_b} J''_{B,F}\left(\frac{m_k^2}{T^2}\right)\right],
\end{align}

\noindent where $J'_{B,F}(y^2)$ and $J''_{B,F}(y^2)$ are the first and second derivatives of the $J_{B,F}(y^2)$ functions, defined in  \Eq{JBF}, that are straightforward to evaluate numerically. Therefore, the gap equation is a function of the following variables
\begin{align}
    \bullet \hspace{3mm} &  m_{0,k}^2(h_1,h_2) \, \, , \qquad ~~~ ~~~~k = h, H, G_0, A,G^\pm, H^\pm 
    \\
    \bullet \hspace{3mm} &  \frac{d m_{0,k}^2}{d\phi_a}(h_1,h_2) \, \, , \qquad ~~~\phi_a = h_1, h_2, a_1, a_2, \phi_1^\pm, \phi_2^\pm
    \\
    \bullet \hspace{3mm} &  \frac{d^2 m_{0,k}^2}{d\phi_a d\phi_b}(h_1,h_2) \, \, , \quad ~~~\phi_{a,b} = h_1, h_2, a_1, a_2, \phi_1^\pm, \phi_2^\pm
    \\
    \bullet \hspace{3mm} &  \theta_T(h_1,h_2,T) \, \, .
\end{align}

\noindent Notice that we need to calculate the field derivatives of $m_{0,k}^2(h_1,h_2)$ over all fields dof, including the CP even and charged ones. This leads to difficulty since we only have the CP even field dependence on $m_{0,k}$. To overcome this issue, we can use the Feynmann-Hellmann theorem, often used in quantum mechanics, to allow us to compute derivatives of mass eigenvalues from information coming from the mass matrices. Next, we describe how to obtain each variable of the gap equation.

In the 2HDM, the field-dependent mass matrix is an $8\times 8$ matrix which factorizes into block diagonal form if we keep only the $h_1$ and $h_2$ field directions,
\begin{equation}
    \mathbb{M}_0^2(\phi)\Big|_{h_1,h_2} =
    \left(
    \begin{array}{c|c|c|c}
        \begin{array}{cc}
            M^2_{H_{11}} & M^2_{H_{12}}
            \\
            M^2_{H_{12}} & M^2_{H_{22}}
        \end{array}
        & \mathbb{0} & \mathbb{0} & \mathbb{0}
        \\[4mm] \hline 
        \mathbb{0} & 
        \begin{array}{cc}
            M^2_{A_{11}} & M^2_{A_{12}}
            \\
            M^2_{A_{12}} & M^2_{A_{22}}
        \end{array}
        & \mathbb{0} & \mathbb{0} 
        \\[4mm] \hline 
        \mathbb{0} & \mathbb{0} &
        \begin{array}{cc}
            M^2_{H^{+}_{11}} & M^2_{H^{+}_{12}}
            \\
            M^2_{H^{+}_{12}} & M^2_{H^{+}_{22}}
        \end{array}
        & \mathbb{0} 
        \\[2mm] \hline
        \mathbb{0} & \mathbb{0} & \mathbb{0} &
        \begin{array}{cc}
            M^2_{H^{-}_{11}} & M^2_{H^{-}_{12}}
            \\
            M^2_{H^{-}_{12}} & M^2_{H^{-}_{22}}
        \end{array}
    \end{array} 
    \right)    \, \, \, ,
    \label{eq:mass_matrix_2HDM}
\end{equation}
each entry in the matrix is a function of only $h_1$ and $h_2$ with the expressions given in  Appendix~\ref{app:fielddepmasses}. The field-dependent mass eigenvalues are given by 
\begin{align}
    m_{k,0}^2 (h_1,h_2)&= \Big[ U^{-1}(\theta_0) ~\mathbb{M}_0^2(\phi)\Big|_{h_1,h_2} ~U(\theta_0) \Big]_{kk},
    \\
    \label{eq:mass_eigen} & = \text{diag}\big(m_{0,h}^2, m_{0,H}^2, m_{0,G_0}^2, m_{0,A}^2, m_{0,G^+}^2, m_{0,H^+}^2, m_{0,G^-}^2, m_{0,H^-}^2 \big)_{kk},
\end{align}

\noindent while the mixing angles of each block, $\theta_0=\{\theta_{0,H},\theta_{0,A},\theta_{0,H^\pm}\}$, and mixing matrix are
\begin{align}
    \label{eq:mix_angles} & \theta_{0,i} = \frac{1}{2}\arcsin{\frac{2M_{i,12}^2}{\sqrt{(M_{i,11}-M_{i,22})^2 - 4M_{i,12}^2}}},
    \\
    &U(\theta) = \text{diag}\big( U_{2\times2}(\theta_{H}),~U_{2\times2}(\theta_A),~U_{2\times2}(\theta_{H^\pm})\big),
    \\[2mm]
    &U_{2\times2}(\theta_i) = 
    \begin{pmatrix}
        -\sin\theta_i & \cos\theta_i 
        \\
        \cos\theta_i & \sin\theta_i
    \end{pmatrix}, \quad i={H,A,H^{\pm}}.
\end{align}

Now, to get the derivatives of the mass eigenvalues \eqref{eq:mass_eigen}, the first step is to find the derivatives of the mass matrix,
\begin{align}
    \frac{d \mathbb{M}_0^2(\phi)}{d\phi_a} \Bigg|_{h_1,h_2}, \quad \frac{d^2 \mathbb{M}_0^2(\phi)}{d\phi_a d\phi_b} \Bigg|_{h_1,h_2}, \qquad \phi_{a,b} = h_{1},h_2,a_{1},a_2, \phi_{1}^\pm, \phi_2^\pm,
    \label{eq:d_mass_matrix}
\end{align}
\noindent where the field dependence of the CP odd and charged fields must be set to zero only after calculating the derivative. The resulting matrices are block diagonal as \Eq{eq:mass_matrix_2HDM}. The Feynman-Hellmann theorem, described in Appendix~\ref{app:Feyn}, allows us to calculate the derivative of the mass eigenvalues from the derivatives of the mass matrices \eqref{eq:d_mass_matrix} and the mixing angles \eqref{eq:mix_angles}. The expressions for the first and second derivatives are

\begin{align}
    \frac{dm_k^2}{d\phi_a} = \left( U^{-1}(\theta) \cdot \frac{d \mathbb{M}_0^2}{d\phi_a}\Big|_{h_1,h_2} \cdot U(\theta) \right)_{kk} \, \, ,
    \label{eq:FH_dm}
\end{align}
\begin{align}
    \nonumber \frac{d^2 m_k^2}{d\phi_a d\phi_b} =& \left( U^{-1}(\theta) \cdot \frac{d^2 \mathbb{M}_0^2}{d\phi_a d\phi_b}\Big|_{h_1,h_2} \cdot U(\theta) \right)_{kk} + \left( \left[U^{-1}(\theta) \cdot \frac{d \mathbb{M}_0^2}{d\phi_a}\Big|_{h_1,h_2} \cdot U(\theta) \right] \cdot \mathbb{A}_b\right)_{kk} 
    \\
    &\hspace{3cm}+\left( \left[U^{-1}(\theta) \cdot \frac{d \mathbb{M}_0^2}{d\phi_b}\Big|_{h_1,h_2} \cdot U(\theta) \right] \cdot \mathbb{A}_a\right)_{kk} \, \, ,
    \label{eq:FH_d2m}
\end{align}

\noindent and the auxiliary matrices $\mathbb{A}$ are given by
\begin{align}
    (\mathbb{A}_c)_{pq} = 
    \begin{cases}
        0 \, \, , \qquad \text{if } p=q,
        \\
        0 \, \, , \qquad \text{if } p\neq q ~\text{ but }~ m_p^2 = m_q^2,
        \\
        \frac{1}{m_p^2 - m_q^2}\left[U^{-1}(\theta) \cdot \frac{d \mathbb{M}_0^2}{d\phi_c}\Big|_{h_1,h_2} \cdot U(\theta) \right]_{pq}\, \, ,  \qquad \text{else}.
    \end{cases}
\end{align}

With these ingredients, we can calculate the matrix associated with the self-energy and diagonalize it to find the temperature-dependent mixing angles $\theta_T = \{\theta_{T,H}, ~\theta_{T,A}, ~\theta_{T,H^\pm}\}$. The iterative gap equation can be written as a function of $x_{i} \equiv M^2(h_1,h_2,T)\big|_{\text{iteration}=i}$,
\begin{align}
    &x_1 = m_0^2(h_1,h_2) \, \, ,
    \\
    &x_{i+1} = x_i + U^{-1}(\theta_T) ~\Pi(x_i, x_{i,a}, x_{i,ab}) ~U(\theta_T)  \, \, ,
\end{align}

\noindent where $x_{i,a}=x_{i,a}(x_i,\theta_T)=\frac{d x_i}{d\phi_a}$ and $x_{i,ab}=x_{i,ab}(x_i,\theta_T)=\frac{d^2 x_i}{d\phi_a d\phi_b}$ and $\Pi_i$ is the self-energy contribution coming from the second derivatives of $V_{\rm CW}$ and $V_{\rm CT}$. We can insert the found values of $\{x_i,x_{i,a},x_{i,ab},\theta_T\}$ in the right-hand side of \Eq{eq:gap_eq_3} to find the first iteration of the thermal mass $x_{i+1}$. With the resulting thermal mass and mixing angles, we can insert it again at the right-hand side of \Eq{eq:gap_eq_3} to find the second iteration value and continue to higher iterations. Notice that the replacement should also happen in \Eq{eq:FH_dm} and \Eq{eq:FH_d2m} to obtain the higher iterations of the mass derivatives.

\begin{figure}
    \centering
    \includegraphics[width=1\linewidth]{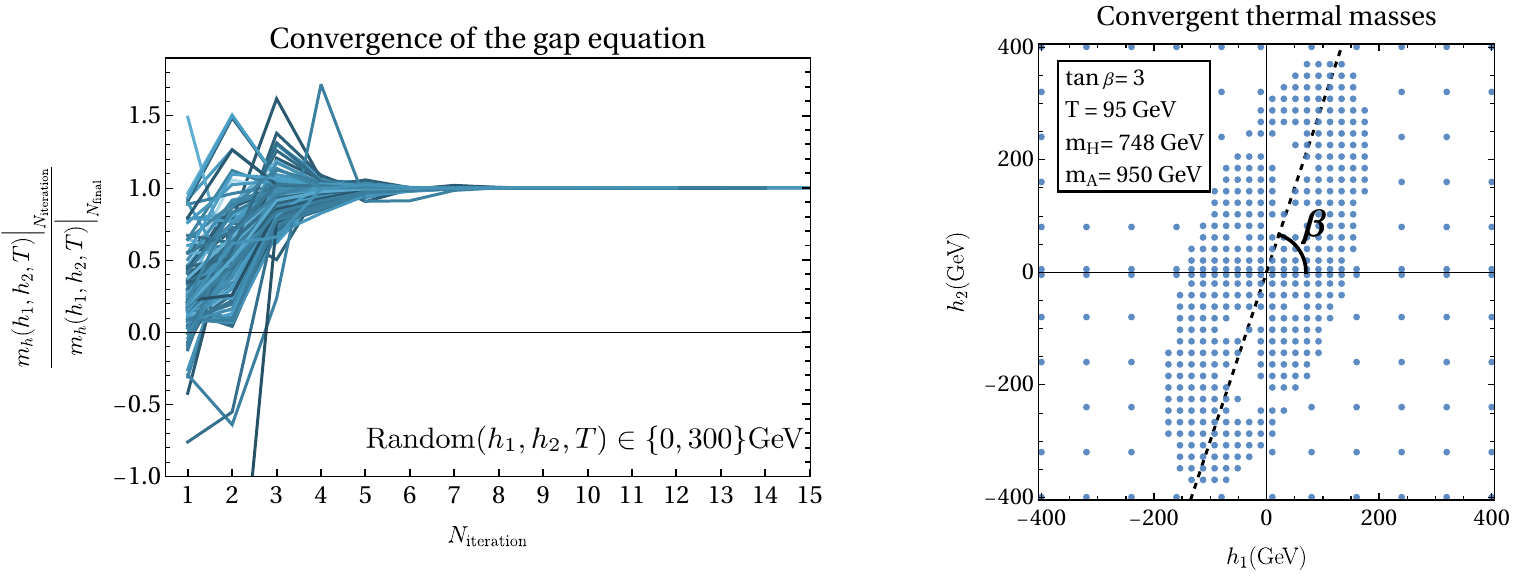}
    \caption{Left: Convergence of the thermal mass obtained from the gap equation as a function of the number of iterations. The thermal mass at each iteration is normalized to its final converged value. The lines represent only the convergent cases from a random sample of $h_1$, $h_2$, and $T$ values ranging from 0 to 300 GeV. Right: Convergent thermal mass points in the $h_1$-$h_2$ plane. Some points along the Higgs direction fail to converge.}
    \label{fig:gap_iter}
\end{figure}

The numerical solution of the gap equation is challenging due to its non-linear structure, which often leads to instabilities. Properly implementing on-shell renormalization conditions is crucial for obtaining well-behaved solutions. The iterative process can yield spurious or divergent results without this careful treatment. A further complication arises from the Goldstone catastrophe. Since our calculation requires the first and second derivatives of the Coleman-Weinberg potential, the presence of massless Goldstone bosons near the minimum induces divergent behavior. To mitigate this issue, we introduce a small infrared cutoff of 1~GeV for the Goldstone masses, ensuring numerical stability, also discussed in Appendix~\ref{CTcoeff}.
However, this infrared regulator must be carefully monitored as small Goldstone masses can lead to ill-behaved thermal corrections in later iterations. To control convergence, we set a maximum number of iterations and a precision cutoff for the iterative procedure, using a moving average over recent iterations to assess stability. We terminate the iteration once the precision reaches $1\%$ for the convergent points. We discard the point if the gap equation does not converge within 40 iterations. The distribution of convergent solutions is shown in the left plot of Fig.~\ref{fig:gap_iter}.

To compute the thermal masses, we specify the values of $(h_1,h_2,T)$ along with the input parameters of the 2HDM. We generate multiple points in the $(h_1,h_2)$ plane for each temperature, focusing on a denser grid around the potential minima. Along the light Higgs direction, where non-trivial extrema appear, the gap equation struggles to find solutions, leading to a failure rate of 5–10\%. This highlights the need for a carefully chosen field resolution to determine thermal masses and effective potential accurately. The right panel of Fig.~\ref{fig:gap_iter} illustrates the typical behavior of the points in the convergent gap equation.  
With the methodology described above, we compute the thermal masses of all scalar dof in the 2HDM. The solutions to the full gap equation exhibit notable features, the most important being the self-consistent inclusion of Boltzmann suppression effects. Unlike standard high-temperature approximations, our approach fully accounts for the thermal dependence of the distribution functions, ensuring accuracy at all temperatures.

\begin{figure}
    \centering
    \includegraphics[width=\linewidth]{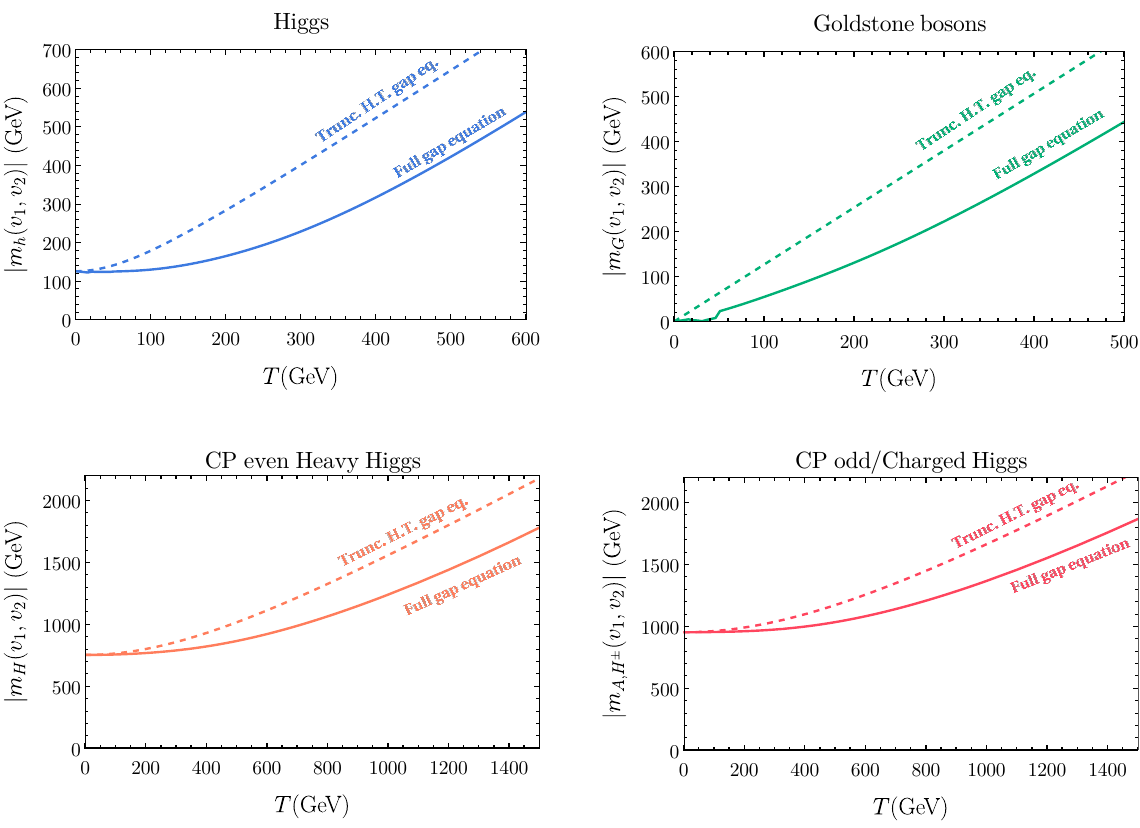}
    \caption{Thermal masses for each scalar dof obtained from the full gap equation after the iteration procedure. The dashed lines represent the high-temperature approximation for comparison.}
    \label{fig:thermalmass}
\end{figure}
\begin{figure}[p]
    \centering
    \includegraphics[width=0.9\linewidth]{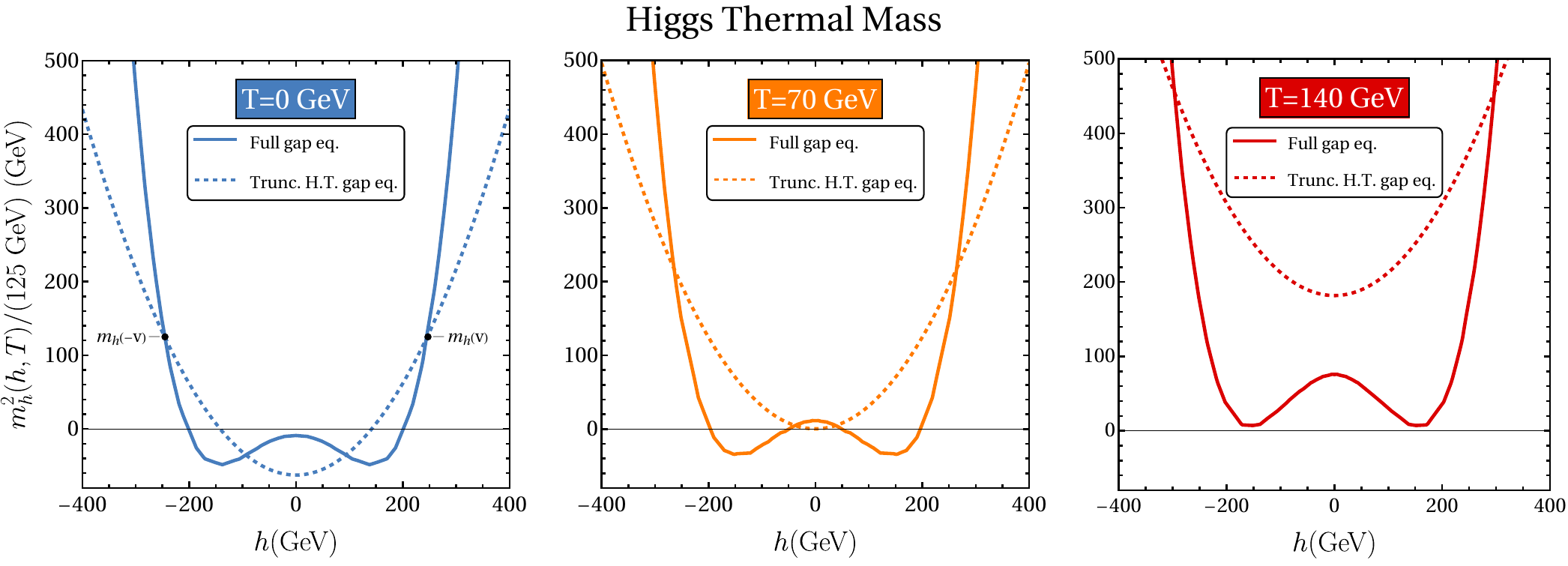} \\[3mm]
    \includegraphics[width=0.9\linewidth]{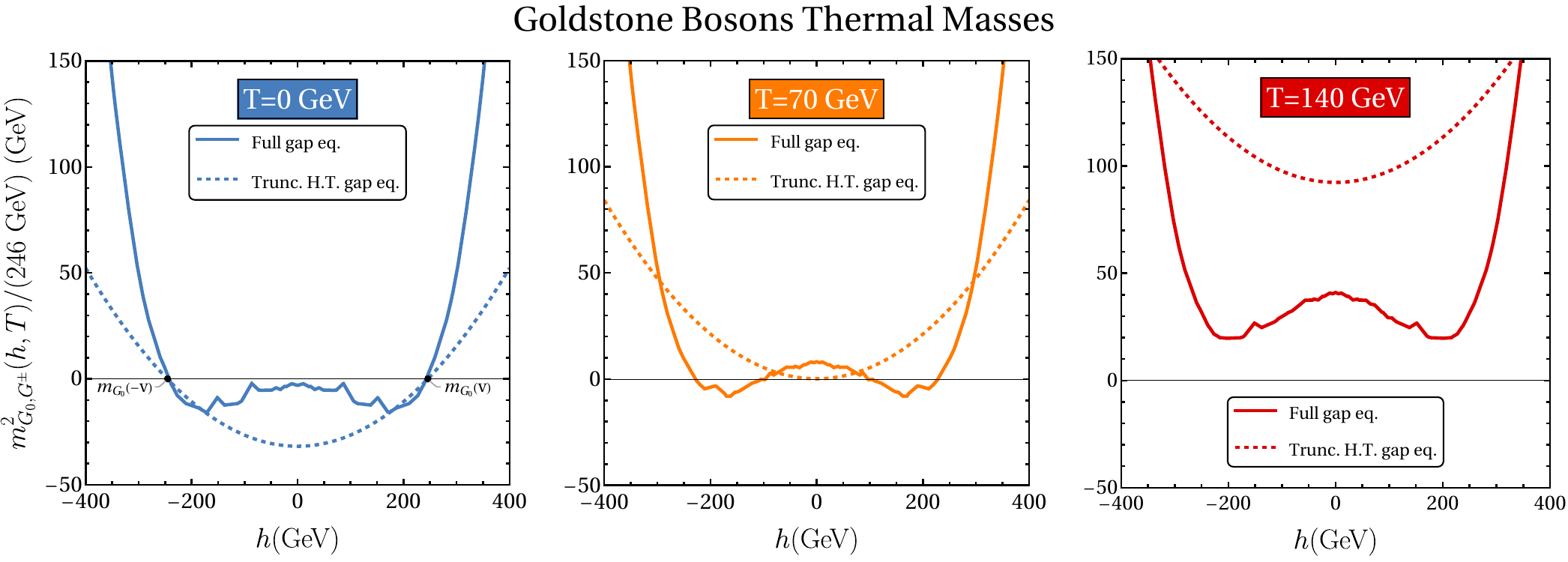} \\[3mm]
    \includegraphics[width=0.9\linewidth]{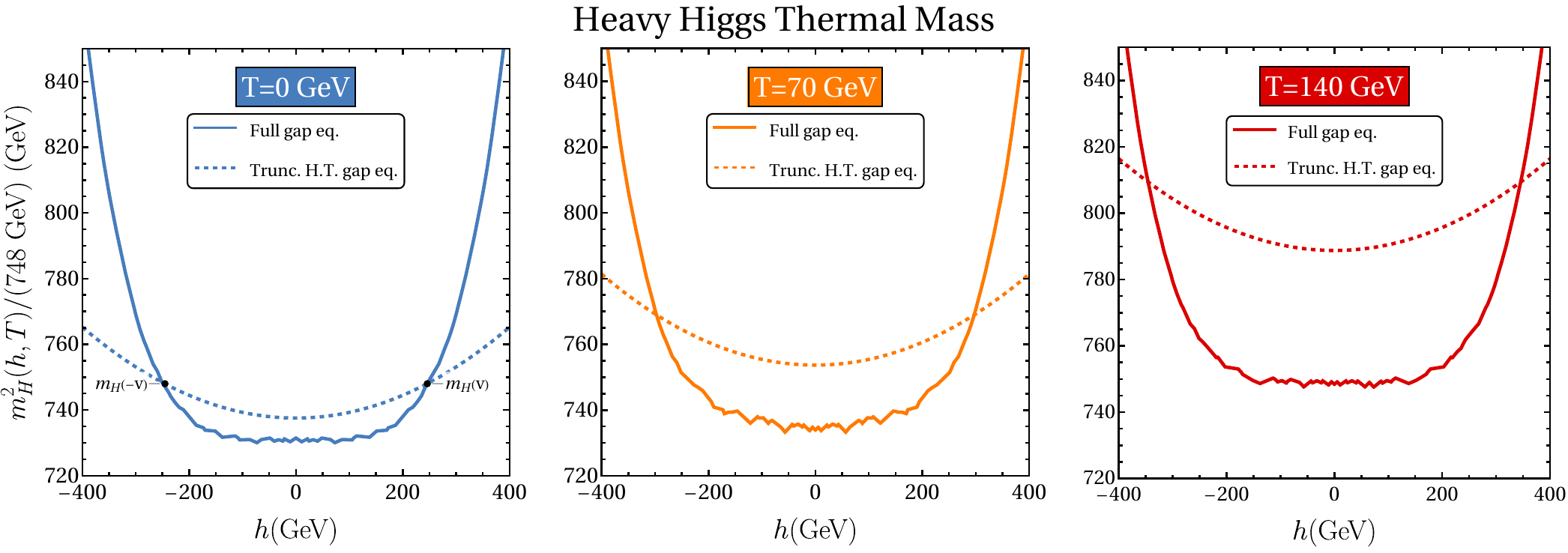} \\[3mm]
    \includegraphics[width=0.9\linewidth]{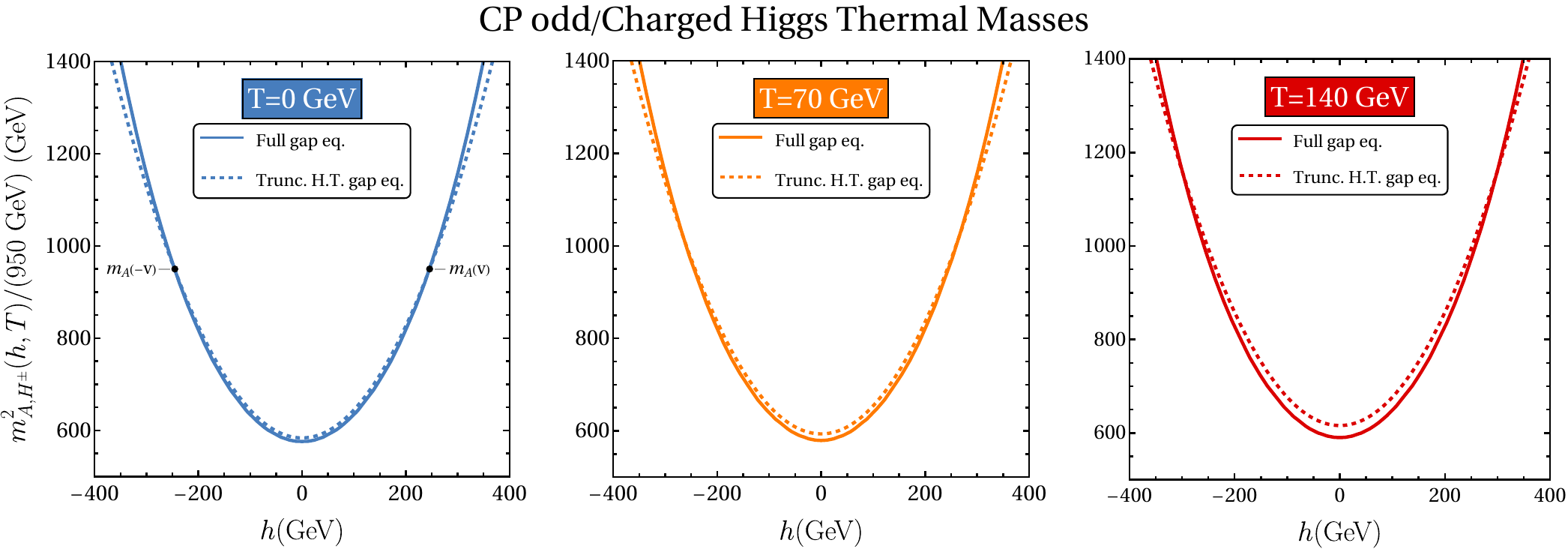}
    \caption{Thermal mass of the scalar dof as a function of the light Higgs field direction $h$, with all other scalar field directions set to zero. The three panels correspond to increasing temperatures. The solid lines are the thermal masses obtained by iterating the gap equation, while the dashed line is the truncated thermal mass in the high-temperature approximation. At $T=0$, the black dots indicate the physical masses at $v=246\GeV$ for this benchmark. The wiggles in the plot are artifacts from numerical resolution, not real physical features.}
    \label{fig:hxmh}
\end{figure}

To illustrate the differences in computing thermal masses for various dof, we compare results obtained using the truncated high-temperature approximation with those derived from solving the full gap equations without relying on this approximation. Figures~\ref{fig:thermalmass} and \ref{fig:hxmh} present these comparisons. 
In Fig.~\ref{fig:thermalmass}, we show the thermal masses of various dof for the benchmark point (BP) BP1, presented in Tab.~\ref{tab:BP-sym-NR}, as functions of temperature for field values $h_1 = v_1$ and $h_2 = v_2$, where $(v_1, v_2) = (77, \, 231)~\GeV$. It is evident that the thermal masses derived from the full gap equations deviate significantly from those obtained via the high-temperature approximation. In particular, heavy dof begins influencing the thermal masses only at sufficiently high temperatures, leading to an overall suppression compared to the naive high-temperature result. Nevertheless, the expected $c T^2$ scaling behavior remains valid at high temperatures, albeit with a modified coefficient `$c$'. 
Furthermore, the full gap equations introduce a non-trivial field dependence in the thermal masses, which is absent in conventional high-temperature treatments. To highlight this effect, Fig.~\ref{fig:hxmh} shows the variation of thermal masses for different dof of the 2HDM along the SM-like Higgs boson field direction, $h$, with all other scalar field directions set to zero, at three fixed temperatures ($T = 0$ (left), 70~GeV (middle) and 140 GeV (right)). For better visualization, the thermal masses squared are normalized to their zero-temperature values at the electroweak minimum. The solid lines represent results from the full gap equations, while the dashed lines correspond to the truncated high-temperature approximation. 
Significant deviations are observed across various field values, even at zero temperature. This arises because the full gap equations incorporate corrections from the Coleman-Weinberg potential as well as higher-order effects through resummation, whereas the truncated high-temperature approximation no longer remains valid in this regime. Importantly, we find that both methods yield the same thermal masses at zero temperature at the electroweak minimum, an outcome of properly incorporating counterterms in the potential, as discussed in Appendix.~\ref{CTcoeff}. 
Moreover, at finite temperatures and certain field regions, the full gap equations exhibit more pronounced deviations, which are particularly relevant for the study of FOEWPT. These deviations become more significant at large field values and high temperatures. Thus, for a precise description of the finite-temperature potential, it is crucial to accurately determine thermal masses by solving the full gap equations, as done in this work. These differences also manifest in the shape and behavior of the effective potential in the PD resummation scheme, in contrast to other resummation prescriptions, as we discuss in the next section.
\subsection{The Electroweak phase transition using Partial Dressing}
\label{EWPTPD}
As discussed, full and PD are the main methods to include higher-order effects and go beyond the high-temperature approximations. In full dressing, one replaces the field-dependent mass $m^2(h_1,h_2)$ with the full thermal mass $M^2(h_1,h_2,T)$ everywhere in the potential $V(\phi)$. This change dresses both propagators and vertices. As a result, some diagrams, such as parts of the daisy and super daisy series, are counted twice. PD avoids this issue by dressing only the propagators. In practice, one first computes the derivative of the effective potential with respect to the field (the tadpole) and dress the field-dependent masses:  
\begin{equation}
\frac{\partial V_1}{\partial h_1}\Bigg|_{m_i^2 \rightarrow M_i^2(h_1,h_2,T)}, \qquad \frac{\partial V_1}{\partial h_2}\Bigg|_{m_i^2 \rightarrow M_i^2(h_1,h_2,T)}  \, \, .
\end{equation}

\noindent After that, one integrates the dressed derivative with respect to $\phi$ to recover the effective potential. This method resums only the self-energy corrections while leaving the vertices undressed. This prevents the double counting that would occur if both propagators and vertices were dressed.

For the two-Higgs-doublet model (2HDM), the first derivative of the potential is taken along several field directions. Since we focus on the CP-even behavior, we only consider the derivatives with respect to $h_1$ and $h_2$. First, the gradient of the one-loop potential is computed, and then the field-dependent masses and mixing angles are replaced by the thermal mass and thermal mixing angles. Following \cite{Bahl:2024ykv}, the potential is obtained by integrating the gradient of the resummed tadpole term along a path $\mathcal{C}$ in field space:  

\begin{equation}
V_{PD} = \int_\mathcal{C} d\vec{s} \cdot \left.\vec{\nabla}V_1\right|_{m_i^2 \rightarrow M_i^2(h_1,h_2,T)}.
\end{equation}

\noindent A simple choice for the path is a straight-line, parametrized by $\vec{s}(t) = (h_1 t, h_2 t)$ with $t \in [0,1]$. The potential then becomes:  
\begin{equation}
V_{PD}(h_1^*,h_2^*) = \int_0^1 dt \, \left( h_1^* \frac{\partial V_1}{\partial h_1}\Bigg|_{(h_1^*t,h_2^* t)} + h_2^* \frac{\partial V_1}{\partial h_2}\Bigg|_{(h_1^* t,h_2^* t)} \right).
\end{equation}
Note that one must replace not only the thermal mass $M_i(h_1,h_2,T)$ but also its first derivatives obtained from the gap equation and the thermal mixing angles.

The parameter space allowed in the $(m_H, m_A)$ plane is scanned to analyze the phase transition behavior across this region.
Since solving the gap equation in the PD scheme is numerically demanding, we restricted the scan to 130 $(m_H, m_A)$ points within the range $m_H \in \{350,900\} \GeV$ and $m_A \in \{600,1000\} \GeV$. 
We limited the scanned parameter space to mass relations that satisfy the tree-level positivity conditions of the potential, as it is observed that the resummation does not alter the metastability or instability of the electroweak vacuum in these regions. For each $(m_H, m_A)$ point, we simulated 30 temperature points and interpolated the results to determine the critical temperatures with high precision. Additionally, to achieve a good resolution of the minimum, we performed a scan over 270 points in the $(h_1, h_2)$ plane. To further refine the resolution, we use a denser grid inside an ellipse that contains the minima, as shown in the right panel of the figure~\ref{fig:gap_iter}. The analysis is implemented in \texttt{Mathematica (v13)}~\cite{Mathematica:2022} script mode to parallelize the calculation.
\begin{figure}
    \centering
    \includegraphics[width=0.7\linewidth]{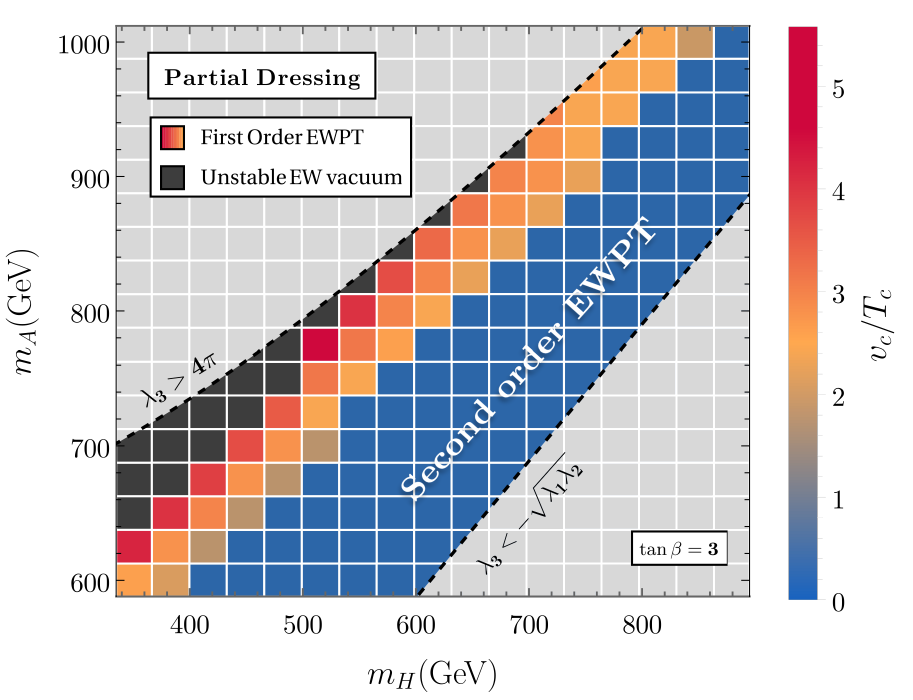}
    \caption{Phase transition behavior in the $m_H-m_A$  plane under the PD resummation scheme. The blue region indicates a second-order phase transition. The transition becomes first order in the yellow to red region with the strength $v_c/T_c$. The black region is excluded as the electroweak minima is metastable/unstable with the global minima at the origin. We remove the points above the dashed line of $\lambda_3 > 4\pi$ where perturbativity of the theory breaks down and below $\lambda_3 <-\sqrt{\lambda_1 \lambda_2}$ where the potential no longer remains bounded from below. The parameter scan covers 130 points in $(m_H, m_A)$, 30 values of $T$ for each $(m_H,m_A)$, 270 points in $(h_1, h_2)$ for each $(T,m_H,m_A)$, and $N_{\text{iter}}$, the number of iterations of the gap equation, varying between 12 and 40.}
    \label{fig:partial_dressing_scan}
\end{figure}

Figure~\ref{fig:partial_dressing_scan} presents the classification of the phase transition behavior across the scanned region, considering the PD resummation prescription in the evolution of the effective potential.
Regions where the EW minimum is not the global minimum at zero temperature or where perturbativity and bounded-from-below conditions are violated are highlighted using different colors. 
Black indicates metastable or unstable EW minima, gray top points above the dashed line represent regions that \( \lambda_3 > 4\pi \) violate perturbativity, and gray bottom points below the dashed line indicate \( \lambda_3 < -\sqrt{\lambda_1 \lambda_2} \), signaling a breakdown of the condition of bounded-from-below for the potential.
Blue indicates the second-order phase transition region.
The strength of the FOEWPT, evaluated at the critical temperature level ($v_c/T_c)$, is shown using a color palette.
Across the entire scanned region, we find that EW symmetry is restored at high temperatures, with no indication of symmetry non-restoration under the PD resummation scheme.
To classify the FOPT points by their strength, it is more appropriate to calculate the nucleation temperature $T_n$, defined as the temperature at which the bubble nucleation rate becomes comparable to the Hubble scale. In addition, to identify the region where the system remains trapped at the origin, known as the vacuum-trapped scenario, it is important to estimate the nucleation rate.
To compute $T_n$, we reduce the two-dimensional field dependence of the potential to a single dependence along the physical Higgs direction. This approximation is well justified since extrema occurs only in this direction, and the potential increases monotonically along the other field direction. The nucleation temperature is then determined using the condition defined in~\Eq{nucleation}
%
%
with the help of the publicly available toolbox \texttt{CosmoTransitions}~\cite{Wainwright:2011kj} that provides a reliable estimate for $T_n$, ensuring that bubble nucleation is efficient enough to complete the phase transition within the timescale set by cosmic expansion. 
\begin{figure}
    \centering
    \includegraphics[width=1\linewidth]{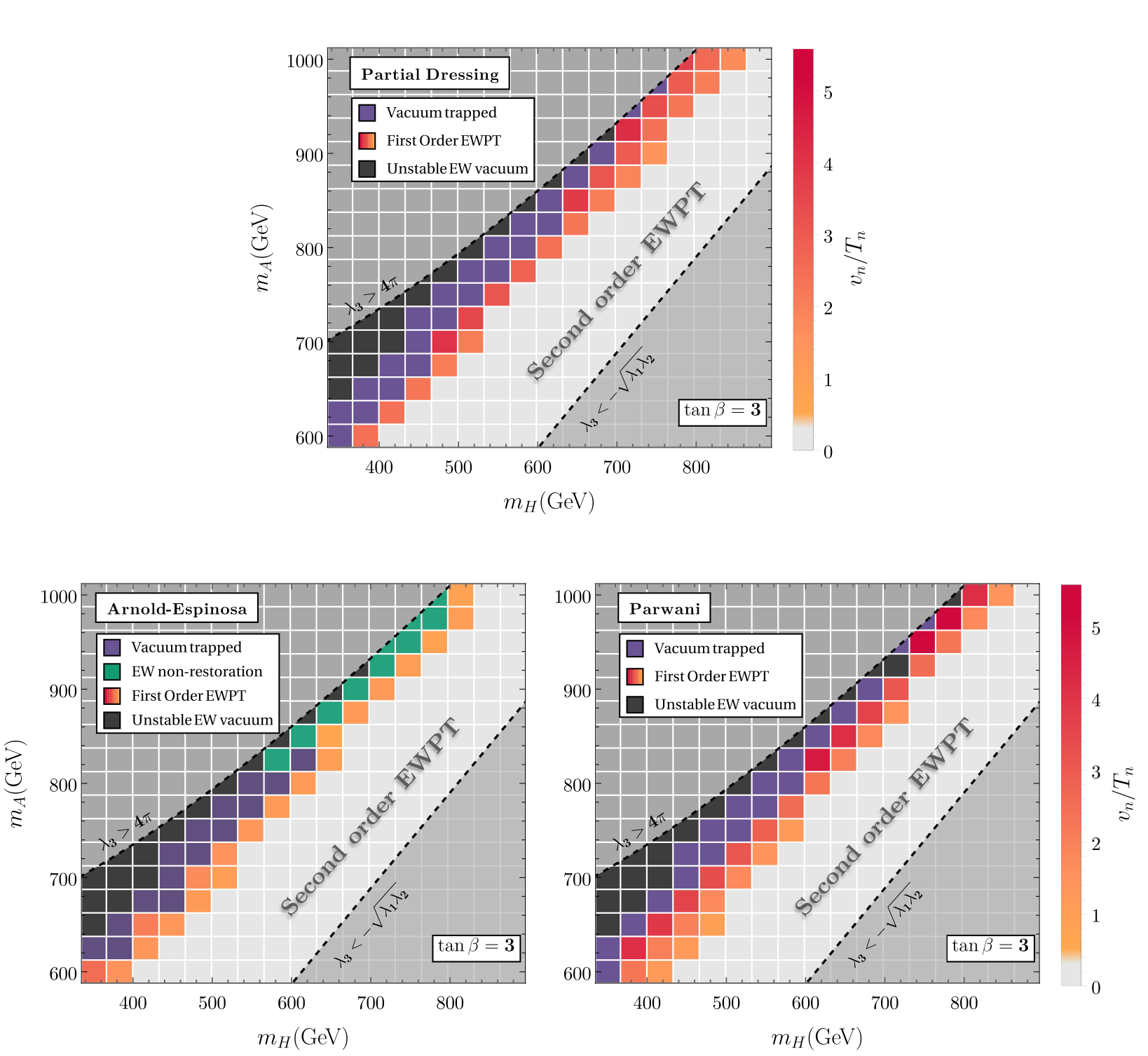}
    \caption{The strength of the FOEWPT, measured by $v_n/T_n$, in the $m_H-m_A$ plane of the 2HDM. 
    In the top panel, we show the results using the PD resummation scheme. The bottom panels present results for the AE (left) and Parwani (right) schemes. In the AE scheme, EWSNR appears as an effect of the resummation procedure rather than a negative thermal mass. Regions labeled ``Vacuum trapped" indicate parameter space where the system does not tunnel to the EW minima.}
    \label{fig:Tnuc_scan}
\end{figure}

In Fig.~\ref{fig:Tnuc_scan}, the phase transition behaviors are compared at the nucleation temperature level, as derived from the PD (top), AE (bottom-left), and Parwani (bottom-right) resummation schemes. We validate our results for the AE and Parwani schemes, obtained through a \texttt{Mathematica}-based analysis, by comparing them with those from \texttt{Cosmotransitions}.
In these plots, the strength of the phase transition at the nucleation level, i.e., $\xi_n = v_n/T_n$, is presented via color palette. 
In the region where $\lambda_3, \lambda_4, \lambda_5$ are relatively large, a barrier exists between the origin and the EW minimum at the zero temperature. This feature appears already at the one-loop effective action at finite temperature and enhances the strength of the phase transition.
However, most parameter points with a zero-temperature barrier are excluded, as they lead to vacuum trapping—the Universe gets stuck at the origin because the tunneling rate is too low to trigger bubble nucleation effectively. Purple points indicate this vacuum-trapped region.
Meanwhile, the AE predicts EWSNR even at high temperatures for certain large $m_H, m_A$ values, shown by the green points in the bottom-left plot, due to an unsuppressed cubic contribution in its prescription. Interestingly, PD and Parwani do not support this conclusion. In Sec.~\ref{SNRsec}, we examine the results for the EWSNR at high temperatures of each resummation method in more detail and discuss the fate of the transition in this region of parameter space.  
 
The strength of the phase transition varies significantly between different resummation methods. This is especially relevant for the AE approach, which predicts a much weaker phase transition than the other methods. In contrast, PD and Parwani produce similar qualitative features, although their quantitative predictions differ. We explore these differences in more detail in Sect.~\ref{sec:LHCandGWs}, particularly in the context of predicting the GW signal from an FOEWPT. 
We also observe a significantly stronger FOEWPT (that is, larger $\xi_n$) in the Parwani prescription compared to the PD scheme. This can be attributed to the fact that in the Parwani prescription, the thermal masses of various dof are determined using a high-temperature approximation, which tends to overestimate them. This overestimation becomes particularly significant when the phase transition is strong, as the $\vevs$ are of the same order as the temperature scale, rendering the high-temperature approximation invalid. In contrast, the PD prescription incorporates the gap equation, leading to a more accurate estimation of the thermal masses. Consequently, the thermal corrections in the PD scheme are relatively smaller than those in the Parwani scheme. As a result, the phase transition in the PD scheme is weaker than that in the Parwani scheme.


Before ending this section, we want to comment on the reliability of the thermal perturbative series and the inclusion of missing higher-order diagrams. 
One common issue of all these resummation schemes is that they miss some higher-order diagrams. Although the PD scheme provides a more efficient way of incorporating higher-order effects, it still misses contributions coming from sunset diagrams. These diagrams first appear at two loops and consist of three propagators forming a loop around the internal propagator. The diagrams involve two independent loop momenta and cannot be obtained by dressing propagators alone. A full two-loop implementation would require an explicit evaluation of thermal loop integrals, which is difficult without the high-temperature approximation.
\\[1mm]
Part of the parameter space of interest for our work corresponds to quartic couplings that are naively large, sometimes approaching $4\pi$. A general concern is that these large couplings could make the missing higher-order diagrams important and undermine the reliability of the calculation. However, the physical states interact through combinations of such couplings. The CP even states are accompanied by the combination $\lambda_3+\lambda_4+\lambda_5$ or the other $\lambda_{1,2}$ couplings which are generically small for the benchmark points we consider. The CP odd and charged states, however, can have combinations that are indeed large, e.g. $\lambda_3-\lambda_4-\lambda_5$. We verified that at two-loops, all sunset diagrams that have potentially large couplings involve two heavy scalars in the loop. Therefore, our results for the thermal corrections should be reliable, as the large coupling two-loop sunset diagrams are expected to be Boltzmann suppressed at the scales relevant for the phase transition. 
A full treatment of these effects, including the explicit evaluation of two-loop effects, is beyond the scope of this work, and it is left for future study. 

\subsection{High-temperature behavior: Non-restoration vs. restoration}
\label{SNRsec}
An intriguing phenomenon, known as EWSNR at high temperatures well above the EW scale, can emerge in the early Universe within certain BSM scenarios.
This possibility is of particular interest for EWBG since EWSNR or delayed restoration can persist well above the electroweak scale. In such scenarios, new physics at a higher scale is required to eventually restore the symmetry via a first-order phase transition. Importantly, the delayed restoration framework can help alleviate stringent CP-violation constraints, as the relevant CP-violating sources are typically associated with heavy new-physics states whose effects on low-energy observables are suppressed~\cite{Weinberg:1974hy,Kirzhnits:1976ts,Mohapatra:1979qt,Salomonson:1984px,Bimonte:1995sc,Bimonte:1995xs,Dvali:1996zr,Pietroni:1996zj,Patel:2013zla,Kilic:2015joa,Ramsey-Musolf:2017tgh,Glioti:2018roy,Baldes:2018nel,Meade:2018saz,Carena:2019une,Kozaczuk:2019pet,Matsedonskyi:2020mlz,Carena:2021onl,Biekotter:2021ysx,Biekotter:2022kgf,Chang:2022psj,Carena:2022qpf,Carena:2022yvx}. These features make EWSNR an attractive avenue for exploring high-scale BSM physics, and motivate a careful re-examination of this phenomenon using different theoretical approaches.

\begin{figure}[t]
     \centering
     \includegraphics[width=0.32\linewidth]{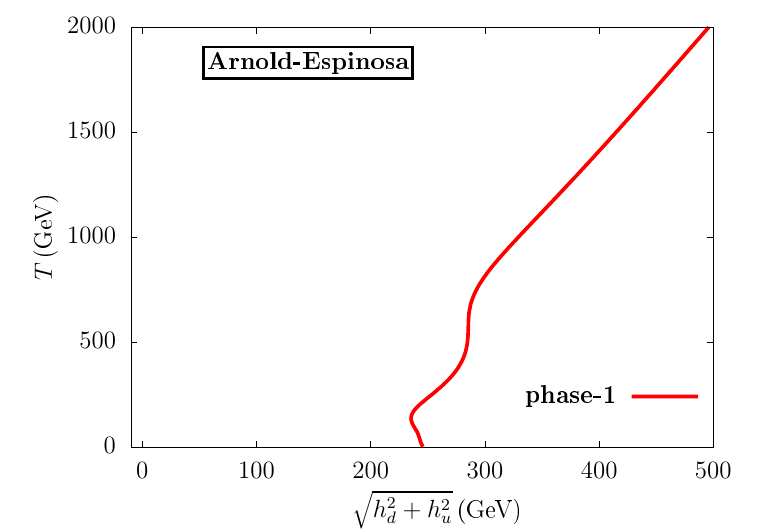}
     \includegraphics[width=0.32\linewidth]{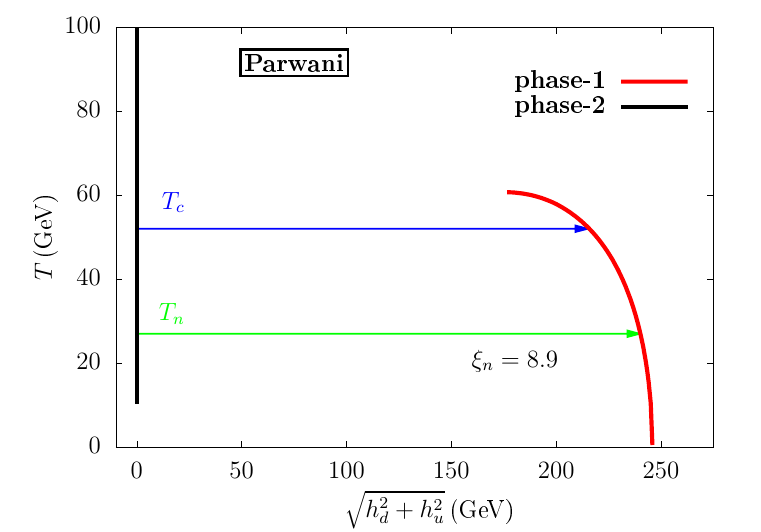}
     \includegraphics[width=0.32\linewidth]{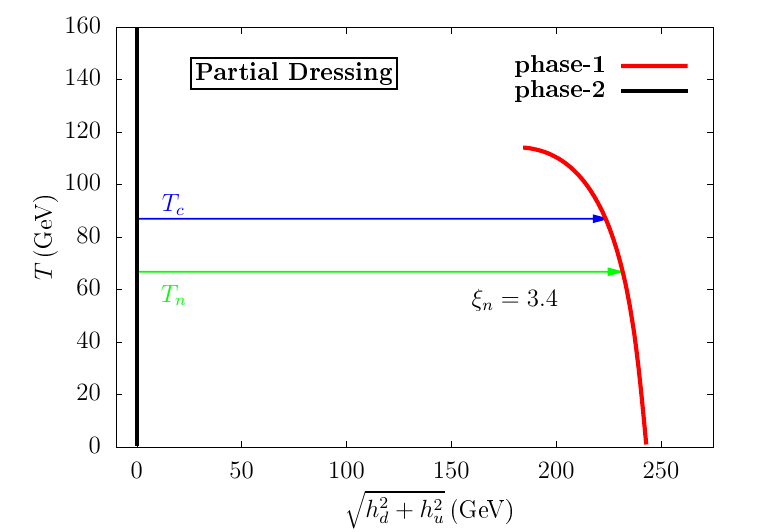}
     \caption{Phase flows for the benchmark scenario BP1, presented in Table~\ref{tab:BP-sym-NR}, considering the AE (left), Parwani (middle) and PD (right) prescriptions are shown. These plots indicate that the AE prescription indicates symmetry non-restoration at high temperatures, whereas the Parwani and PD prescriptions exhibit symmetry restoration and an SFOEWPT in the early Universe. Each color represents a distinct minimum of the potential (phase), and the lines depict phase evolution along the field direction $\sqrt{h_1^2+h_2^2}$ with temperature. Arrows indicate the transition path from the false vacuum to the true vacuum, calculated at  $T_c$ and $T_n$. For an SFOEWPT, the strength of the phase transition, $\xi_n$,  is also mentioned. }
     \label{fig:phasediagrams}
 \end{figure}

Recently, this phenomenon has been investigated in the context of the 2HDM in Ref.~\cite{Biekotter:2022kgf}, where EWSNR was observed at high temperatures in a specific corner of the 2HDM parameter space. In this section, we revisit that parameter space and analyze the finite-temperature potential using different thermal resummation prescriptions. Our results reveal that EW
 symmetry is restored under both the Parwani and PD resummation schemes, even for parameter points that exhibit EWSNR when analyzed using the AE thermal resummation prescription.
To illustrate this, we provide a detailed discussion based on a BP, shown in Table~\ref{tab:BP-sym-NR}.
The phase transition details of different resummation prescriptions are listed in Table~\ref{tab:TcTn}.
Figure~\ref{fig:phasediagrams} illustrates the evolution of the phases (each minimum) with temperature for BP1, with the results shown for the Parwani, AE, and PD prescriptions in the left, middle, and right plots, respectively.
These plots demonstrate that an SFOEWPT occurs in the Parwani and PD schemes, both of which exhibit symmetry restoration at high temperatures. Notably, the results for the Parwani and PD prescriptions show significant differences in \(T_c\)), \(T_n\)), $\xi_n$ and $\alpha$. Specifically, for the Parwani scheme, \(T_c = 52~\mathrm{GeV}\) and \(T_n = 27~\mathrm{GeV}\), whereas for the PD scheme, \(T_c = 88~\mathrm{GeV}\) and \(T_n = 67~\mathrm{GeV}\). In contrast, the AE scheme predicts neither an FOPT nor symmetry restoration at high temperatures.

\begin{table}[t]
    \renewcommand{\arraystretch}{1.2}
    \centering
    \small{\begin{tabular}{|c|c|c|c|c|c|c|c|c|c}
    \hline 
    BP No. & \multicolumn{8}{|c|}{Parameters}
    \\
    \hline  \hline 
    & $m_h$ & \multirow{1}{*}{$m_{H}$} & \multirow{1}{*}{$m_{A}$} & $m_{H^{\pm}}$  & $v$ & $m_{12}^2$ & $\tanb$ & $c_{\alpha-\beta}$ 
    \\
    \hline 
    BP1 & $125\GeV$ & $748\GeV$ & $950\GeV$ & $950\GeV$ & $246\GeV$& $m_H^2 s_\beta c_\beta$ & $3$ & $0$
    \\
    \hline \hline  
    \multirow{1}{*}{}& \multirow{1}{*}{$-m_{11}^2$} & \multirow{1}{*}{$-m_{22}^2$} & \multirow{1}{*}{$m_{12}^2$} & \multirow{1}{*}{$\lambda_1$} & \multirow{1}{*}{$\lambda_2$} & \multirow{1}{*}{$\lambda_3$} & \multirow{1}{*}{$\lambda_4$} & \multirow{1}{*}{$\lambda_5$}
    \\
    \hline
    BP1 & $(704 \GeV)^2$ & $(219\GeV)^2$ & $(410\GeV)^2$ & $0.26$ & $0.26$ & $11.6$ & $-5.7$ & $-5.7$
    \\
    \hline
\end{tabular}}
   \caption{Benchmark scenario, BP1, demonstrating high-temperature symmetry non-restoration with the AE prescription. In contrast, the Parwani and PD prescriptions not only restore symmetry at high temperatures but also exhibit an SFOEWPT. Details are presented in Table~\ref{tab:TcTn}. The phase evolution diagrams for this BP are presented in Figure~\ref{fig:phasediagrams}.}
   \label{tab:BP-sym-NR}
   \vspace{-0.3cm}
\end{table}
\begin{table}[t!]
\renewcommand{\arraystretch}{1.8}
\setlength{\tabcolsep}{3pt}
   \centering
\small{\begin{tabular}{|c|c||c|c|c|c|c|c|c|c|}
\hline
BP & Resummation & $T_c$ & $T_n$ & ${\{h_1, h_2\}}_{\text{false}}$ & \multirow{2}{*}{$\xrightarrow[\text{type}]{\text{Transition}}$} & ${\{h_1, h_2\}}_{\text{true}}$ & $\xi_n$ & $\alpha$ & $\beta/H$  \\
No. & Scheme & (GeV) & (GeV) & (GeV) &  &  (GeV) & & & \\
\hline 
\hline
\multirow{3}{*}{BP1} &{Parwani} & 52 & 27 & \{0, 0\} & FO & \{228, 76\} & 8.9 & 2.1 & 175.5 \\   
\cline{2-10} 
 & {PD} & 88 & 67 & \{0, 0\} & FO & \{219, 73\} & 3.4 & 0.14 & 521 \\  
\cline{2-10} 
& \multicolumn{1}{|c|}{Arnold Espinosa} & \multicolumn{8}{c|}{Symmetry Non-Restoration at high temperature} \\
\hline
\end{tabular}}
   \caption{Phase transition characteristics  of the benchmark scenarios BP1, presented in Table~\ref{tab:BP-sym-NR}, considering Parwani, PD and AE prescriptions. Values of $T_c$, $T_n$, the corresponding field values at the false and true phases and the strength of the phase transition, ($\xi_n = \frac{(\sqrt{(h_{1_{\text{true}}}-h_{1_{\text{false}}})^2+(h_{2_{\text{true}}}-h_{2_{\text{false}}})^2})}{T_n}$) represented for different resummation schemes. `FO indicates that the phase transition is first-order type. The quantities $\alpha$ and $\beta/H$, which are required to estimate the GW spectrum from the FOPT, are also listed.}
  \label{tab:TcTn}
\end{table}
\begin{figure}[t]
    \centering
    \includegraphics[width=1\linewidth]{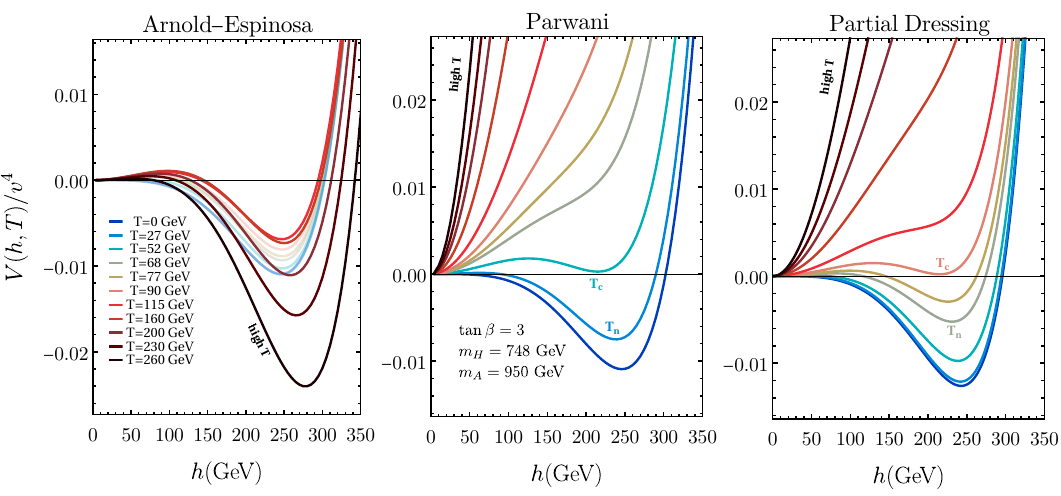}
    \caption{Comparison of the effective potential of BP1, presented in (Table \ref{tab:BP-sym-NR}), for AE (left), Parwani (middle) and PD (right) resummation schemes. The choice of prescription significantly affects the behaviour of the potential. The AE method leads to EWSNR due to the unsuppressed daisy terms. 
    For better visualization, the lower temperature lines of the left plot are shown with reduced opacity to distinguish them from the high-temperature ones. In contrast, the Parwani  and PD  approaches predict symmetry restoration, though they yield different critical and nucleation temperatures as shown in Table \ref{tab:TcTn}.}
    \label{fig:potentials}
\end{figure}


%

The symmetry non-restoration phenomenon is primarily attributed to the additional daisy-resummation terms in the AE prescription, as shown in~\Eq{Vdaisy}, which are found to drive the EW symmetry non-restoration at high temperatures.
In certain regions of parameter space, the field-dependent masses of some dogs can become significantly larger than the temperature. Consequently, these dofs should experience Boltzmann suppression in their contributions to the effective potential at finite temperature. However, in the AE scheme, the additional daisy-resummation terms are derived under the high-temperature approximation. As a result, the AE resummation scheme is not applicable in this parameter space. In this scheme, the heavy dofs not only lack the expected Boltzmann suppression but also contributes additional cubic terms to the effective potential, which should instead be suppressed.
It can be observed in the middle plot of Figure~\ref{fig:phasediagrams} that the field values at the minimum increase with temperature, which, in turn, keeps the field-dependent masses of certain dof heavy in this scheme.
In contrast, the Parwani and PD schemes account for the Boltzmann suppression of these heavy dofs and do not introduce artificial cubic terms in the potential, maintaining consistency in their treatment of the effective potential.

Symmetry non-restoration scenarios with large negative thermal corrections, which can lead to a negative total thermally improved mass at high temperatures, have been studied in the literature~\cite{Aoki:2023lbz, Meade:2018saz, Baldes:2018nel, Biekotter:2021ysx, Biekotter:2022kgf}. However, in the AE scheme, symmetry non-restoration with positive thermal corrections can be observed because of the added daisy resummation term. In contrast, symmetry restoration occurs in the same scenario within the Parwani and PD schemes. To describe this in detail, we provide a simple calculation considering a one-dimensional potential in Appendix~\ref{snr-toy}.
In the Parwani scheme, along with the cubic term, the fourth power of the mass terms in the potential is also resummed, which helps restore symmetry at high temperatures. 
Among all these resummation prescriptions, the PD scheme is a more refined thermal resummation approach, which also exhibits symmetry restoration at high temperatures.
It is important to point out that, beyond this BP, we observe that all the symmetry non-restoration points reported in Ref.~\cite{Biekotter:2022kgf} are restored at high temperatures in the Parwani and PD schemes, as shown in Fig.~\ref{fig:Tnuc_scan}. Thus, we finally conclude that we do not observe the symmetry non-restoration behaviour at high temperatures in 2HDM in our choice of parameter space when analyzing the potential with improved thermal resummation prescription. 
\subsection{Impact of thermal resummation on gravitational waves predictions}
\label{GWimpact}
The advent of GW astronomy, marked by the first direct detection of GW from binary black hole mergers~\cite{LIGOScientific:2016aoc}, has opened a new window into the Universe. More recently, the NanoGrav~\cite{NANOGrav:2023gor} and EPTA~\cite{EPTA:2023fyk} collaborations have reported the first detection of a stochastic GW background, further broadening the scope of GW searches. One particularly exciting prospect is the potential detection of stochastic GW originating from an FOPT in the early Universe, which could offer crucial insights into BSM physics. Unlike GW from astrophysical sources, these stochastic backgrounds exhibit random, unpolarized fluctuations that can be characterized through their two-point correlation function, linked to the power spectral density $\Omega_{\text{GW}}{\rm h}^2$.
The ``cross-correlation" technique can be used across multiple detectors to detect such stochastic GW~\cite{Caprini:2015zlo, Cai:2017cbj, Caprini:2018mtu, Romano:2016dpx, Christensen:2018iqi}. Several upcoming space-based interferometers, including LISA~\cite{LISA:2017pwj}, ALIA~\cite{Gong:2014mca}, TAIJI~\cite{Hu:2017mde}, the Big Bang Observer (BBO)~\cite{Corbin:2005ny}, and Ultimate (U)-DECIGO~\cite{Kudoh:2005as}, are expected to be operational within the next decade.
Each of these experiments targets distinct sensitivity regions in terms of peak intensity and frequency. Notably, they collectively cover a frequency range from approximately \(\sim10^{-4}\,\mathrm{Hz}\) to \(\sim10^{1}\,\mathrm{Hz}\), which is particularly compelling since stochastic GW generated by an FOPT at the electroweak scale are expected to fall within this range. These detectors thus offer promising prospects for probing the dynamics of the early Universe and exploring new physics scenarios associated with the electroweak scale~\cite{Athron:2023xlk, Guo:2020grp, Balazs:2023kuk, Vaskonen:2016yiu, Ghosh:2022fzp, Chatterjee:2022pxf, Roy:2022gop, Alves:2018jsw, Biekotter:2021ysx, Carena:2019une, Goncalves:2021egx}.

The production of stochastic GW from a cosmological FOPT primarily occurs through three mechanisms: collisions of bubble walls, sound waves in the plasma, and magnetohydrodynamic (MHD) turbulence. GW from bubble collisions arises from the stress-energy tensor of expanding bubble walls, which can be approximated using the envelope method~\cite{Kosowsky:1991ua, Kosowsky:1992vn}. While analytical expressions for the GW spectrum exist in this framework~\cite{Jinno:2016vai}, lattice simulations provide more refined spectral predictions that surpass the envelope approximation and are now widely adopted~\cite{Hindmarsh:2015qta, Hindmarsh:2017gnf}. The bulk motion of the plasma during the transition induces velocity perturbations that generate sound waves, which persist long after bubble collisions and dominate GW production~\cite{Hindmarsh:2013xza, Giblin:2013kea,Hindmarsh:2015qta}. Lattice results indicate that the GW contribution from these long-lived sound waves significantly exceeds that from bubble collisions~\cite{Hindmarsh:2015qta, Hindmarsh:2017gnf, Hindmarsh:2013xza}. Several models, including the sound shell model~\cite{Hindmarsh:2016lnk, Hindmarsh:2019phv} and bulk flow model~\cite{Jinno:2019jhi, Konstandin:2017sat}, along with their extensions~\cite{Cai:2023guc}, have been developed to describe the sound wave contribution accurately. Additionally, plasma percolation can induce turbulence, particularly MHD turbulence, due to the ionized nature of the medium, providing another GW source~\cite{Caprini:2006jb, Kahniashvili:2008pf, Kahniashvili:2008pe, Kahniashvili:2009mf, Caprini:2009yp, Kisslinger:2015hua, Yang:2021uid, Di:2020kbw, Dolgov:2002ra}. 
As upcoming experiments probe the frequency range relevant to electroweak-scale phase transitions, precise modeling of these GW sources will be crucial for interpreting potential signals. Here, we sum the contributions of sounds waves and MHD turbulence in the production of GW.
A detailed discussion on the production of GW from such an FOPT in the early Universe is presented in Appendix~\ref{GWs_section}.
\begin{table}[t]
    \renewcommand{\arraystretch}{1.2}
    \centering
    \small{\begin{tabular}{|c|c|c|c|c|c|c|c|c|c}
    \hline 
    BP No. & \multicolumn{8}{|c|}{Parameters}
    \\
    \hline  \hline 
    & $m_h$ & \multirow{1}{*}{$m_{H}$} & \multirow{1}{*}{$m_{A}$} & $m_{H^{\pm}}$  & $v$ & $m_{12}^2$ & $\tanb$ & $c_{\alpha-\beta}$ 
    \\
    \hline 
    BP2 & $125\GeV$ & $482.5\GeV$ & $700\GeV$ & $700\GeV$ & $246\GeV$& $m_H^2 s_\beta c_\beta$ & $3$ & $0$
    \\
    \hline \hline  
    \multirow{1}{*}{}& \multirow{1}{*}{$-m_{11}^2$} & \multirow{1}{*}{$-m_{22}^2$} & \multirow{1}{*}{$m_{12}^2$} & \multirow{1}{*}{$\lambda_1$} & \multirow{1}{*}{$\lambda_2$} & \multirow{1}{*}{$\lambda_3$} & \multirow{1}{*}{$\lambda_4$} & \multirow{1}{*}{$\lambda_5$}
    \\
    \hline
    BP2 & $(449 \GeV)^2$ & $(124 \GeV)^2$ & $(264\GeV)^2$ & $0.26$ & $0.26$ & $8.76$ & $-4.25$ & $-4.25$
    \\
    \hline
\end{tabular}}
   \caption{Benchmark scenario, BP2, demonstrating an FOEWPT with all three resummation schemes, Parwani, PD, and AE. Phase transition details are presented in Table~\ref{tab:TcTnBP2}.}
   \label{tab:BP2}
   \vspace{-0.3cm}
\end{table}
\begin{table}[t!]
\renewcommand{\arraystretch}{1.8}
\setlength{\tabcolsep}{3pt}
   \centering
\small{\begin{tabular}{|c|c||c|c|c|c|c|c|c|c|}
\hline
 BP & Resummation & $T_c$ & $T_n$ & ${\{h_1, h_2\}}_{\text{false}}$ & \multirow{2}{*}{$\xrightarrow[\text{type}]{\text{Transition}}$} & ${\{h_1, h_2\}}_{\text{true}}$ & $\xi_n$ & $\alpha$ & $\beta/H$  \\
No. & Scheme & (GeV) & (GeV) & (GeV) &  &  (GeV) & & & \\
\hline 
\hline
\multirow{3}{*}{BP2}  & {Parwani} & 77 & 67 & \{0, 0\} & FO & \{210, 70\} & 3.3 & 0.104 & 1444 \\   
\cline{2-10}
 & {PD} & 92 & 56 & \{0, 0\} & FO & \{225, 75\} & 4.3 & 0.117 & 232 \\  
\cline{2-10}
 &\multicolumn{1}{|c|}{Arnold Espinosa} & 144 & 130 & \{0, 0\} & FO & \{210, 70\} & 1.7 & 0.014 & 1917 \\
\hline
\end{tabular}}
   \caption{Phase transition characteristics  of the benchmark scenarios BP2, presented in Table~\ref{tab:BP2}, considering Parwani, PD and AE prescriptions. All of these resummation schemes predict FOEWPT for this BP. Values of $T_c$, $T_n$, the corresponding field values at the false and true phases, $\xi_n$,$\alpha$, and $\beta/H$ represented for different resummation schemes.}
  \label{tab:TcTnBP2}
\end{table}

To match the progress in experiments, growing attention is being given to reducing uncertainties in GW predictions from an FOPT, demanding significant theoretical improvements. Achieving higher precision requires better modeling of bubble dynamics, sound wave contributions, and MHD turbulence effects in the plasma. Additionally, accurately determining the finite-temperature effective potential is essential for understanding phase transition dynamics~\cite{Athron:2022jyi, Gould:2021oba}. Further uncertainties arise from the choices of renormalization scale and gauge, bounce action calculations~\cite{Andreassen:2016cvx, Dunne:2005rt, Ivanov:2022osf, Athron:2023rfq} and nucleation rates~\cite{Ekstedt:2021kyx, Athron:2022mmm}. In this work, we have analyzed the impact of different thermal resummation schemes on the effective potential and now focus on their effects on GW production.

\begin{figure}[t]
    \centering
    \includegraphics[width=0.49\linewidth]{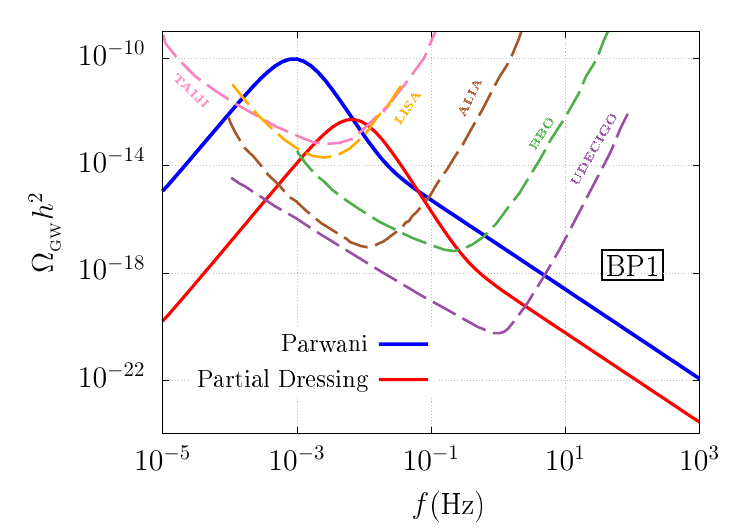}
        \includegraphics[width=0.49\linewidth]{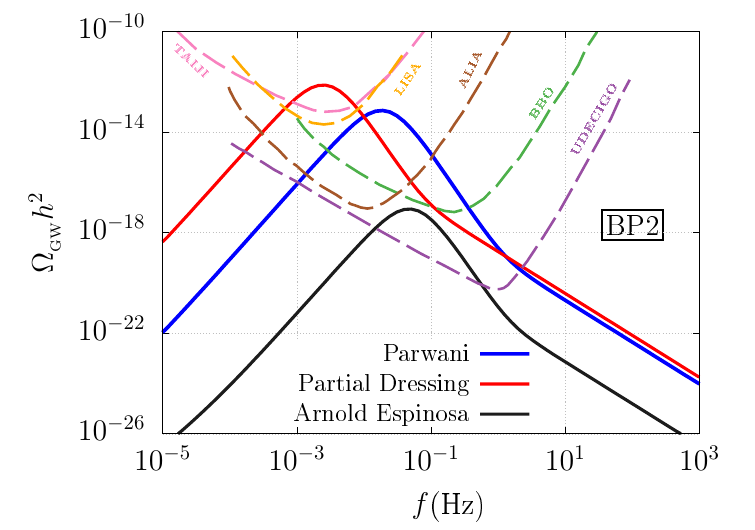}
    \caption{GW energy density spectrum with respect to frequency for the benchmark scenarios, BP1 and BP2, illustrated
against the experimental sensitivity curves of various proposed GW detectors such as LISA, TAIJI,
BBO, ALIA, and U-DECIGO. 
The red, blue, and black solid line indicates the overall GW energy density estimated considering the PD, Parwani, and AE prescriptions, respectively.
The peak frequency for BP1 is \( f_{\text{peak}} = 0.0066 \)~Hz and \( 0.0009 \)~Hz, and the peak amplitude is \( \Omega_{\text{GW}} h^2_{\text{peak}} = 5.2 \times 10^{-13} \) and \( 9.5 \times 10^{-11} \), for the PD and Parwani prescriptions, respectively. In the AE prescription, BP1 does not produce a GW spectrum, as it predicts symmetry non-restoration at high temperatures and no FOEWPT. In the case of BP2, the peak frequency is \( f_{\text{peak}} = 0.0183 \)~Hz, \( 0.0025 \)~Hz and \( 0.0471 \)~Hz, and the peak amplitude is \( \Omega_{\text{GW}} h^2_{\text{peak}} = 7.1 \times 10^{-14} \), \( 7.4 \times 10^{-13} \) and \( 8.7 \times 10^{-18} \), for the Parwani, PD and AE prescriptions, respectively.
}
    \label{fig:gwBP}
\end{figure}

To illustrate the impact of thermal resummations on the prediction of  GW production from an FOPT, we select two benchmark scenarios, BP1 and BP2. As discussed in the previous section, BP1 is selected to demonstrate that the AE prescription predicts symmetry non-restoration at high temperatures, whereas the Parwani and PD prescriptions exhibit symmetry restoration at high temperatures and also predict an FOEWPT. Additionally, we introduce another benchmark scenario, BP2, detailed in Tab.~\ref{tab:BP2}, for which all three resummation prescriptions predict an FOEWPT, as shown in Tab.~\ref{tab:TcTnBP2}. 
The corresponding GW energy density spectrum ($\Omega_{\text{GW}}{\rm h}^2$)  as a function of frequency ($f$) for the benchmark scenarios BP1 and BP2 are displayed in Fig.~\ref{fig:gwBP}.
The left plot of Fig.~\ref{fig:gwBP} shows that the PD prescription predicts a lower GW amplitude compared to the Parwani scheme. Specifically, the difference in peak amplitudes, $(\Omega_{\text{GW}}{\rm h}^2)_{\rm peak}$, is approximately a factor of 220, while the peak frequency, $f_{\rm peak}$, differs by about a factor of 3. Although both prescriptions indicate that the spectrum lies within the sensitivity region of LISA, the signal-to-noise ratio for the PD scheme would be significantly smaller than that of the Parwani scheme. This highlights the substantial impact that different resummation prescriptions can have on the predicted GW spectrum, potentially altering the detection prospects of a given BP at various proposed GW detectors.
For instance, in the case of BP2 (right plot of Fig.~\ref{fig:gwBP}), the PD prescription predicts a higher GW amplitude than both the Parwani and AE prescriptions. Under this benchmark scenario, the PD scheme suggests that the signal falls within LISA's sensitivity, whereas the Parwani (AE) prescription predicts an amplitude lower by one (four) orders of magnitude, making detection at LISA unlikely. This further underscores the importance of resummation choices in evaluating the detectability of stochastic GW signals from an FOPT.

%
\begin{figure}[t]
    \centering
    \includegraphics[width=0.49\linewidth]{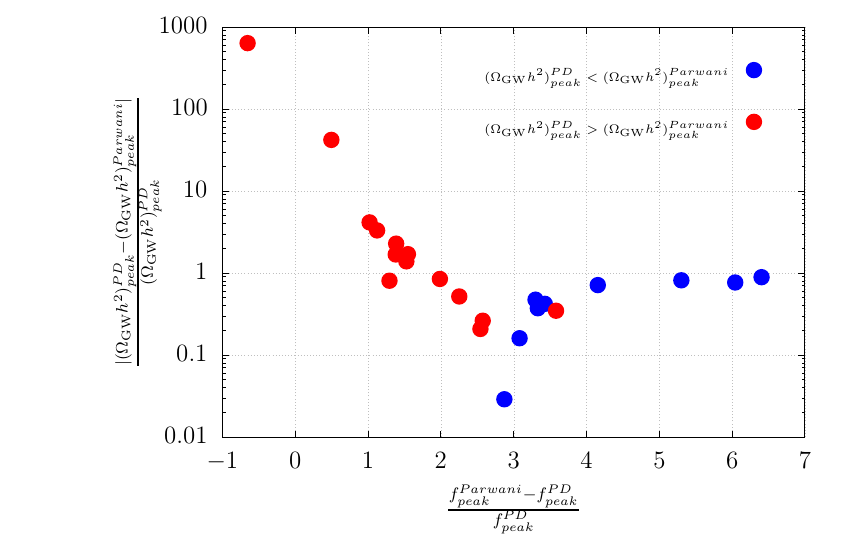}
    \includegraphics[width=0.49\linewidth]{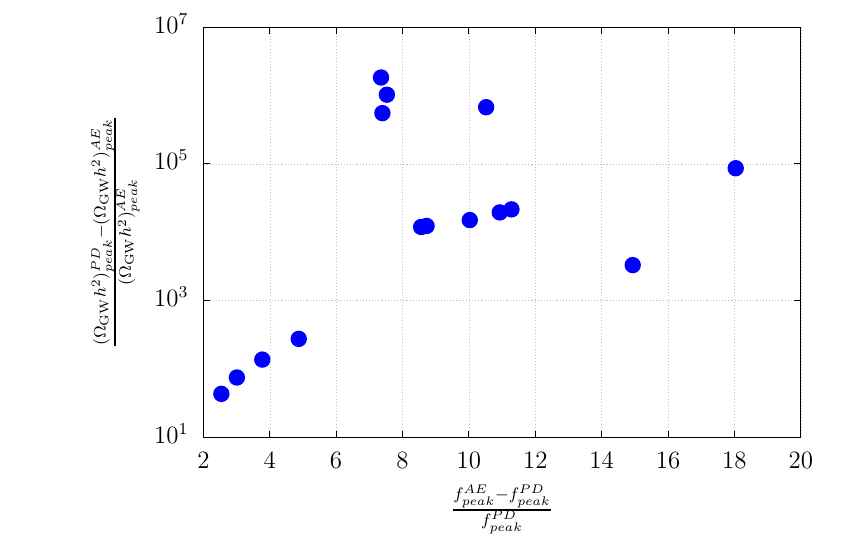}
    \caption{Variation of $(\Omega_{\text{GW}}{\rm h}^2)_{\rm peak}$ and $f_{\rm peak}$ across different resummation prescriptions. The left plot illustrates the absolute variation in $(\Omega_{\text{GW}}{\rm h}^2)_{\rm peak}$ and $f_{\rm peak}$ between the PD and Parwani prescriptions, expressed relative to their values in the PD scheme. Blue and red points indicate which scheme predicts a larger GW amplitude. The right plot presents a similar comparison between the PD and AE prescriptions. For improved visualization, the frequency scaling is done relative to the PD scheme in this plot, whereas the amplitude scaling is referenced to the AE scheme.}
    \label{fig:gwpeakvariation}
\end{figure}

To quantify the dependence of the predicted GW spectrum on the different resummation schemes across the parameter space described in Sec.~\ref{2hdm}, we present the absolute variation of $(\Omega_{\text{GW}}{\rm h}^2)_{\rm peak}$ and $f_{\rm peak}$	  for each scheme in Fig.~\ref{fig:gwpeakvariation}. 
The left plot of Fig.~\ref{fig:gwpeakvariation} illustrates the absolute difference between the predictions of the Parwani and PD prescriptions, while the right plot shows the corresponding differences between the AE and PD prescriptions. 
The left plot indicates that the Parwani and PD schemes show discrepancies, with variations in both peak amplitude and peak frequency typically ranging from one to two orders of magnitude. The points in red (blue) indicate that the GW amplitude estimated from the PD scheme is larger (smaller) than that estimated from the Parwani scheme.
In contrast, the right plot suggests that the AE prescription deviates significantly from the PD scheme. The peak amplitude uncertainty between the AE and PD prescriptions can range from one to six orders of magnitude, while the peak frequency varies by approximately up to a factor of 20. Furthermore, this plot reveals that the AE prescription consistently predicts a significantly smaller $(\Omega_{\text{GW}}{\rm h}^2)_{\rm peak}$ and a relatively larger $f_{\rm peak}$ compared to the PD prescription.
These findings suggest that the uncertainty is relatively small when comparing the Parwani and PD prescriptions. Since the PD scheme provides a more refined approach to thermal resummation, we consider it to yield more reliable GW predictions. However, in this work, we have limited our calculations to the one-loop level within the PD scheme. It would be interesting to investigate how the inclusion of higher-loop corrections affects the predicted GW spectrum. As discussed earlier, multiple sources of uncertainty can influence precise GW predictions, and further studies in this direction are necessary for a more accurate theoretical understanding.
In the next section, we will use these results and apply them to compute the GW signals for different characteristic benchmark scenarios. 
\subsection{Collider and GW Probes of FOEWPT-Favored Regions}
\label{sec:LHCandGWs}
The parameter space of the 2HDM that allows for an FOEWPT can be explored through various searches at the LHC~\cite{Dorsch:2013wja, Dorsch:2014qja, Basler:2016obg, Dorsch:2016tab, Bernon:2017jgv, Dorsch:2017nza, Andersen:2017ika, Kainulainen:2019kyp, Su:2020pjw, Aoki:2021oez, Goncalves:2021egx, Biekotter:2022kgf} and potentially via the detection of a stochastic GW signal in future GW observatories~\cite{Aoki:2021oez, Goncalves:2021egx, Biekotter:2022kgf}. Among various collider search strategies for heavy scalars in the 2HDM, one of the most distinctive signatures of an FOEWPT scenario is the production of the CP-odd scalar, \( A \), followed by its decay into a \( Z \) boson and the heavy CP-even scalar, \( H \)~\cite{Dorsch:2014qja}. Previous LHC searches have analyzed this channel, considering leptonic decays of the \( Z \) boson and the \( H \) decays into bottom-quark and tau-lepton pairs. As we have already pointed out in Sec.~\ref{sec:LHCbounds}, the direct searches for heavy Higgs bosons have already excluded \( m_H \lesssim 350 \) GeV, even in the low-\(\tan\beta\) regime~\cite{CMS:2016xnc, ATLAS:2018oht, CMS:2019ogx, Biekotter:2021ysx, Bagnaschi:2018ofa, ATLAS:2021upq, CMS:2018rmh, CMS:2019bnu, ATLAS:2020zms}. Once \( m_H \) exceeds the di-top threshold, its branching fractions into bottom-quark and tau-lepton pairs decrease significantly at moderately low \(\tan\beta\), reducing the sensitivity of previous LHC searches to the FOEWPT-favored region considered in this study.
Recent studies suggest that the High-Luminosity LHC (HL-LHC) can probe up to \( m_H \lesssim 550 \) GeV and \( m_A \lesssim 750 \) GeV with an integrated luminosity of 3~\(\text{ab}^{-1}\)~\cite{Biekotter:2022kgf}. 
While studying phase transitions, we observe that varying $\tanb$ does not affect the dynamics of the phase transition, and the region in the $m_A-m_H$ plane that we identified in this work as being favored by an FOPT remains unchanged. However, the existing collider constraints are modified, potentially influencing search strategies for the remaining allowed regions. We plan to investigate this further in future work.
%
\begin{figure}[t]
    \centering
    \includegraphics[width=0.50\textwidth,height=0.35\linewidth]{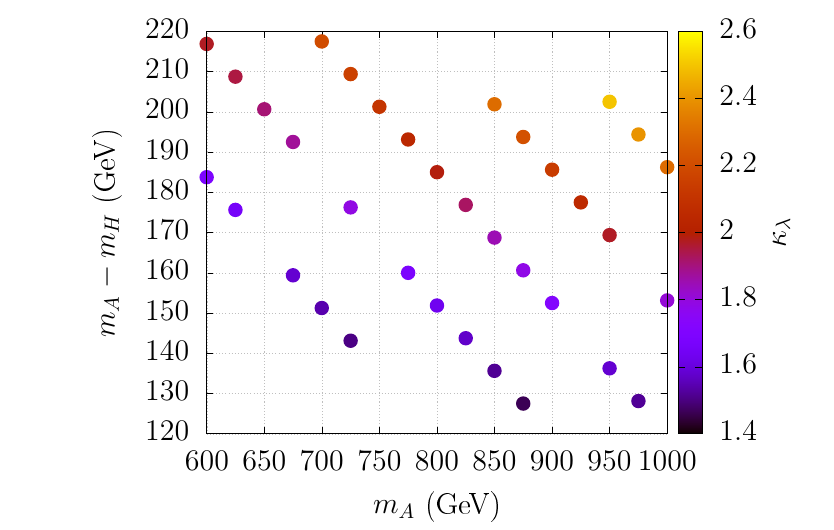}
        \includegraphics[width=0.49\linewidth]{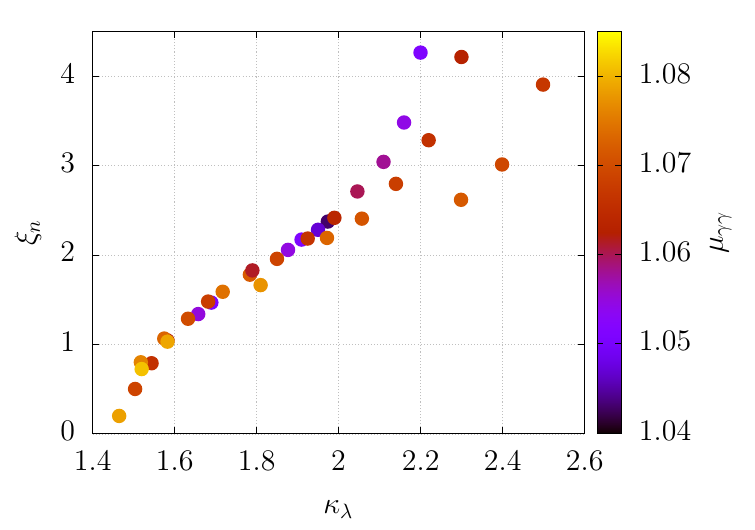}
\caption{Left: FOEWPT-favored points in the PD scheme shown in the $m_A$ vs. $m_A - m_H$ plane, with the trilinear self-coupling of the observed Higgs boson, $\kappa_{\lambda}$, represented by the color palette (see text for details). Right: Correlation between $\kappa_{\lambda}$ and the phase transition strength, $\xi_n$, with the color palette indicating the Higgs di-photon decay signal strength, $\mu_{\gamma \gamma}$.}
    \label{fig:collider}
\end{figure}

Alongside searches for heavy Higgs bosons of 2HDM at the LHC, measuring the trilinear self-coupling $\lambda_{hhh}$ of the observed Higgs boson provides an additional avenue to probe the FOEWPT scenario, as it is often associated with an enhanced $\lambda_{hhh}$~\cite{Noble:2007kk, Huang:2015tdv, Biekotter:2022kgf}. The left plot of Fig.~\ref{fig:collider} presents the variation of $\kappa_{\lambda}$ in the $m_A$ vs. $m_A - m_H$ plane for parameter points exhibiting an FOEWPT within the PD prescription, where $\kappa_{\lambda} = \lambda_{hhh} / \lambda_{hhh}^{\text{SM}}$. Here, $\lambda_{hhh}^{\text{SM}}$ represents the one-loop corrected SM prediction, while $\lambda_{hhh}$ denotes the corresponding trilinear self-coupling of the SM-like Higgs boson in 2HDM. Notably, ATLAS and CMS analyses project their results based on the tree-level value of $\lambda_{hhh}^{\text{SM}}$, which can lead to deviations at the $\sim 10\%$ level~\cite{Dorsch:2017nza}. The plot reveals that $\kappa_{\lambda}$ increases with the mass splitting for a fixed $m_A$.
The right plot of Fig.~\ref{fig:collider} illustrates the correlation between the strength of the phase transition, $\xi_n$, and $\kappa_{\lambda}$, showing the expected trend of $\kappa_{\lambda}$ increasing with $\xi_n$. The color palette in the same plot represents the variation of the signal strength parameter for the $h \to \gamma \gamma$ decay, $\mu_{\gamma\gamma}$, as defined in~\Eq{mugammagamma}.

ATLAS and CMS currently place upper limits on $\kappa_{\lambda}$ at 6.3~\cite{ATLAS:2022jtk} and 6.5~\cite{CMS:2022dwd}, respectively, at the confidence level of 95\%, based on analyzes incorporating single Higgs and di-Higgs production while assuming other couplings remain at their SM values. The HL-LHC is expected to probe $\kappa_{\lambda}$ down to approximately 2.2~\cite{Goncalves:2018qas, Kling:2016lay, Cepeda:2019klc}, making it possible to explore a subset of the FOEWPT-favored parameter space through $\kappa_{\lambda}$ measurements.
Furthermore, while all parameter points satisfy the current ATLAS~\cite{ATLAS:2022tnm} and CMS~\cite{CMS:2021kom} bounds on $\mu_{\gamma\gamma}$, which are $1.04^{+0.10}_{-0.09}$ and $1.12^{+0.09}_{-0.09}$, respectively, the HL-LHC is expected to improve sensitivity to an uncertainty level of 2\%~\cite{Mlynarikova:2023bvx}. This suggests that the entire FOEWPT-favored parameter space could be probed through precision measurements of the di-photon decay of the observed Higgs boson at the HL-LHC.
%
\begin{figure}[t]
    \centering
    \includegraphics[width=0.65\linewidth]{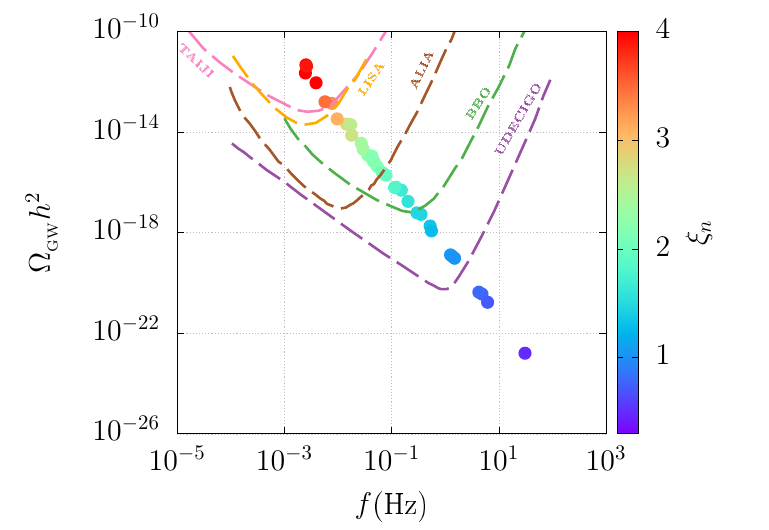}
    \caption{Variation of the GW peak amplitude  ($\Omega_{\text{GW}} h^2)_{\text{peak}}$ with the peak frequency ($f_{\text{peak}}$) in the GW power spectrum ($\Omega_{\text{GW}}h^2$) - frequency ($f$) plane for the region of parameter space that exhibits an FOEWPT, as shown in Fig.~\ref{fig:Tnuc_scan}, using the PD prescription. The
palette color shows the variation of the strength of the phase transition ($\xi_n$). The different colored lines represent the experimental sensitivity curves of future proposed GW detectors, including LISA, TAIJI, BBO, ALIA, and U-DECIGO.}
    \label{fig:gwpeak}
\end{figure}

In addition to the ongoing search for BSM physics at the LHC, future proposed GW experiments could provide sensitivity to certain regions of the parameter space in various BSM scenarios that exhibit an FOPT around the electroweak scale, as it leads to a GW spectrum around the mHz to Hz frequency range, after redshifting the signal to the present time~\cite{Grojean:2006bp, Roshan:2024qnv}. 
To investigate the stochastic GWs spectral signal region arising from our scenario, we present the variation of the peak amplitude, \((\Omega_{\text{GW}} h^2)_{\text{peak}}\), with the peak frequency, \(f_{\text{peak}}\), of the GW generated by an FOEWPT considering the PD prescription, as it is the most refined approach. This is depicted in the \(\Omega_{\text{GW}} h^2\)-\(f\) plane in Fig.~\ref{fig:gwpeak}, for the points exhibiting an FOEWPT, as identified in Fig.~\ref{fig:Tnuc_scan} using the PD prescription.
It is important to note that \(f_{\text{peak}}\) and \((\Omega_{\text{GW}} h^2)_{\text{peak}}\) are primarily determined by the sound wave contribution (as described from~\Eq{soundgw} to~\Eq{fsw}), with the turbulence contribution (described from~\Eq{GWturb} to ~\Eq{eq:h-star}) playing a relatively minor role in estimating the peak amplitude.  
The strength of the phase transition, \(\xi_n\), is represented by the color palette in the plot. The color variation reveals a clear trend: as the strength of the FOEWPT increases, the peak of GW amplitude grows while the peak frequency decreases. This behavior can be understood from the fact that, in our scenario, a larger \(\xi_n = v_n/T_n\) corresponds to a lower nucleation temperature, \(T_n\). A lower \(T_n\) leads to a smaller \(\beta/H\), which shifts \(f_{\text{peak}}\) to lower values (see~\Eq{fsw}) and \((\Omega_{\text{GW}} h^2)_{\text{peak}}\) to higher values (see~\Eq{soundgw}). 
From this plot, it can be inferred that the full spectral distributions of the stochastic GW generated from these scanned points are unlikely to fall within the expected sensitivity range of the upcoming GW detector LISA. However, based on the points displayed, it can be reasonably anticipated that the majority portion of the parameter space will fall within the sensitivity range of the proposed future U-DECIGO experiment. Additionally, other proposed experiments, such as BBO and ALIA, may partially probe this parameter space.

From this discussion, it is evident that this region of the parameter space can be explored through a complementary approach, combining collider analyses at the HL-LHC with stochastic GW searches at proposed future GW detectors~\cite{Biekotter:2022kgf}.
For instance, BP2 is expected to be accessible via the $A \rightarrow Z h$ search at the HL-LHC, whereas BP1 would remain unconstrained by the same search. However, BP1 can still be probed by studying the Higgs trilinear self-coupling, as its corresponding $\kappa_{\lambda}$ exceeds 2.2, making it within reach of HL-LHC sensitivity. Additionally, the di-photon decay channel of the Higgs boson could provide further insights into these scenarios.
The absence of any new physics signals in these channels would place significant constraints on the prospects of detecting a stochastic GW signal at proposed GW detectors such as LISA. Nevertheless, as discussed in the previous section,  the prediction of stochastic GWs from an FOEWPT is subject to various uncertainties arising from different sectors~\cite{Athron:2022jyi}, even when employing more refined thermal resummation schemes, such as the PD prescription used in Fig.~\ref{fig:Tnuc_scan}. Therefore, to enhance our understanding of GW production from an FOEWPT and its correlation with collider signals, further theoretical refinements and improvements are essential.
\section{Conclusion}
\label{conclusion}
A precise description of the effective potential at finite temperature is crucial for accurately predicting an FOEWPT phenomenon in the early Universe. This can have far-reaching physical implications, such as explaining the observed baryon asymmetry via the EWBG mechanism and generating a stochastic GW spectrum. In this work, we investigate the impact of various resummation prescriptions on the effective potential at finite temperature and their influence on the dynamics of the EWPT in the 2HDM.
In particular, we explore the PD scheme, a more refined resummation method that provides a consistent treatment of higher-order thermal corrections without relying on the high-temperature limit, for the first time in a realistic model like the 2HDM. 
We demonstrate how to explicitly implement PD in scenarios with multiple mixing scalar fields, providing a detailed discussion on solving the gap equation using the iterative method. Furthermore, we investigate the Parwani and AE resummation prescriptions, which are more commonly used in the literature and rely on the high-temperature approximation, which may significantly break down in the regime where $v_c/T_c \sim \mathcal{O}(1)$. We compare the results obtained from these different resummation schemes to assess their impact on the phase transition dynamics. Here are the key differences:
\begin{itemize}
\item Field-dependent thermal masses of various dof are obtained by solving the full gap equation without relying on the high-temperature approximation. These thermal masses can differ significantly from those derived using the truncated solution under the high-temperature approximation, as the heavy modes contributions should experience Boltzmann suppression. These differences become particularly important in certain field regions and temperature ranges that are highly relevant in the context of an SFOEWPT.
\item The FOEWPT-allowed parameter space can vary depending on the choice of resummation scheme. In particular, significant deviations are observed when comparing results from the AE scheme with those from the Parwani and PD schemes. 
 While the overall FOEWPT-allowed regions from the Parwani and PD schemes may appear similar, their predictions can differ significantly in certain regions of parameter space. For instance, PD may predict an FOEWPT where Parwani instead leads to vacuum trapping.
Additionally, the strength of an FOEWPT can vary significantly depending on the resummation scheme, with the largest deviations observed when comparing the results from the AE scheme with those of the PD or Parwani schemes.

\item At high temperatures, the PD and Parwani resummation prescriptions predict EW symmetry restoration. In contrast, the AE resummation scheme suggests symmetry non-restoration in certain regions of the parameter space where thermal mass corrections to various dof remain effectively positive.
\item The prediction of stochastic GW from an FOPT can vary significantly depending on the choice of resummation scheme in the finite-temperature effective potential. Notably, we find that the AE prescription consistently predicts a substantially lower peak amplitude and a relatively higher peak frequency compared to the PD  scheme within our chosen parameter space. The uncertainty in peak amplitude between these two schemes can span up to six orders of magnitude, while the peak frequency can differ by a factor of up to 20. In contrast, the discrepancies between the Parwani and PD prescriptions are significant but remain relatively small compared to the AE scheme.
\end{itemize}
Various proposed GW experiments, such as LISA, BBO, ALIA, TAIJI, and UDECIGO, are designed to probe different sensitivity regions in terms of peak amplitude and frequency. Consequently, a precise prediction of the GW spectrum from an FOPT is crucial to assessing their potential to explore the BSM parameter space. 
Further theoretical refinements can enhance the precision of phase transition predictions. For instance, incorporating two-loop effects without relying on the high-temperature approximation can help reduce uncertainties arising from higher-order corrections.
Beyond refining the description of the effective potential at finite temperatures, further theoretical advancements in multiple directions are necessary to improve the accuracy of GW spectrum predictions from an FOPT.

PD is a more refined resummation scheme, offering greater reliability compared to other resummation methods.
It provides a self-consistent treatment of temperature effects, properly incorporating higher-order daisy and superdaisy contributions without relying on the high-temperature approximation. In contrast, both AE and Parwani introduce uncontrolled approximations that break down at intermediate temperatures. AE’s negative daisy contributions and lack of higher-order corrections can lead to spurious effects like EWSNR, while Parwani’s inclusion of unsuppressed thermal corrections in the CW potential can prevent proper decoupling of heavy modes. PD naturally resolves these issues, making it the more reliable resummation scheme.
Therefore, in this work, we explore the 2HDM parameter space using this scheme to robustly determine the regions that favor an FOEWPT and to examine key physical phenomena, such as symmetry restoration at high temperatures. We also compare our findings with existing results in the literature, which are based on the AE and/or Parwani methods.
Finally, we discuss potential future experimental strategies for probing the FOEWPT-favored parameter space. These include direct searches for $A \rightarrow Z H$, precise measurements of the Higgs self-coupling $\kappa_{\lambda}$, and the di-photon decay rate of the observed Higgs boson at the HL-LHC. Additionally, we explore the potential of future GW experiments to provide complementary insights into this parameter space.
\acknowledgments
The authors thank Henning Bahl for useful discussions. PB acknowledge the financial support provided by CAPES grant 88887.816450/2023-00.  
PB thanks CW and the hospitality
from the EFI at the University of Chicago and the HEP division at ANL where part of this work was done.
SR and CW have been supported by the U.S.~Department of Energy under contracts No.\ DEAC02-06CH11357 at Argonne National Laboratory.
SR would like to thank the University of Chicago and Fermilab, where a significant part of this work has been done. The work of CW\ at the University of Chicago has been supported by the DOE grant DE-SC0013642. C.W.\ would like to thank the Aspen Center for Physics, which is supported by National Science Foundation grant No.~PHY-1607611, where part of this work has been done. 
%
\appendix
\label{appendix}
\section*{Appendix}
\label{appendix}
\section{Relations among the masses and various Lagrangian parameters}
\label{modelrealtions}
The mass-squared matrices of the Higgs fields of the scalar potential, defined in equation~\ref{Tree-Level-Potential}, are given by, 
\bea
&&\label{Z2_basis_diag}
\begin{pmatrix}h_1 & h_2\end{pmatrix} \begin{pmatrix} m_{12}^2 \tb+ \lambda_1 v^2 \cb^2  &\quad - m^2_{12} + {\lambda_{345} \over 2} v^2 \stwob \\[5pt]
- m^2_{12} + {\lambda_{345} \over 2} v^2 \stwob & \quad m_{12}^2/ \tb+ \lambda_2 v^2 \sb^2  \\[5pt] \end{pmatrix} \begin{pmatrix}h_1\\ h_2\end{pmatrix}\,,\label{eq:mass_matrix}\\
&&\begin{pmatrix}a_1 & a_2\end{pmatrix} \left[ m_{12}^2 - {1\over 2} \lambda_5 v^2 \stwob \right] \begin{pmatrix} \tb & \quad -1 \\[5pt]
-1 & \quad  1/\tb \end{pmatrix} \begin{pmatrix}a_1\\ a_2\end{pmatrix}\,,\\
&&\begin{pmatrix}\phi^+_1 & \phi^+_2\end{pmatrix} \left[ m_{12}^2 - {1\over 4} (\lambda_4+ \lambda_5) v^2 \stwob \right] \begin{pmatrix} \tb & \quad -1 \\[5pt]
-1 & \quad  1/\tb \end{pmatrix} \begin{pmatrix}\phi^-_1 \\ \phi^-_2\end{pmatrix}\,.
\eea
The mass eigenstates are obtained from the original fields by the rotation matrices:
\begin{eqnarray}
\left(\begin{array}{c}H \\ h \end{array}\right) =  \left(\begin{array}{cc}\cos\alpha & \sin\alpha \\ -\sin\alpha & \cos\alpha \end{array}\right)  \left(\begin{array}{c} h_1 \\ h_2 \end{array}\right) , \\
\left(\begin{array}{c}G^0 \\ A \end{array}\right) =  \left(\begin{array}{cc}\cos\beta & \sin\beta \\ -\sin\beta & \cos\beta \end{array}\right)  \left(\begin{array}{c} a_1 \\ a_2 \end{array}\right) , \\
\left(\begin{array}{c}G^{\pm} \\ H^{\pm} \end{array}\right) =  \left(\begin{array}{cc}\cos\beta & \sin\beta \\ -\sin\beta & \cos\beta \end{array}\right)  \left(\begin{array}{c} \phi^{\pm}_1 \\ \phi^{\pm}_2 \end{array}\right),
\end{eqnarray}
where $G^0$ and $G^\pm$ are Goldstone bosons which are absorbed as longitudinal components of the $Z$ and $W^\pm$ bosons. 
The remained physical states are two neutral
CP-even states $h$ and $H$, one neutral pseudoscalar $A$, and a pair of charged scalars $H^{\pm}$. 
Their mass-squared relations are given by 
\begin{align}
m^2_{H,h} &= \frac{1}{2}\left[M^2_{P, 11} + M^2_{P,22}\pm \sqrt{(M^2_{P,11}-M^2_{P,22})^2+4 (M^2_{P,12})^2 } \right]  \label{eq:hmass} \ ,\\
m_A^2 &= \frac{m_{12}^2}{s_\beta c_\beta} - \lambda_5 v^2 \label{eq:Amass} \ , \\
m_{H^{\pm}}^2 &= \frac{m_{12}^2}{s_\beta c_\beta} - \frac{1}{2} (\lambda_4+\lambda_5) v^2 \ ,\label{eq:Hcmass}
\end{align}
where $M^2_{{P}_{ij}}$ relations, used in equation~\ref{eq:hmass}, are the components of the $CP$-even mass-squared matrix defined in~\Eq{eq:mass_matrix}. From equations~(\ref{eq:Amass}) and (\ref{eq:Hcmass}), the condition $m_{H^{\pm}}^2 = m_A^2$ implies that $\lambda_4 = \lambda_5$.

The parameters $v_1$ and $v_2$ can be expressed in terms of $v$ ($=246$~GeV) and $\tan\beta$, i.e., $v_1 = v \cos\beta$ and $v_2 = v \sin\beta$. The  Lagrangian coupling parameters
$\lambda_{i} (i=1,2,...,5)$ can be expressed in terms of the physical masses ($m_{h, H, A, H^{\pm}}$), mixing angles ($\alpha$, $\beta$) and $m_{12}^2$. The relations between these two equivalent sets of parameters are given below
\begin{subequations}
\label{conversionrelation}
 \begin{align}
     \lambda_1v^2 &= \frac{1}{c_\beta^2}\left(s_\alpha^2 m_h^2 + c_\alpha^2 m_H^2 - m_{12}^2\tan\beta\right), \\
     \lambda_2v^2 &= \frac{1}{s_\beta^2}\left(c_\alpha^2 m_h^2 + s_\alpha^2 m_H^2 - m_{12}^2/\tan\beta\right),\\
     \lambda_3v^2 &= 2m_{H^\pm}^2+\frac{s_{2\alpha}}{s_{2\beta}}(m_H^2-m_h^2)-\frac{m_{12}^2}{s_\beta c_\beta},\\
     \lambda_4v^2 &= m_A^2 - 2m_{H^\pm}^2+\frac{m_{12}^2}{s_\beta c_\beta},\\
     \lambda_5v^2 &= \frac{m_{12}^2}{s_\beta c_\beta}-m_A^2 \, .
 \end{align}
\end{subequations}
\section{UV-finite counterterm}
\label{CTcoeff}
The one-loop CW correction to the tree-level potential modifes the masses and mixing angles of various scalar dof of the model.
It is essential to consider these loop corrections when testing the parameter space of the model with experimental constraints.
To facilitate an efficient scan of the parameter space, it is advantageous to directly use the loop-corrected masses and mixing angles as inputs. 
We choose a renormalisation prescription by which we enforce the one-loop corrected masses and mixing matrix elements to be equal to the tree-level ones, satisfying relations defined in \ref{CTcond.}.
 The added counterterm potential $V_{\rm CT}$ is parameterized as,
\begin{align}
V_{\rm CT}&= \delta m_{11}^{2}\left|\Phi_{1}\right|^{2}+ \delta m_{22}^{2}\left|\Phi_{2}\right|^{2}-\delta m_{12}^{2}\left(\Phi_{1}^{\dagger}\Phi_{2}+\text{h.c.}\right)+\frac{\delta \lambda_{1}}{2}\left(\Phi_{1}^{\dagger}\Phi_{1}\right)^{2}+\frac{\delta \lambda_{2}}{2}\left(\Phi_{2}^{\dagger}\Phi_{2}\right)^{2} \notag \\
&+\delta \lambda_{3}\left(\Phi_{1}^{\dagger}\Phi_{1}\right)\left(\Phi_{2}^{\dagger}\Phi_{2}\right)+ \delta \lambda_{4}\left(\Phi_{1}^{\dagger}\Phi_{2}\right)\left(\Phi_{2}^{\dagger}\Phi_{1}\right)+\frac{\delta \lambda_{5}}{2}\left[\left(\Phi_{1}^{\dagger}\Phi_{2}\right)^{2}+
\mathrm{h.c.}\right] ,
\label{VCT}
\end{align}
where,
\begin{subequations}
\small{\begin{align}
\delta_{\lambda_{3}}= \frac{1}{v_1 v_2} ( DV[h_1 h_2] -DV[a_0 G_0]) \, \, , \quad\quad \quad\quad \quad\quad \quad\quad\quad\quad
\\
\delta{\lambda_{5}}= \frac{1}{v_2^2} ( DV[G^{\pm} G^{\mp}] -DV[G_0 G_0])\, \, , \quad\quad \quad\quad \quad\quad \quad\quad \quad\quad
\\
\delta{m_{12}^2}=  2 \frac{v_1}{v_2} ( DV[G^{\pm} G^{\mp}] - DV[G_0 G_0]) - DV[a_0 G_0] \, \, , \quad\quad \quad
\\
\delta{\lambda_{2}}= \frac{1}{v_2^3} ((v_2  DV[h_2 h_2] - DV[h_2]) + \delta m_{12}^2 v_1)\, \, , \quad\quad \quad\quad \quad\quad 
\\
\delta{\lambda_{1}}= \frac{1}{v_1^3} ((v_1  DV[h_1 h_1] - DV[h_1]) + \delta m_{12}^2 v_2)\, \, , \quad\quad \quad\quad \quad\quad 
\\
\delta{m_{11}^2}= - ( \frac{3}{2}v_1^2 \delta \lambda_1 + \frac{v_2^2}{2} (\delta \lambda_3 + \delta \lambda_5) + DV[h_1 h_1]   )\, \, , \quad\quad \quad\quad 
\\
\delta{m_{22}^2}= - ( \frac{3}{2}v_2^2 \delta \lambda_2 + \frac{v_1^2}{2} (\delta \lambda_3 + \delta \lambda_5) + DV[h_2 h_2]   )\, \, , \quad\quad \quad\quad 
\end{align}}
\end{subequations}
$DV[\phi_i] = \frac{\partial V_\text{CW}}{\partial \phi_{_i}}$ and $DV[\phi_i \phi_j] = \frac{\partial^2 V_\text{CW}}{\partial \phi_{_i}  \partial \phi_{_j}}$ where $\phi_{_i}, \phi_{_j} = \{h_1, h_2, a_1, a_2, \phi_1^{\pm}, \phi_2^{\pm}\}$. All the derivatives are taken at the true EW minima, i.e., $h_2 = v_2$, $h_1 = v_1$ and all other field directions are zero. 
The Goldstone modes exhibit vanishing masses at the true EW minimum at $T = 0$ due to the choice of the Landau gauge in this analysis. This results in an infrared (IR) divergence, as noted in Refs.~\cite{Martin:2014bca,Elias-Miro:2014pca}, which arises from the second derivatives employed in the renormalization conditions described in the preceding equations. To mitigate this issue, an IR regulator can be applied by modifying the Goldstone mode masses as $m_G^2 \to m_G^2 + \mu_{\rm IR}^2$. 
For numerical calculations, choosing $\mu_{\rm IR}^2 = 1\,{\rm GeV}^2$  is sufficient, as implemented in Refs.~\cite{Baum:2020vfl, Chatterjee:2022pxf, Ghosh:2022fzp, Roy:2022gop}. We also examine the variations of 
$\mu_{\rm IR}^2$ from 1~$\rm{GeV}^2$ to 100~$\rm{GeV}^2$ and find that the results remain mostly unchanged.
\vspace{0.1cm}
\section{Tree-level field dependent masses of the degrees of freedom}
\label{app:fielddepmasses}

The field-dependent tree-level mass-squared matrices for the $(2\times2)$, symmetric matrix ($\mathbb{M}_{H}^2$) for the $CP$-even scalars, 
in the basis $\{h_1, h_2\}$, is given by:
\begin{subequations}
\label{cpevenmasssq}
\small{\begin{align}
\mathbb{M}_{{H}_{11}}^2 = - m_{11}^2 +  \frac{3}{2} \lambda_1 h_1^2 + \frac{1}{2} \lambda_{345} h_2^2 \, , \quad\quad \quad\quad \quad\quad \\
\mathbb{M}_{{H}_{22}}^2 = - m_{22}^2 +  \frac{3}{2} \lambda_2 h_2^2 + \frac{1}{2} \lambda_{345} h_1^2 \, , \quad\quad \quad\quad \quad\quad \\
\mathbb{M}_{{H}_{12}}^2 =  M_{{H}_{21}}^2 = - m_{12}^2 + \lambda_{345} h_1 h_2 \, ,\label{Eq:DaisyCoeffs_H} \quad\quad \quad\quad \quad\quad
\end{align}}
\end{subequations}
where $\lambda_{345} = \lambda_3 + \lambda_4 + \lambda_5$.
The $(2\times2)$ symmetric mass squared matrix ($\mathbb{M}_{A}^2$) for the $CP$-odd scalars, 
in the basis $\{h_1, h_2\}$, is given by:
\begin{subequations}
\label{cpoddmasssq}
\small{\begin{align}
\mathbb{M}_{{A}_{11}}^2 = - m_{11}^2  + \frac{1}{2} \lambda_{345}^{'} h_2^2 \, , \quad\quad \quad\quad \quad\quad \quad\quad  \\
\mathbb{M}_{{A}_{22}}^2 = - m_{22}^2 + \frac{1}{2} \lambda_{345}^{'} h_1^2 \, , \quad\quad \quad\quad \quad\quad \quad\quad\\
\mathbb{M}_{{A}_{12}}^2 =  M_{{A}_{21}}^2 = - m_{12}^2 + \lambda_{5} h_1 h_2 \, ,\label{Eq:DaisyCoeffs_A} \quad\quad \quad\quad \quad 
\end{align}}
\end{subequations}
where $\lambda_{345}^{'} = \lambda_3 + \lambda_4 - \lambda_5$. The $(2\times2)$ symmetric mass squared matrix ($\mathbb{M}_{H^{\pm}}^2$) for the chared scalars, 
in the basis $\{h_1, h_2\}$, is given by:
\begin{subequations}
\label{chargmasssq}
\small{\begin{align}
\mathbb{M}_{H^{\pm}_{11}}^2 = - m_{11}^2 +  \frac{1}{2} \lambda_1 h_1^2 + \frac{1}{2} \lambda_{3} h_2^2 \, , \quad\quad \quad\quad \quad\quad \quad\quad \\
\mathbb{M}_{H^{\pm}_{22}}^2 = - m_{22}^2 +  \frac{1}{2} \lambda_2 h_2^2 + \frac{1}{2} \lambda_{3} h_1^2 \, , \quad\quad \quad\quad \quad\quad \quad\quad\\
\mathbb{M}_{H^{\pm}_{12}}^2 =  M_{H^{\pm}_{21}}^2 = - m_{12}^2 + \frac{1}{2} \lambda_{45} h_1 h_2 \, ,\label{Eq:DaisyCoeffs_HP} \quad\quad \quad\quad \quad\quad
\end{align}}
\end{subequations}
where $\lambda_{45} = \lambda_4 + \lambda_5$.
In the fermionic sector, we only consider the top quark dof and its field-dependent mass is given by
\begin{equation}
\label{eq:FermionMasses}
\begin{gathered}
m_t  =   \frac{y_t}{\sqrt{2}\sin\beta} h_2 \,  \quad.
\end{gathered}
\end{equation}
The field dependent mass of the charged gauge boson, $W^{\pm}$, is
\vspace{0.1cm}
\begin{equation}
\begin{gathered}
m_{W^{^\pm}}^2  =  \tfrac{1}{4} g_2^2 \left(h_1^2  +   h_2^2 \right) \, \, .
\end{gathered}
\end{equation}
The field dependent mass-squared matrix  of the neutral electroweak gauge bosons $W^3$ and $B$ of $SU(2)$ and $U(1)$ gauge groups, respectively, is given by,
\vspace{0.1cm}
\begin{equation}
\begin{gathered}
m_{W^{^3}}^2  =  \tfrac{1}{4} g_2^2 \left(h_1^2  +  h_2^2 \right), \, \, \\
m_{B}^2  =  \tfrac{1}{4} g_1^2 \left(h_1^2  +  h_2^2 \right), \, \,  \\
m_{W^{^3}{B}}^2  =  - \tfrac{1}{4} g_1 g_2 \left(h_1^2  +  h_2^2 \right) .
\end{gathered}
\end{equation}
The $Z$-boson and the photon ($\gamma$) field dependent masses can befound via diagonalising this mass matrix and it is given by,
\vspace{0.1cm}
\begin{equation}
\begin{gathered}
m_Z^2  =  \tfrac{1}{4} (g_2^2 + g_1^2) \left(h_1^2  +  h_2^2 \right), \, \, \quad \quad \quad
m_{\gamma}^2  =  0 \, \, .
\end{gathered}
\end{equation}

%
%
\section{Truncated full dressing thermal mass at high temperature approximation}
Solution of the gap equation, defined in~\Eq{eq:gap_2hdm}, at the high temperature limit and remain at the leading order is defined as the Truncated full dressing (TFD) thermal mass ($\Pi_i$), defined in~\Eq{eq:PiT2}. Substituting the eigenvalues of the $(m_X^2 + \Pi_X)$ directly into the effective potential is called the TFD resummation prescription, where $m_{X}^2 , (X = P, A, H^{\pm})$ are the tree-level mass-squared matrices defined in eqs.~ \ref{cpevenmasssq} to \ref{chargmasssq}.
The TFD thermal mass functions are given by,
$\Pi_X, (X=P,A,\pm)$ given by
\begin{align}
\label{piT}
\Pi_X & = \begin{pmatrix}
\Pi^{X}_{11} &  \quad \Pi^{X}_{12}  \\[5pt]
\Pi^{X}_{12} &  \quad \Pi^{X}_{22}  
\end{pmatrix} {\frac{T^{2}}{24}} \ ,
\end{align}
where,
\vspace{0.1cm}
\begin{equation}
\begin{gathered}
\Pi^{X}_{11} = c_{\rm SM}- 6 y_t^2 + 6 \lambda_1 +4 \lambda_3 + 2 \lambda_4,, \, \, \\
\Pi^{X}_{22}  = c_{\rm SM} + 6 \lambda_2 +4 \lambda_3 + 2 \lambda_4      \, , \quad \quad \quad
\end{gathered}
\end{equation}
The SM contributions (considering top quark, $SU(2)_L$ and $U(1)_Y$ gauge fields), defined as $c_{\rm SM}$, is given by,
\begin{eqnarray}	
c_{\rm SM} = \frac{9}{2}g^2 + \frac{3}{2}g'^2 + 6y_t^2 .
\end{eqnarray}
The subscripts $\{1,2\}$ denote the states $\{h_{d}, h_{u}\}$, respectively.
The imposed discrete $Z_2$-symmetry keeps the off-diagonal thermal mass terms vanishingly small at the leading order.
Additionally, as noted by ~\cite{Blinov:2015vma}, subleading thermal corrections to off-diagonal self-energies terms are suppressed by extra powers of coupling constants and electroweak $vevs$.
It is worth highlighting that TFD thermal masses are independent of $\lambda_5$, where a potential $CP$ phase might appear. 

The longitudinal modes of the gauge bosons also receive thermal corrections. TFD thermal masses of the $W^{\pm}_L$, $W^{3}_L$ and $B_L$ (`$L$' corresponds to the longitudinal mode) are given by,
\begin{equation}
\begin{gathered}
\Pi_{W^{\pm}_L}^2  =  2 g_2^2 T^2, \, \,
\end{gathered}
\end{equation}
\begin{equation}
\begin{gathered}
\label{w3bLTM}
m_{W^{^3}}^2  =  2 g_2^2 T^2, \, \, \\
m_{{B}_L}^2  =  2 g_1^2 T^2 \, \, .
\end{gathered}
\end{equation}
The TFD thermal mass-squared of $W^{\pm}_L$ is given by, $m_{W^{\pm}_L}^2 = \tfrac{1}{4} g_2^2 \left(h_1^2  +   h_2^2 \right) + 2 g_2^2 T^2$.
The TFD thermal mass-squared of $Z_L$ and $\gamma_L$ can be obtained via diagonalizing the gaube boson mass matrix considering the correction defined in~\Eq{w3bLTM}. These are given by,
\begin{equation}
\label{ZLGLTmass}
m_{Z_L,\gamma_L}^2 = \frac{1}{8} (g_1^2+g_2^2) (h_1^2+h_2^2) + (g_1^2 + g_2^2 )T^2 \pm \delta, 
\end{equation}
where 
\begin{equation}
\delta^2 =\frac{1}{64}  (g_1^2 + g_2^2 )^2(h_{d}^2 + h_{u}^2+8T^2)^2
- g_1^2 g_2^2 T^2 ( h_{d}^2 + h_{u}^2 + 4 T^2). 
\end{equation}
These TFD thermal mass-squared relations, defined from~\Eq{piT} to~\Eq{ZLGLTmass} of various dof of 2HDM are used to estimate the effective potential at finite temperature using the so called AE and Parwani prescriptions, described in Secs.~\ref{AEpores} and~\ref{Parwanipres}, respectively. The PD prescription requires estimating the thermal mass by solving the gap equations. The details of this solution are provided in the following section.
%
%
%
\section{Stochastic GW Production from an FOPT}\label{GWs_section}
%
Various numerical simulations have been already performed to predict the GW spectrum form an FOPT at the early Universe based on the knowledge of the  following key parameters: $T_n, \alpha, \beta/H_n, g_{*}, v_w$~\cite{Caprini:2019egz, Caprini:2015zlo, Hindmarsh:2017gnf,  Hindmarsh:2019phv, Roshan:2024qnv, Cutting:2019zws}.

As mentioned previously in the main text of the paper,
$T_n$ denotes the bubble nucleation temperature of the phase transition. This characterizes the onset of the phase transition when approximately one bubble per Hubble volume forms.
This temperature is typically determined by solving the nucleation condition,
\begin{small}
\beq
\label{nucleation}
\int_{T_n}^{\infty}\frac{dT}{T}\frac{\Gamma(T)}{H(T)^4} \simeq 1 \, \, . 
\eeq
\end{small}
where, $\Gamma(T)$ is the tunneling probability from the false vacuum to the true vacuum per unit time per unit volume~\cite{Turner:1992tz}.
To compute this, it is necessary to solve for the bounce solution of the so-called Euclidean action ($S_3(T)$)~\cite{Linde:1981zj}. For this purpose, we employed the publicly available toolbox \texttt{CosmoTransitions}~\cite{Wainwright:2011kj}.
\Eq{nucleation} describes the condition under which the nucleation probability of a single bubble within a horizon volume becomes approximately unity. This translates to the criterion $S_3(T)/T \approx 140$. Solving this equation allows one to determine $T_n$, which corresponds to the maximum temperature at which $S_3/ T \lesssim 140$~\cite{Apreda:2001us}.

The dimensionless parameter $\alpha$ is defined as the latent heat ($\epsilon$) released during the phase transition to the radiation energy density ($\rho^*_{\text{rad}}$)~\cite{Espinosa:2010hh}:

\begin{small}
\beq
\label{alpha}
\alpha = \, \frac{\epsilon}{\rho^{*}_{\text{rad}}} = \, \frac{1}{\rho^{*}_{\text{rad}}} \left[ T \frac{\mathrm{d} \Delta V(T)}{\mathrm{d} T} - \Delta V(T) \right] \Bigg|_{T_*},
\eeq
\end{small}
where $T_*$ is the temperature at which the phase transition completes, corresponding approximately to $T_n$ in the absence of significant reheating.\footnote{More precisely, one may estimate the completion temperature of the phase transition using the so-called \emph{percolation temperature}, defined as the temperature at which a chosen fraction of the Universe’s volume (commonly $1/e \approx 37\%$) has converted to the true vacuum. In scenarios with substantial supercooling, the percolation and nucleation temperatures can differ significantly. A detailed computation of the percolation temperature, however, lies beyond the scope of this work.} Hence,
$\rho^*_{\text{rad}}  =  g_* \pi^2 T_*^4/30$, where $g_*$ denotes the number of relativistic dof at $T=T_*$. Here, we consider $g_* \sim 100$ for $T_n$ around the EW scale. 
The potential energy difference between the false and true vacua is defined as $\Delta V(T) = \, V_{\text{false}}(T)-V_{\text{true}}(T)$.

The characteristic time scale of the phase transition is captured by the inverse duration parameter $\beta$, which is defined as
\begin{small}
\beq
\label{beta}
\beta  =   -\frac{d S_3}{dt}\Bigr|_{t_*}  \, \simeq \, \dfrac{\dot{\Gamma}}{\Gamma} = \, H_* T_*  \frac{d(S_3/T)}{dT} \Bigr|_{T_*}  \, ,
\eeq
\end{small}
where $H_*$ is the Hubble rate at $T_*$. Another key quantity $v_w$  represents the velocity of the expanding bubble walls.

The energy released during the phase transition is distributed between plasma kinetic energy, which induces bulk fluid motion and generates GW, and thermal energy, which reheats the plasma. The fraction of the released energy converted into fluid motion is characterized by the efficiency factor $\kappa_v$, given by~\cite{Espinosa:2010hh}:
\begin{small}
\begin{equation}\label{kappav}
\kappa_v (\alpha) \, \simeq \, \left[  \dfrac{\alpha} {0.73+0.083\sqrt{\alpha}+\alpha} \right]\,\, \,.
\end{equation}
\end{small}
This semi-analytical formula for the efficiency factor is valid for wall speeds $v_w \sim 1$ and requires modifications for lower wall velocities~\cite{Espinosa:2010hh}.
Additionally, a portion of this kinetic energy contributes to Magneto-Hydrodynamic (MHD) turbulence in the plasma, quantified via 
$\kappa_{\text{turb}}$, which is typically estimated as $\kappa_\text{turb} \approx (5\sim 10) \, \kappa_v$ from numerical simulations~\cite{Hindmarsh:2015qta}. For this work,   we adopt a fiducial value of $\kappa_{\text{turb}} = 0.1$.
With these parameters established, we are now equipped to compute the resulting GW energy density spectrum.

The contribution of the GW spectrum from sound waves, denoted as $\Omega_{\text{sw}}h^2$, can be approximated by the following empirical formula~\cite{Hindmarsh:2019phv}:
\begin{small}
\begin{align}
\label{soundgw}
\Omega_{\text{SW}}h^2  =  2.65 \times 10^{-6} \Upsilon(\tau_{SW}) \left(\frac{\beta}{H_{\star}}\right)^{-1} v_{w} 
\left(\frac{\kappa_{v} \alpha}{1+\alpha}\right)^2
\left(\frac{g_*}{100}\right)^{-\frac{1}{3}} \left(\frac{f}{f_{\text{SW}}}\right)^3 
\left[\frac{7}{4 + 3 \left(\frac{f}{f_{\text{SW}}}\right)^2}\right]^{\frac{7}{2}}.
\end{align}
\end{small}
The characteristic peak frequency associated with the sound wave contribution is given by:
\begin{small}
\begin{equation}
\label{fsw}
f_{\text{SW}} \, = \, 1.9\times10^{-5}\hspace{1mm} \text{Hz} \, \left( \dfrac{1}{v_{w}}\right)\left(\dfrac{\beta}{H_{\star}} \right) \left(\dfrac{T_n}{100 \hspace{1mm} \text{GeV}} \right) \left(\dfrac{g_*}{100}\right)^{\frac{1}{6}}\,\,.
\end{equation}
\end{small}
By evaluating~\Eq{soundgw} at $f=f_{\text{SW}}$, we obtain the peak amplitude of the GW power spectrum contribution from sound waves, denoted as $\Omega_{\text{SW}}h^2_{\text{peak}}$.
In \Eq{soundgw}, $\Upsilon(\tau_{SW})$ represents the suppression factor in $\Omega_{\text{sw}}h^2$ due to the consideration of finite lifetime of the sound waves~\cite{Guo:2020grp, Hindmarsh:2020hop}. This is given by, 
\begin{small}
\beq
\Upsilon(\tau_{SW}) = 1 - \frac{1}{\sqrt{1 + 2 \tau_{\text{sw}} H_{\ast}}} \, \, .
\label{eq:upsilon}
\eeq
\end{small}
Here, the lifetime $\tau_{\text{sw}}$ characterizes the time scale over which turbulence emerges and can be approximated as~\cite{Pen:2015qta,Hindmarsh:2017gnf}:
\begin{small}
\begin{eqnarray}
\tau_{\text{sw}} \sim \frac{R_{\ast}}{\bar{U}_f} \, \, ,
\end{eqnarray}
\end{small}
where $R_{\ast} = (8\pi)^{1/3} v_w /\beta$ represents the typical separation between bubbles~\cite{Hindmarsh:2019phv, Guo:2020grp}., and   $\bar{U}_f$ denotes the root-mean-squared (RMS) velocity of the fluid and from hydrodynamic analyses it is given by, $\bar{U}_f = \sqrt{3 \kappa_v \alpha/4}$~\cite{Hindmarsh:2019phv,Weir:2017wfa}. 
At $\tau_{\text{sw}} \rightarrow \infty$, $\Upsilon(\tau_{SW})$ approaches the asymptotic value $1$.

The contribution of MHD turbulence to the GW spectrum is modeled by the following relation~\cite{Caprini:2015zlo}:
\begin{small}
\begin{align}
\label{GWturb}
\Omega_{\text{turb}}h^2 &= 3.35 \times 10^{-4} \left(\frac{H_*}{\beta}\right) \left(\frac{\kappa_{\text{turb}} \alpha}{1+\alpha}\right)^{\frac{3}{2}} 
\left(\frac{100}{g_s}\right)^{\frac{1}{3}} v_w  \frac{(f / f_{\text{turb}})^3}{\left[1 + (f / f_{\text{turb}})\right]^{\frac{11}{3}} \left(1 + 8 \pi f / h_{\star}\right)} \, \, .
\end{align}
\end{small}
The corresponding peak frequency for this contribution is expressed as,
\begin{small}
\beq
\label{peakfreqturb}
f_{\text{turb}} \, = \,  2.7 \times 10^{-5}~\text{Hz} \, \, \frac{1}{v_w}\left(\frac{\beta}{H_*}\right)\left(\frac{T_*}{100~\text{GeV}}\right)\left(\frac{g_s}{100}\right)^{\frac{1}{6}},
\eeq
\end{small}
with the parameter
\begin{small}
\begin{equation}\label{eq:h-star}
h_{*} \,= \, 16.5\times10^{-6}\hspace{1mm} \text{Hz}  \left(\dfrac{T_n}{100 \hspace{1mm} \text{GeV}} \right)  \left(\dfrac{g_*}{100}\right)^{\frac{1}{6}}\,\,.
\end{equation}
\end{small}
Before concluding our discussion of GW production from an FOPT, it is essential to address the role of the bubble-wall velocity, $v_w$, and its implications for both GW signals and EWBG.
As discussed above, larger $v_w$ generally enhances the GW signal.
 However, for the successful generation of the observed matter–antimatter asymmetry via EWBG, the wall velocity must be subsonic. This presents a challenge, as large $v_w$ values that support detectable GW signals may simultaneously hinder the production of the observed baryon asymmetry. Recent studies suggest that $v_w$ does not by itself control EWBG; rather, the plasma velocity profile around the bubble wall plays a critical role~\cite{No:2011fi}. A comprehensive study of transport dynamics in the vicinity of the wall is therefore required to evaluate these effects accurately, which we defer to future work.\footnote{For recent developments on estimating $v_w$, see Refs.~\cite{Laurent:2022jrs, Ekstedt:2024fyq, Carena:2025flp}. In certain BSM scenarios with weakly interacting species coupling to the wall, they observe $v_w \sim 0.6$ with only mild variation. Predictions for the GW signal at different $v_w$ can be seen in Ref.~\cite{Biekotter:2022kgf}.} For the purposes of this study, we assume that expanding bubbles achieve a relativistic terminal velocity in the plasma, approximately $v_w \sim 1$. When comparing results obtained using different resummation schemes, one should in principle estimate $v_w$ for each scheme, since it can introduce an additional source of uncertainty in the predicted GW amplitude. In this work we neglect this dependence, adopt a fixed $v_w$ for our comparison, and leave a detailed study to future work.

\section{Feynman-Hellmann trick and the iterative method for solving the gap equation}
\label{app:Feyn}
Solving the gap equation keeping all of the field dependence of the CP odd and charged scalars is a very challenging task. Because of this, it is useful to use a trick to calculate derivatives of the mass eigenvalues from the original mass matrix and the mixing angles. In quantum mechanics, this is often associated to the Feynman-Hellmann theorem that relates the derivatives of the total energy with respect to some parameter to the expectation value of the derivatives of the Hamiltonian operator. Here discuss the Feynman-Hellmann theorem and how to write the gap equation only as a function of the mass eigenvalues and mixing angles. This last step is important to be able to solve the gap equation iteratively.

We start with the eigenvalue equation,
\begin{equation}
    \mathbb{M}^2 \ket{n} = m_n^2 \ket{n}, \qquad n=h,H,A_0,G_0,H^\pm, G^\pm
    \label{eq:eigenvalue_eq}
\end{equation}

\noindent where $\mathbb{M}^2$ is the field dependent mass matrix, $m_n^2$ is the field dependent mass eigenvalue and $\ket{n}$ are the mass eigenstates. Taking one derivative with respect to one of the field dof $\phi_a = \{h_1, h_2, a_1, a_2, \phi_1^\pm, \phi_2^\pm\}$ of \Eq{eq:eigenvalue_eq}, we have
\begin{equation}
    \frac{d \mathbb{M}^2}{d\phi_a} \ket{n} + \mathbb{M}^2 \frac{d\ket{n}}{d\phi_a} = \frac{d m_n^2}{d\phi_a} \ket{n} + m_n^2 \frac{d\ket{n}}{d\phi_a}.
    \label{eq:first_derivatives_n}
\end{equation}

\noindent we can take the matrix elements of this equation between the state $\bra{l}$, we have
\begin{equation}
    \bra{l}\frac{d \mathbb{M}^2}{d\phi_a} \ket{n} = \frac{d m_n^2}{d\phi_a} \delta_{ln} + (m_n^2 -m_l^2)\bra{l}\frac{d}{d\phi_a}\ket{n}.
    \label{eq:first_derivatives_ln}
\end{equation}

\noindent Taking $l=n$, we arrive for an expression for the first derivative of the mass eigenvalue in terms of the derivative of the mass matrix,
\begin{equation}
    \frac{dm_n^2}{d\phi_a} = \left\langle n\right| \frac{d \mathbb{M}^2}{d\phi_a} \left|n\right\rangle.
    \label{eq:FH_first_derivative}
\end{equation}

\noindent This expression is particularly useful since we can take the field derivatives with respect to all field dof easily, then set them to zero and keep only the CP even components. We can write \Eq{eq:FH_first_derivative} in matrix form by relating the mass eingenbasis with the interaction basis by using the rotation matrices $U(\theta)$,
\begin{equation}
    \frac{dm_n^2}{d\phi_a} = \left( U^{-1}(\theta) \cdot \frac{d \mathbb{M}^2}{d\phi_a}\Big|_{h_1,h_2} \cdot U(\theta) \right)_{nn} \, \, . 
    \label{eq:FH_first_derivative1}
\end{equation}

\noindent Notice that after keeping only the $h_1,h_2$ components of $U$ and $\frac{d \mathbb{M}^2}{d\phi_a}$, the derivative of the mass eigenvalues are also dependent only on $h_1,h_2$. Importantly, we are able to keep the information on the other CP odd and charged field derivatives, i.e. $\frac{dm_n^2}{da_1},\frac{dm_n^2}{da_2},\dots$~. The rotation matrices are easily obtained from the original mass matrix.

Now, taking $l\neq n$ in \Eq{eq:first_derivatives_ln} and for non-degenerate mass eigenvalues $m_l^2 \neq m_n^2$, we have
\begin{equation}
    \bra{l}\frac{d}{d\phi_a}\ket{n} = \frac{1}{m_n^2 -m_l^2}\bra{l}\frac{d \mathbb{M}^2}{d\phi_a} \ket{n}, \qquad \text{for }m_l^2 \neq m_n^2 \, \, .
    \label{eq:exp1}
\end{equation}
while for the degenerate case we simply have
\begin{equation}
    \bra{l}\frac{d \mathbb{M}^2}{d\phi_a} \ket{n} = 0, \qquad \text{for }m_l^2 = m_n^2.
    \label{eq:exp2}
\end{equation}
These expressions will allow us to simplify the equation for the second derivative of the mass eigenvalues.
Taking one more field derivative of \Eq{eq:first_derivatives_n}, we arrive at
\begin{align}
    &\frac{d^2 \mathbb{M}^2}{d\phi_a d\phi_b} \ket{n} + \frac{d \mathbb{M}^2}{d\phi_a} \frac{d\ket{n}}{d\phi_b} + \frac{d\mathbb{M}^2}{d\phi_b} \frac{d\ket{n}}{d\phi_a} +\mathbb{M}^2 \frac{d^2\ket{n}}{d\phi_a d\phi_b} 
    \\ \nonumber
    &\hspace{3cm} = \frac{d^2 m_n^2}{d\phi_a d\phi_b} \ket{n} + \frac{d m_n^2}{d\phi_a} \frac{d\ket{n}}{d\phi_b} + \frac{d m_n^2}{d\phi_b} \frac{d\ket{n}}{d\phi_a} + m_n^2 \frac{d^2\ket{n}}{d\phi_a d\phi_b} \, \, .
\end{align}

\noindent Again, taking the matrix element with the state $\bra{l}$, we arrive at
\begin{align}
    &\bra{l}\frac{d^2 \mathbb{M}^2}{d\phi_a d\phi_b} \ket{n} + \sum_m \bra{l}\frac{d \mathbb{M}^2}{d\phi_a} \ket{m} \bra{m} \frac{d}{d\phi_b}\ket{n} + \sum_m \bra{l} \frac{d\mathbb{M}^2}{d\phi_b} \ket{m} \bra{m} \frac{d}{d\phi_a} \ket{n}
    \\ \nonumber
    &\hspace{1cm} = \frac{d^2 m_n^2}{d\phi_a d\phi_b} \delta_{nl}+ \frac{d m_n^2}{d\phi_a} \bra{l}\frac{d}{d\phi_b}\ket{n} + \frac{d m_n^2}{d\phi_b} \frac{d\ket{n}}{d\phi_a} + (m_n^2 - m_l^2) \bra{l} \frac{d^2}{d\phi_a d\phi_b}\ket{n}.
\end{align}

\noindent Where we have inserted the identity operators $\mathbb{1} = \sum_m \ket{m} \bra{m}$ to separate each matrix element. Using the previous expressions, Eqs.~\eqref{eq:FH_first_derivative1}, \eqref{eq:exp1} and \eqref{eq:exp2}, and taking $l=n$, we can write the final form of the second derivative of the mass eigenvalues
\begin{align}
    \nonumber \frac{d^2 m_n^2}{d\phi_a d\phi_b} =& \left( U^{-1}(\theta) \cdot \frac{d^2 \mathbb{M}^2}{d\phi_a d\phi_b}\Big|_{h_1,h_2} \cdot U(\theta) \right)_{nn} + \left( \left[U^{-1}(\theta) \cdot \frac{d \mathbb{M}^2}{d\phi_a}\Big|_{h_1,h_2} \cdot U(\theta) \right] \cdot \mathbb{A}_b\right)_{nn}
    \\
    &\hspace{3cm}+\left( \left[U^{-1}(\theta) \cdot \frac{d \mathbb{M}^2}{d\phi_b}\Big|_{h_1,h_2} \cdot U(\theta) \right] \cdot \mathbb{A}_a\right)_{nn} \, \, ,
\end{align}

\noindent where the auxiliary matrix $\mathbb{A}_{a,b}$ is given by
\begin{align}
    (\mathbb{A}_c)_{pq} = 
    \begin{cases}
        0 \qquad \text{if } p=q,
        \\
        0 \qquad \text{if } p\neq q \, \, , ~\text{ but }~ m_p^2 = m_q^2,
        \\
        \frac{1}{m_p^2 - m_q^2}\left[U^{-1}(\theta) \cdot \frac{d \mathbb{M}^2}{d\phi_c}\Big|_{h_1,h_2} \cdot U(\theta) \right]_{pq} \, \, , \qquad \text{else}.
    \end{cases}
    \label{eq:FH_second_derivative}
\end{align}
Eqs.~\eqref{eq:FH_first_derivative1} and \eqref{eq:FH_second_derivative} are useful because they allow to write the gap equation exclusively as a function of the mass eigenvalues and mixing angles. We only need to calculate the mixing angles of the CP even, CP odd and charged blocks of the $(h_1,h_2)$ field dependent mass matrix and evaluate the derivatives of the mass matrices, which have simple analytic expressions.

\section{Thermal Resummation and Symmetry (non)-Restoration behaviour in a Toy Model}
\label{snr-toy}
We consider a toy model with a one-dimensional potential. The tree-level potential is given by:
\begin{equation}
    V_0 = \frac{\lambda}{4} \phi^4.
\end{equation}
The field-dependent mass squared of the scalar field $\phi$ is:
\begin{equation}
    m_0^2 = 3\lambda\phi^2.
\end{equation}
Note that the global minimum of this potential is at $\phi = 0$.

As discussed in Sec.~\ref{Pot}, in the high temperature approximation, the logarithmic terms cancel between the Coleman-Weinberg (CW) correction, as defined in~\Eq{eq:CW_potential}, and the finite-temperature contributions, as shown in~\Eq{vt1loop}. Consequently, at the high temperature limit the finite-temperature potential can be expressed as:
\begin{align}
\label{v1dtot}
    V_{\text{tot}} &\approx V_0 + \frac{T^2}{24} m_0^2 - \frac{T}{12\pi} (m_0^2)^{3/2} - \frac{3}{2}\frac{m_0^4}{64\pi^2} + \frac{m_0^4}{64\pi^2}\log\left(\frac{a_b T^2}{\mu_R^2}\right),
\end{align}
where $a_b = 16\pi^2 \exp(3/2 - 2\gamma_E)$ and $\gamma_E \approx 0.577$ is the Euler-Mascheroni constant.

Let us assume that the Truncated thermal correction to the mass is given by $c T^2$ at the high temperature approximation. The thermally improved mass of $\phi$ then becomes:
\begin{equation}
    m^2 = m_0^2 + c T^2.
\end{equation}

In the AE prescription, as discussed in Sec.~\ref{AEpores}, $m_0^2$ is replaced by $m^2$ only in the term proportional to $(m_0^2)^{3/2}$ in~\Eq{v1dtot}. This term can be expressed as:
\begin{align}
    (m_0^2 + c T^2)^{3/2} &\rightarrow c^{3/2} T^3 \left(1 + \frac{m_0^2}{c T^2}\right)^{3/2} \\
    &\approx c^{3/2} T^3 \left(1 + \frac{3}{2}\frac{m_0^2}{c T^2}\right) = c^{3/2} T^3 + \frac{3\sqrt{c}}{2} m_0^2 T.
\end{align}

Here, we assume $c T^2 \gg m_0^2$ in the high temperature approximation and retain only the leading-order term.
To investigate symmetry non-restoration, we focus on the coefficient of the $\phi^2$ term, which is proportional to $m_0^2 (= 3\lambda\phi^2)$. We denote this coefficient by $S_{m_0^2}(\text{AE})$, which can be expressed as:
\begin{align}
\label{m3term}
    S_{m_0^2}(\text{AE}) &= \frac{T^2}{24} + \frac{3\sqrt{c}}{2} T \left(-\frac{T}{12\pi}\right) \\
    &= \frac{T^2}{24} \left(1 - \frac{3\sqrt{c}}{\pi}\right). \nonumber
\end{align}

Therefore, if $c > \frac{\pi^2}{9}$, the coefficient $S_{m_0^2}(\text{AE})$ can become negative, causing the $\phi^2$ term to break symmetry at high temperatures. This provides a simple example of how the daisy-resummed term in the AE prescription can lead to symmetry non-restoration under large thermal corrections.

In the Parwani prescription, as discussed in Sec.~\ref{Parwanipres}, $m_0^2$ is replaced by $m^2$ throughout the potential. In the potential defined in~\Eq{v1dtot}, the replacement of $m_0^2 \rightarrow m^2$ in the $\frac{T^2}{24} m_0^2$ term is inconsequential for studying symmetry non-restoration, as the correction remains field-independent in the total potential. 
The primary distinction of the Parwani prescription, compared to the AE prescription at the one-loop level, lies in the resummation of the quartic power of the mass terms. Specifically, the $\phi^2$-dependent term arises as $2 \, c \, T^2 m_0^2$ from the expansion of $(m_0^2 + c T^2)^2$. Consequently, the relevant contribution from the quartic term in the potential, as defined in~\Eq{v1dtot}, is expressed as: 
\begin{align}
\label{m4term}
   \frac{m_0^4}{64\pi^2} \bigg(- \frac{3}{2} + \log\left(\frac{a_b T^2}{\mu_R^2}\right)\bigg) \rightarrow \frac{ c \, T^2 m_0^2}{32\pi^2} \bigg(- \frac{3}{2} + \log\left(\frac{a_b T^2}{\mu_R^2}\right)\bigg).
\end{align}

Combining the contributions from~\Eq{m3term} and~\Eq{m4term}, the coefficient of the $\phi^2$ term, denoted as $S_{m_0^2}(\text{PW})$, is given by:
\begin{align}
\label{pwsm0}
    S_{m_0^2}(\text{PW}) &= \frac{T^2}{24} + \frac{3\sqrt{c}}{2} T \left(-\frac{T}{12\pi}\right) 
    + \frac{c \, T^2}{32\pi^2} \bigg(-\frac{3}{2} + \log\left(\frac{a_b T^2}{\mu_R^2}\right)\bigg) \nonumber \\
    &= \frac{T^2}{24} \bigg(1 - \frac{3\sqrt{c}}{\pi} + \frac{3 c}{4\pi^2} \bigg(-\frac{3}{2} + \log\left(\frac{a_b T^2}{\mu_R^2}\right)\bigg)\bigg) \,.
\end{align}
Note that \(\log(a_b) = \log(16\pi^2 \exp(3/2 - 2\gamma_E)) \sim 5.4\). Consequently, the coefficient of the term proportional to `$c$' can be expressed as 
$
\frac{3}{4 \pi^2}(3.9 + \log(T^2/\mu_R^2)).
$
Importantly, the coefficient of `$c$' increases with temperature due to the \(\log(T)\) dependence in its expression. It can be shown that for all positive values of `$c$', \(S_{m_0^2}(\text{PW})\) can always remain positive at sufficiently high temperatures. Consequently, the \(\phi^2\) term does not become negative, ensuring that symmetry is not broken at high temperatures.~\footnote{This occurs due to the inclusion of the resummation of the fourth power term in~\Eq{v1dtot} as part of the Parwani prescription.}
This conclusion can be verified using the following relations.

Define the coefficient of the term proportional to `\(c\)' in~\Eq{pwsm0} as $d$, where 
$d \,=  \frac{3}{4\pi^2} \bigg(-\frac{3}{2} + \log\left(\frac{a_b T^2}{\mu_R^2}\right)\bigg)$,
and let \(c = x^2\). Then, \(S_{m_0^2}(\text{PW})\) can be expressed as 
\begin{align}
\label{conditionSNR}
\left(\frac{T^2 d}{24}\right)^{-1}  S_{m_0^2}({\rm PW}) = \, x^2 - \frac{3}{\pi d} x + \frac{1}{d} 
&= \bigg(x - \frac{3}{2\pi d}\bigg)^2 + \frac{1}{d^2} \bigg(d - \frac{9}{4 \pi^2}\bigg) \,.
\end{align}
From this expression, it is clear that the first term on the right-hand side is always positive. If \(d > \frac{9}{4\pi^2}\), the second term is also positive. Hence, a sufficient condition for symmetry restoration at high temperatures is \(d > \frac{9}{4\pi^2}\), which implies $T > e^{-0.45} \mu_R $ criteria, following the above mentioned relation of `$d$'. Note that in this  one-dimensional toy model, this condition is independent of the thermal correction coefficient `$c$'. 
Although, the overall thermal correction coefficient `$c$' can be modified at high temperatures when computed by solving the full gap equations. However, incorporating the thermal mass at the tadpole level within the PD scheme yields a condition similar to~\Eq{conditionSNR}, which also suggests symmetry restoration at high temperatures. Thus, while both the Parwani and PD resummation prescriptions predict symmetry restoration at high temperatures, the AE approach predicts symmetry non-restoration.
This behavior in the AE prescription arises because it only resums the Matsubara zero modes at the one-loop level, whereas the other prescriptions resum contributions from all modes, enabling symmetry restoration. 
Consequently, it is plausible that symmetry restoration may also emerge within the AE scheme once two-loop or higher-order corrections are included, as these would improve the treatment of the non-zero modes, and we plan to investigate this in future work.

\bibliography{EWNR.bib}

\providecommand{\href}[2]{#2}\begingroup\raggedright\begin{thebibliography}{100}

\bibitem{ATLAS:2012yve}
{\scshape ATLAS} collaboration, \emph{{Observation of a new particle in the search for the Standard Model Higgs boson with the ATLAS detector at the LHC}}, \href{https://doi.org/10.1016/j.physletb.2012.08.020}{\emph{Phys. Lett. B} {\bfseries 716} (2012) 1} [\href{https://arxiv.org/abs/1207.7214}{{\ttfamily 1207.7214}}].

\bibitem{CMS:2012qbp}
{\scshape CMS} collaboration, \emph{{Observation of a New Boson at a Mass of 125 GeV with the CMS Experiment at the LHC}}, \href{https://doi.org/10.1016/j.physletb.2012.08.021}{\emph{Phys. Lett. B} {\bfseries 716} (2012) 30} [\href{https://arxiv.org/abs/1207.7235}{{\ttfamily 1207.7235}}].

\bibitem{Curtin:2014jma}
D.~Curtin, P.~Meade and C.-T.~Yu, \emph{{Testing Electroweak Baryogenesis with Future Colliders}}, \href{https://doi.org/10.1007/JHEP11(2014)127}{\emph{JHEP} {\bfseries 11} (2014) 127} [\href{https://arxiv.org/abs/1409.0005}{{\ttfamily 1409.0005}}].

\bibitem{Papaefstathiou:2020iag}
A.~Papaefstathiou and G.~White, \emph{{The electro-weak phase transition at colliders: confronting theoretical uncertainties and complementary channels}}, \href{https://doi.org/10.1007/JHEP05(2021)099}{\emph{JHEP} {\bfseries 05} (2021) 099} [\href{https://arxiv.org/abs/2010.00597}{{\ttfamily 2010.00597}}].

\bibitem{Ramsey-Musolf:2019lsf}
M.J.~Ramsey-Musolf, \emph{{The electroweak phase transition: a collider target}}, \href{https://doi.org/10.1007/JHEP09(2020)179}{\emph{JHEP} {\bfseries 09} (2020) 179} [\href{https://arxiv.org/abs/1912.07189}{{\ttfamily 1912.07189}}].

\bibitem{LISA:2017pwj}
{\scshape LISA} collaboration, \emph{{Laser Interferometer Space Antenna}},  \href{https://arxiv.org/abs/1702.00786}{{\ttfamily 1702.00786}}.

\bibitem{Gong:2014mca}
X.~Gong et~al., \emph{{Descope of the ALIA mission}}, \href{https://doi.org/10.1088/1742-6596/610/1/012011}{\emph{J. Phys. Conf. Ser.} {\bfseries 610} (2015) 012011} [\href{https://arxiv.org/abs/1410.7296}{{\ttfamily 1410.7296}}].

\bibitem{Hu:2017mde}
W.-R.~Hu and Y.-L.~Wu, \emph{{The Taiji Program in Space for gravitational wave physics and the nature of gravity}}, \href{https://doi.org/10.1093/nsr/nwx116}{\emph{Natl. Sci. Rev.} {\bfseries 4} (2017) 685}.

\bibitem{Corbin:2005ny}
V.~Corbin and N.J.~Cornish, \emph{{Detecting the cosmic gravitational wave background with the big bang observer}}, \href{https://doi.org/10.1088/0264-9381/23/7/014}{\emph{Class. Quant. Grav.} {\bfseries 23} (2006) 2435} [\href{https://arxiv.org/abs/gr-qc/0512039}{{\ttfamily gr-qc/0512039}}].

\bibitem{Kudoh:2005as}
H.~Kudoh, A.~Taruya, T.~Hiramatsu and Y.~Himemoto, \emph{{Detecting a gravitational-wave background with next-generation space interferometers}}, \href{https://doi.org/10.1103/PhysRevD.73.064006}{\emph{Phys. Rev. D} {\bfseries 73} (2006) 064006} [\href{https://arxiv.org/abs/gr-qc/0511145}{{\ttfamily gr-qc/0511145}}].

\bibitem{Sesana:2019vho}
A.~Sesana et~al., \emph{{Unveiling the gravitational universe at $\mu$-Hz frequencies}}, \href{https://doi.org/10.1007/s10686-021-09709-9}{\emph{Exper. Astron.} {\bfseries 51} (2021) 1333} [\href{https://arxiv.org/abs/1908.11391}{{\ttfamily 1908.11391}}].

\bibitem{Caprini:2019egz}
C.~Caprini et~al., \emph{{Detecting gravitational waves from cosmological phase transitions with LISA: an update}}, \href{https://doi.org/10.1088/1475-7516/2020/03/024}{\emph{JCAP} {\bfseries 03} (2020) 024} [\href{https://arxiv.org/abs/1910.13125}{{\ttfamily 1910.13125}}].

\bibitem{Yagi:2011wg}
K.~Yagi and N.~Seto, \emph{{Detector configuration of DECIGO/BBO and identification of cosmological neutron-star binaries}}, \href{https://doi.org/10.1103/PhysRevD.83.044011}{\emph{Phys. Rev. D} {\bfseries 83} (2011) 044011} [\href{https://arxiv.org/abs/1101.3940}{{\ttfamily 1101.3940}}].

\bibitem{Punturo:2010zz}
M.~Punturo et~al., \emph{{The Einstein Telescope: A third-generation gravitational wave observatory}}, \href{https://doi.org/10.1088/0264-9381/27/19/194002}{\emph{Class. Quant. Grav.} {\bfseries 27} (2010) 194002}.

\bibitem{Hild:2010id}
S.~Hild et~al., \emph{{Sensitivity Studies for Third-Generation Gravitational Wave Observatories}}, \href{https://doi.org/10.1088/0264-9381/28/9/094013}{\emph{Class. Quant. Grav.} {\bfseries 28} (2011) 094013} [\href{https://arxiv.org/abs/1012.0908}{{\ttfamily 1012.0908}}].

\bibitem{Sakharov:1967dj}
A.D.~Sakharov, \emph{{Violation of CP Invariance, C asymmetry, and baryon asymmetry of the universe}}, \href{https://doi.org/10.1070/PU1991v034n05ABEH002497}{\emph{Pisma Zh. Eksp. Teor. Fiz.} {\bfseries 5} (1967) 32}.

\bibitem{Cohen:1993nk}
A.G.~Cohen, D.B.~Kaplan and A.E.~Nelson, \emph{{Progress in electroweak baryogenesis}}, \href{https://doi.org/10.1146/annurev.ns.43.120193.000331}{\emph{Ann. Rev. Nucl. Part. Sci.} {\bfseries 43} (1993) 27} [\href{https://arxiv.org/abs/hep-ph/9302210}{{\ttfamily hep-ph/9302210}}].

\bibitem{Rubakov:1996vz}
V.A.~Rubakov and M.E.~Shaposhnikov, \emph{{Electroweak baryon number nonconservation in the early universe and in high-energy collisions}}, \href{https://doi.org/10.1070/PU1996v039n05ABEH000145}{\emph{Usp. Fiz. Nauk} {\bfseries 166} (1996) 493} [\href{https://arxiv.org/abs/hep-ph/9603208}{{\ttfamily hep-ph/9603208}}].

\bibitem{Trodden:1998ym}
M.~Trodden, \emph{{Electroweak baryogenesis}}, \href{https://doi.org/10.1103/RevModPhys.71.1463}{\emph{Rev. Mod. Phys.} {\bfseries 71} (1999) 1463} [\href{https://arxiv.org/abs/hep-ph/9803479}{{\ttfamily hep-ph/9803479}}].

\bibitem{Riotto:1998bt}
A.~Riotto, \emph{{Theories of baryogenesis}},  in \emph{{ICTP Summer School in High-Energy Physics and Cosmology}}, pp.~326--436, 7, 1998 [\href{https://arxiv.org/abs/hep-ph/9807454}{{\ttfamily hep-ph/9807454}}].

\bibitem{Riotto:1999yt}
A.~Riotto and M.~Trodden, \emph{{Recent progress in baryogenesis}}, \href{https://doi.org/10.1146/annurev.nucl.49.1.35}{\emph{Ann. Rev. Nucl. Part. Sci.} {\bfseries 49} (1999) 35} [\href{https://arxiv.org/abs/hep-ph/9901362}{{\ttfamily hep-ph/9901362}}].

\bibitem{Cline:2006ts}
J.M.~Cline, \emph{{Baryogenesis}},  in \emph{{Les Houches Summer School - Session 86: Particle Physics and Cosmology: The Fabric of Spacetime}}, 9, 2006 [\href{https://arxiv.org/abs/hep-ph/0609145}{{\ttfamily hep-ph/0609145}}].

\bibitem{Morrissey:2012db}
D.E.~Morrissey and M.J.~Ramsey-Musolf, \emph{{Electroweak baryogenesis}}, \href{https://doi.org/10.1088/1367-2630/14/12/125003}{\emph{New J. Phys.} {\bfseries 14} (2012) 125003} [\href{https://arxiv.org/abs/1206.2942}{{\ttfamily 1206.2942}}].

\bibitem{Davoudiasl:2004gf}
H.~Davoudiasl, R.~Kitano, G.D.~Kribs, H.~Murayama and P.J.~Steinhardt, \emph{{Gravitational baryogenesis}}, \href{https://doi.org/10.1103/PhysRevLett.93.201301}{\emph{Phys. Rev. Lett.} {\bfseries 93} (2004) 201301} [\href{https://arxiv.org/abs/hep-ph/0403019}{{\ttfamily hep-ph/0403019}}].

\bibitem{Hooper:2025fda}
D.~Hooper, G.~Krnjaic, D.~Rocha and S.~Roy, \emph{{Gamma-Rays and Gravitational Waves from Inelastic Higgs Portal Dark Matter}},  \href{https://arxiv.org/abs/2507.22975}{{\ttfamily 2507.22975}}.

\bibitem{White:2016nbo}
G.A.~White, \emph{{A Pedagogical Introduction to Electroweak Baryogenesis}}, .

\bibitem{Chatterjee:2022pxf}
A.~Chatterjee, A.~Datta and S.~Roy, \emph{{Electroweak phase transition in the Z$_{3}$-invariant NMSSM: Implications of LHC and Dark matter searches and prospects of detecting the gravitational waves}}, \href{https://doi.org/10.1007/JHEP06(2022)108}{\emph{JHEP} {\bfseries 06} (2022) 108} [\href{https://arxiv.org/abs/2202.12476}{{\ttfamily 2202.12476}}].

\bibitem{Wagner:2023vqw}
C.E.M.~Wagner, \emph{{Electroweak Baryogenesis and Higgs Physics}}, \href{https://doi.org/10.31526/lhep.2023.466}{\emph{LHEP} {\bfseries 2023} (2023) 466} [\href{https://arxiv.org/abs/2311.06949}{{\ttfamily 2311.06949}}].

\bibitem{Schwaller:2015tja}
P.~Schwaller, \emph{{Gravitational Waves from a Dark Phase Transition}}, \href{https://doi.org/10.1103/PhysRevLett.115.181101}{\emph{Phys. Rev. Lett.} {\bfseries 115} (2015) 181101} [\href{https://arxiv.org/abs/1504.07263}{{\ttfamily 1504.07263}}].

\bibitem{Baldes:2018emh}
I.~Baldes and C.~Garcia-Cely, \emph{{Strong gravitational radiation from a simple dark matter model}}, \href{https://doi.org/10.1007/JHEP05(2019)190}{\emph{JHEP} {\bfseries 05} (2019) 190} [\href{https://arxiv.org/abs/1809.01198}{{\ttfamily 1809.01198}}].

\bibitem{Breitbach:2018ddu}
M.~Breitbach, J.~Kopp, E.~Madge, T.~Opferkuch and P.~Schwaller, \emph{{Dark, Cold, and Noisy: Constraining Secluded Hidden Sectors with Gravitational Waves}}, \href{https://doi.org/10.1088/1475-7516/2019/07/007}{\emph{JCAP} {\bfseries 07} (2019) 007} [\href{https://arxiv.org/abs/1811.11175}{{\ttfamily 1811.11175}}].

\bibitem{Croon:2018erz}
D.~Croon, V.~Sanz and G.~White, \emph{{Model Discrimination in Gravitational Wave spectra from Dark Phase Transitions}}, \href{https://doi.org/10.1007/JHEP08(2018)203}{\emph{JHEP} {\bfseries 08} (2018) 203} [\href{https://arxiv.org/abs/1806.02332}{{\ttfamily 1806.02332}}].

\bibitem{Hall:2019ank}
E.~Hall, T.~Konstandin, R.~McGehee, H.~Murayama and G.~Servant, \emph{{Baryogenesis From a Dark First-Order Phase Transition}}, \href{https://doi.org/10.1007/JHEP04(2020)042}{\emph{JHEP} {\bfseries 04} (2020) 042} [\href{https://arxiv.org/abs/1910.08068}{{\ttfamily 1910.08068}}].

\bibitem{Baldes:2017rcu}
I.~Baldes, \emph{{Gravitational waves from the asymmetric-dark-matter generating phase transition}}, \href{https://doi.org/10.1088/1475-7516/2017/05/028}{\emph{JCAP} {\bfseries 05} (2017) 028} [\href{https://arxiv.org/abs/1702.02117}{{\ttfamily 1702.02117}}].

\bibitem{Geller:2018mwu}
M.~Geller, A.~Hook, R.~Sundrum and Y.~Tsai, \emph{{Primordial Anisotropies in the Gravitational Wave Background from Cosmological Phase Transitions}}, \href{https://doi.org/10.1103/PhysRevLett.121.201303}{\emph{Phys. Rev. Lett.} {\bfseries 121} (2018) 201303} [\href{https://arxiv.org/abs/1803.10780}{{\ttfamily 1803.10780}}].

\bibitem{Croon:2019rqu}
D.~Croon, A.~Kusenko, A.~Mazumdar and G.~White, \emph{{Solitosynthesis and Gravitational Waves}}, \href{https://doi.org/10.1103/PhysRevD.101.085010}{\emph{Phys. Rev. D} {\bfseries 101} (2020) 085010} [\href{https://arxiv.org/abs/1910.09562}{{\ttfamily 1910.09562}}].

\bibitem{Hall:2019rld}
E.~Hall, T.~Konstandin, R.~McGehee and H.~Murayama, \emph{{Asymmetric matter from a dark first-order phase transition}}, \href{https://doi.org/10.1103/PhysRevD.107.055011}{\emph{Phys. Rev. D} {\bfseries 107} (2023) 055011} [\href{https://arxiv.org/abs/1911.12342}{{\ttfamily 1911.12342}}].

\bibitem{Chao:2020adk}
W.~Chao, X.-F.~Li and L.~Wang, \emph{{Filtered pseudo-scalar dark matter and gravitational waves from first order phase transition}}, \href{https://doi.org/10.1088/1475-7516/2021/06/038}{\emph{JCAP} {\bfseries 06} (2021) 038} [\href{https://arxiv.org/abs/2012.15113}{{\ttfamily 2012.15113}}].

\bibitem{Ghosh:2022fzp}
P.~Ghosh, T.~Ghosh and S.~Roy, \emph{{Interplay among gravitational waves, dark matter and collider signals in the singlet scalar extended type-II seesaw model}}, \href{https://doi.org/10.1007/JHEP10(2023)057}{\emph{JHEP} {\bfseries 10} (2023) 057} [\href{https://arxiv.org/abs/2211.15640}{{\ttfamily 2211.15640}}].

\bibitem{Roy:2022gop}
S.~Roy, \emph{{Dilution of Dark Matter Relic abundance due to First Order Electroweak Phase Transition}},  \href{https://arxiv.org/abs/2212.11230}{{\ttfamily 2212.11230}}.

\bibitem{Dent:2022bcd}
J.B.~Dent, B.~Dutta, S.~Ghosh, J.~Kumar and J.~Runburg, \emph{{Sensitivity to dark sector scales from gravitational wave signatures}}, \href{https://doi.org/10.1007/JHEP08(2022)300}{\emph{JHEP} {\bfseries 08} (2022) 300} [\href{https://arxiv.org/abs/2203.11736}{{\ttfamily 2203.11736}}].

\bibitem{Borah:2024emz}
P.~Borah, P.~Ghosh and A.K.~Saha, \emph{{Prospecting bipartite Dark Matter through Gravitational Waves}},  \href{https://arxiv.org/abs/2412.17141}{{\ttfamily 2412.17141}}.

\bibitem{Weinberg:1974hy}
S.~Weinberg, \emph{{Gauge and Global Symmetries at High Temperature}}, \href{https://doi.org/10.1103/PhysRevD.9.3357}{\emph{Phys. Rev. D} {\bfseries 9} (1974) 3357}.

\bibitem{Meade:2018saz}
P.~Meade and H.~Ramani, \emph{{Unrestored Electroweak Symmetry}}, \href{https://doi.org/10.1103/PhysRevLett.122.041802}{\emph{Phys. Rev. Lett.} {\bfseries 122} (2019) 041802} [\href{https://arxiv.org/abs/1807.07578}{{\ttfamily 1807.07578}}].

\bibitem{Baldes:2018nel}
I.~Baldes and G.~Servant, \emph{{High scale electroweak phase transition: baryogenesis \textbackslash{}\& symmetry non-restoration}}, \href{https://doi.org/10.1007/JHEP10(2018)053}{\emph{JHEP} {\bfseries 10} (2018) 053} [\href{https://arxiv.org/abs/1807.08770}{{\ttfamily 1807.08770}}].

\bibitem{Cline:1999wi}
J.M.~Cline, G.D.~Moore and G.~Servant, \emph{{Was the electroweak phase transition preceded by a color broken phase?}}, \href{https://doi.org/10.1103/PhysRevD.60.105035}{\emph{Phys. Rev. D} {\bfseries 60} (1999) 105035} [\href{https://arxiv.org/abs/hep-ph/9902220}{{\ttfamily hep-ph/9902220}}].

\bibitem{Baum:2020vfl}
S.~Baum, M.~Carena, N.R.~Shah, C.E.M.~Wagner and Y.~Wang, \emph{{Nucleation is more than critical: A case study of the electroweak phase transition in the NMSSM}}, \href{https://doi.org/10.1007/JHEP03(2021)055}{\emph{JHEP} {\bfseries 03} (2021) 055} [\href{https://arxiv.org/abs/2009.10743}{{\ttfamily 2009.10743}}].

\bibitem{Biekotter:2021ysx}
T.~Biek\"otter, S.~Heinemeyer, J.M.~No, M.O.~Olea and G.~Weiglein, \emph{{Fate of electroweak symmetry in the early Universe: Non-restoration and trapped vacua in the N2HDM}}, \href{https://doi.org/10.1088/1475-7516/2021/06/018}{\emph{JCAP} {\bfseries 06} (2021) 018} [\href{https://arxiv.org/abs/2103.12707}{{\ttfamily 2103.12707}}].

\bibitem{Biekotter:2022kgf}
T.~Biek\"otter, S.~Heinemeyer, J.M.~No, M.O.~Olea-Romacho and G.~Weiglein, \emph{{The trap in the early Universe: impact on the interplay between gravitational waves and LHC physics in the 2HDM}}, \href{https://doi.org/10.1088/1475-7516/2023/03/031}{\emph{JCAP} {\bfseries 03} (2023) 031} [\href{https://arxiv.org/abs/2208.14466}{{\ttfamily 2208.14466}}].

\bibitem{Chang:2022psj}
J.H.~Chang, M.O.~Olea-Romacho and E.H.~Tanin, \emph{{Electroweak asymmetric early Universe via a scalar condensate}}, \href{https://doi.org/10.1103/PhysRevD.106.113003}{\emph{Phys. Rev. D} {\bfseries 106} (2022) 113003} [\href{https://arxiv.org/abs/2210.05680}{{\ttfamily 2210.05680}}].

\bibitem{Aoki:2023lbz}
M.~Aoki, L.~Biermann, C.~Borschensky, I.P.~Ivanov, M.~M\"uhlleitner and H.~Shibuya, \emph{{Intermediate charge-breaking phases and symmetry non-restoration in the 2-Higgs-Doublet Model}}, \href{https://doi.org/10.1007/JHEP02(2024)232}{\emph{JHEP} {\bfseries 02} (2024) 232} [\href{https://arxiv.org/abs/2308.04141}{{\ttfamily 2308.04141}}].

\bibitem{Huang:2017laj}
F.P.~Huang and X.~Zhang, \emph{{Probing the gauge symmetry breaking of the early universe in 3-3-1 models and beyond by gravitational waves}}, \href{https://doi.org/10.1016/j.physletb.2018.11.024}{\emph{Phys. Lett. B} {\bfseries 788} (2019) 288} [\href{https://arxiv.org/abs/1701.04338}{{\ttfamily 1701.04338}}].

\bibitem{Croon:2018kqn}
D.~Croon, T.E.~Gonzalo and G.~White, \emph{{Gravitational Waves from a Pati-Salam Phase Transition}}, \href{https://doi.org/10.1007/JHEP02(2019)083}{\emph{JHEP} {\bfseries 02} (2019) 083} [\href{https://arxiv.org/abs/1812.02747}{{\ttfamily 1812.02747}}].

\bibitem{Hashino:2018zsi}
K.~Hashino, M.~Kakizaki, S.~Kanemura, P.~Ko and T.~Matsui, \emph{{Gravitational waves from first order electroweak phase transition in models with the U(1)$_{X}$ gauge symmetry}}, \href{https://doi.org/10.1007/JHEP06(2018)088}{\emph{JHEP} {\bfseries 06} (2018) 088} [\href{https://arxiv.org/abs/1802.02947}{{\ttfamily 1802.02947}}].

\bibitem{Brdar:2019fur}
V.~Brdar, L.~Graf, A.J.~Helmboldt and X.-J.~Xu, \emph{{Gravitational Waves as a Probe of Left-Right Symmetry Breaking}}, \href{https://doi.org/10.1088/1475-7516/2019/12/027}{\emph{JCAP} {\bfseries 12} (2019) 027} [\href{https://arxiv.org/abs/1909.02018}{{\ttfamily 1909.02018}}].

\bibitem{Dunsky:2021tih}
D.I.~Dunsky, A.~Ghoshal, H.~Murayama, Y.~Sakakihara and G.~White, \emph{{GUTs, hybrid topological defects, and gravitational waves}}, \href{https://doi.org/10.1103/PhysRevD.106.075030}{\emph{Phys. Rev. D} {\bfseries 106} (2022) 075030} [\href{https://arxiv.org/abs/2111.08750}{{\ttfamily 2111.08750}}].

\bibitem{Blasi:2022woz}
S.~Blasi and A.~Mariotti, \emph{{Domain Walls Seeding the Electroweak Phase Transition}}, \href{https://doi.org/10.1103/PhysRevLett.129.261303}{\emph{Phys. Rev. Lett.} {\bfseries 129} (2022) 261303} [\href{https://arxiv.org/abs/2203.16450}{{\ttfamily 2203.16450}}].

\bibitem{Dolan:1973qd}
L.~Dolan and R.~Jackiw, \emph{{Symmetry Behavior at Finite Temperature}}, \href{https://doi.org/10.1103/PhysRevD.9.3320}{\emph{Phys. Rev. D} {\bfseries 9} (1974) 3320}.

\bibitem{Kirzhnits:1974as}
D.A.~Kirzhnits and A.D.~Linde, \emph{{A Relativistic phase transition}}, {\emph{Zh. Eksp. Teor. Fiz.} {\bfseries 67} (1974) 1263}.

\bibitem{Linde:1978px}
A.D.~Linde, \emph{{Phase Transitions in Gauge Theories and Cosmology}}, \href{https://doi.org/10.1088/0034-4885/42/3/001}{\emph{Rept. Prog. Phys.} {\bfseries 42} (1979) 389}.

\bibitem{Linde:1980ts}
A.D.~Linde, \emph{{Infrared Problem in Thermodynamics of the Yang-Mills Gas}}, \href{https://doi.org/10.1016/0370-2693(80)90769-8}{\emph{Phys. Lett. B} {\bfseries 96} (1980) 289}.

\bibitem{Parwani:1991gq}
R.R.~Parwani, \emph{{Resummation in a hot scalar field theory}}, \href{https://doi.org/10.1103/PhysRevD.45.4695}{\emph{Phys. Rev. D} {\bfseries 45} (1992) 4695} [\href{https://arxiv.org/abs/hep-ph/9204216}{{\ttfamily hep-ph/9204216}}].

\bibitem{Arnold:1992rz}
P.B.~Arnold and O.~Espinosa, \emph{{The Effective potential and first order phase transitions: Beyond leading-order}}, \href{https://doi.org/10.1103/PhysRevD.47.3546}{\emph{Phys. Rev. D} {\bfseries 47} (1993) 3546} [\href{https://arxiv.org/abs/hep-ph/9212235}{{\ttfamily hep-ph/9212235}}].

\bibitem{Laine:2016hma}
M.~Laine and A.~Vuorinen, \emph{{Basics of Thermal Field Theory}}, vol.~925, Springer (2016), \href{https://doi.org/10.1007/978-3-319-31933-9}{10.1007/978-3-319-31933-9}, [\href{https://arxiv.org/abs/1701.01554}{{\ttfamily 1701.01554}}].

\bibitem{Farakos:1994kx}
K.~Farakos, K.~Kajantie, K.~Rummukainen and M.E.~Shaposhnikov, \emph{{3-D physics and the electroweak phase transition: Perturbation theory}}, \href{https://doi.org/10.1016/0550-3213(94)90173-2}{\emph{Nucl. Phys. B} {\bfseries 425} (1994) 67} [\href{https://arxiv.org/abs/hep-ph/9404201}{{\ttfamily hep-ph/9404201}}].

\bibitem{Kajantie:1995dw}
K.~Kajantie, M.~Laine, K.~Rummukainen and M.E.~Shaposhnikov, \emph{{Generic rules for high temperature dimensional reduction and their application to the standard model}}, \href{https://doi.org/10.1016/0550-3213(95)00549-8}{\emph{Nucl. Phys. B} {\bfseries 458} (1996) 90} [\href{https://arxiv.org/abs/hep-ph/9508379}{{\ttfamily hep-ph/9508379}}].

\bibitem{Braaten:1995cm}
E.~Braaten and A.~Nieto, \emph{{Effective field theory approach to high temperature thermodynamics}}, \href{https://doi.org/10.1103/PhysRevD.51.6990}{\emph{Phys. Rev. D} {\bfseries 51} (1995) 6990} [\href{https://arxiv.org/abs/hep-ph/9501375}{{\ttfamily hep-ph/9501375}}].

\bibitem{Ekstedt:2022bff}
A.~Ekstedt, P.~Schicho and T.V.I.~Tenkanen, \emph{{DRalgo: A package for effective field theory approach for thermal phase transitions}}, \href{https://doi.org/10.1016/j.cpc.2023.108725}{\emph{Comput. Phys. Commun.} {\bfseries 288} (2023) 108725} [\href{https://arxiv.org/abs/2205.08815}{{\ttfamily 2205.08815}}].

\bibitem{Zhou:2020irf}
R.~Zhou and L.~Bian, \emph{{Gravitational wave and electroweak baryogenesis with two Higgs doublet models}}, \href{https://doi.org/10.1016/j.physletb.2022.137105}{\emph{Phys. Lett. B} {\bfseries 829} (2022) 137105} [\href{https://arxiv.org/abs/2001.01237}{{\ttfamily 2001.01237}}].

\bibitem{Patel:2011th}
H.H.~Patel and M.J.~Ramsey-Musolf, \emph{{Baryon Washout, Electroweak Phase Transition, and Perturbation Theory}}, \href{https://doi.org/10.1007/JHEP07(2011)029}{\emph{JHEP} {\bfseries 07} (2011) 029} [\href{https://arxiv.org/abs/1101.4665}{{\ttfamily 1101.4665}}].

\bibitem{Garny:2012cg}
M.~Garny and T.~Konstandin, \emph{{On the gauge dependence of vacuum transitions at finite temperature}}, \href{https://doi.org/10.1007/JHEP07(2012)189}{\emph{JHEP} {\bfseries 07} (2012) 189} [\href{https://arxiv.org/abs/1205.3392}{{\ttfamily 1205.3392}}].

\bibitem{Blinov:2015sna}
N.~Blinov, J.~Kozaczuk, D.E.~Morrissey and C.~Tamarit, \emph{{Electroweak Baryogenesis from Exotic Electroweak Symmetry Breaking}}, \href{https://doi.org/10.1103/PhysRevD.92.035012}{\emph{Phys. Rev. D} {\bfseries 92} (2015) 035012} [\href{https://arxiv.org/abs/1504.05195}{{\ttfamily 1504.05195}}].

\bibitem{Bahl:2022lio}
H.~Bahl, M.~Carena, N.M.~Coyle, A.~Ireland and C.E.M.~Wagner, \emph{{New tools for dissecting the general 2HDM}}, \href{https://doi.org/10.1007/JHEP03(2023)165}{\emph{JHEP} {\bfseries 03} (2023) 165} [\href{https://arxiv.org/abs/2210.00024}{{\ttfamily 2210.00024}}].

\bibitem{Dine:1992vs}
M.~Dine, R.G.~Leigh, P.~Huet, A.D.~Linde and D.A.~Linde, \emph{{Comments on the electroweak phase transition}}, \href{https://doi.org/10.1016/0370-2693(92)90026-Z}{\emph{Phys. Lett. B} {\bfseries 283} (1992) 319} [\href{https://arxiv.org/abs/hep-ph/9203201}{{\ttfamily hep-ph/9203201}}].

\bibitem{Dine:1992wr}
M.~Dine, R.G.~Leigh, P.Y.~Huet, A.D.~Linde and D.A.~Linde, \emph{{Towards the theory of the electroweak phase transition}}, \href{https://doi.org/10.1103/PhysRevD.46.550}{\emph{Phys. Rev. D} {\bfseries 46} (1992) 550} [\href{https://arxiv.org/abs/hep-ph/9203203}{{\ttfamily hep-ph/9203203}}].

\bibitem{Boyd:1992xn}
C.G.~Boyd, D.E.~Brahm and S.D.H.~Hsu, \emph{{Corrections to the electroweak effective action at finite temperature}}, \href{https://doi.org/10.1103/PhysRevD.48.4952}{\emph{Phys. Rev. D} {\bfseries 48} (1993) 4952} [\href{https://arxiv.org/abs/hep-ph/9206235}{{\ttfamily hep-ph/9206235}}].

\bibitem{Laine:2017hdk}
M.~Laine, M.~Meyer and G.~Nardini, \emph{{Thermal phase transition with full 2-loop effective potential}}, \href{https://doi.org/10.1016/j.nuclphysb.2017.04.023}{\emph{Nucl. Phys. B} {\bfseries 920} (2017) 565} [\href{https://arxiv.org/abs/1702.07479}{{\ttfamily 1702.07479}}].

\bibitem{Boyd:1993tz}
C.G.~Boyd, D.E.~Brahm and S.D.H.~Hsu, \emph{{Resummation methods at finite temperature: The Tadpole way}}, \href{https://doi.org/10.1103/PhysRevD.48.4963}{\emph{Phys. Rev. D} {\bfseries 48} (1993) 4963} [\href{https://arxiv.org/abs/hep-ph/9304254}{{\ttfamily hep-ph/9304254}}].

\bibitem{Curtin:2016urg}
D.~Curtin, P.~Meade and H.~Ramani, \emph{{Thermal Resummation and Phase Transitions}}, \href{https://doi.org/10.1140/epjc/s10052-018-6268-0}{\emph{Eur. Phys. J. C} {\bfseries 78} (2018) 787} [\href{https://arxiv.org/abs/1612.00466}{{\ttfamily 1612.00466}}].

\bibitem{Curtin:2022ovx}
D.~Curtin, J.~Roy and G.~White, \emph{{Gravitational waves and tadpole resummation: Efficient and easy convergence of finite temperature QFT}},  \href{https://arxiv.org/abs/2211.08218}{{\ttfamily 2211.08218}}.

\bibitem{Bahl:2024ykv}
H.~Bahl, M.~Carena, A.~Ireland and C.E.M.~Wagner, \emph{{Improved thermal resummation for multi-field potentials}}, \href{https://doi.org/10.1007/JHEP09(2024)153}{\emph{JHEP} {\bfseries 09} (2024) 153} [\href{https://arxiv.org/abs/2404.12439}{{\ttfamily 2404.12439}}].

\bibitem{Turok:1990zg}
N.~Turok and J.~Zadrozny, \emph{{Electroweak baryogenesis in the two doublet model}}, \href{https://doi.org/10.1016/0550-3213(91)90356-3}{\emph{Nucl. Phys. B} {\bfseries 358} (1991) 471}.

\bibitem{Cline:1996mga}
J.M.~Cline and P.-A.~Lemieux, \emph{{Electroweak phase transition in two Higgs doublet models}}, \href{https://doi.org/10.1103/PhysRevD.55.3873}{\emph{Phys. Rev. D} {\bfseries 55} (1997) 3873} [\href{https://arxiv.org/abs/hep-ph/9609240}{{\ttfamily hep-ph/9609240}}].

\bibitem{Fromme:2006cm}
L.~Fromme, S.J.~Huber and M.~Seniuch, \emph{{Baryogenesis in the two-Higgs doublet model}}, \href{https://doi.org/10.1088/1126-6708/2006/11/038}{\emph{JHEP} {\bfseries 11} (2006) 038} [\href{https://arxiv.org/abs/hep-ph/0605242}{{\ttfamily hep-ph/0605242}}].

\bibitem{Cline:2011mm}
J.M.~Cline, K.~Kainulainen and M.~Trott, \emph{{Electroweak Baryogenesis in Two Higgs Doublet Models and B meson anomalies}}, \href{https://doi.org/10.1007/JHEP11(2011)089}{\emph{JHEP} {\bfseries 11} (2011) 089} [\href{https://arxiv.org/abs/1107.3559}{{\ttfamily 1107.3559}}].

\bibitem{Dorsch:2013wja}
G.C.~Dorsch, S.J.~Huber and J.M.~No, \emph{{A strong electroweak phase transition in the 2HDM after LHC8}}, \href{https://doi.org/10.1007/JHEP10(2013)029}{\emph{JHEP} {\bfseries 10} (2013) 029} [\href{https://arxiv.org/abs/1305.6610}{{\ttfamily 1305.6610}}].

\bibitem{Basler:2016obg}
P.~Basler, M.~Krause, M.~Muhlleitner, J.~Wittbrodt and A.~Wlotzka, \emph{{Strong First Order Electroweak Phase Transition in the CP-Conserving 2HDM Revisited}}, \href{https://doi.org/10.1007/JHEP02(2017)121}{\emph{JHEP} {\bfseries 02} (2017) 121} [\href{https://arxiv.org/abs/1612.04086}{{\ttfamily 1612.04086}}].

\bibitem{Dorsch:2016tab}
G.C.~Dorsch, S.J.~Huber, K.~Mimasu and J.M.~No, \emph{{Hierarchical versus degenerate 2HDM: The LHC run 1 legacy at the onset of run 2}}, \href{https://doi.org/10.1103/PhysRevD.93.115033}{\emph{Phys. Rev. D} {\bfseries 93} (2016) 115033} [\href{https://arxiv.org/abs/1601.04545}{{\ttfamily 1601.04545}}].

\bibitem{Bernon:2017jgv}
J.~Bernon, L.~Bian and Y.~Jiang, \emph{{A new insight into the phase transition in the early Universe with two Higgs doublets}}, \href{https://doi.org/10.1007/JHEP05(2018)151}{\emph{JHEP} {\bfseries 05} (2018) 151} [\href{https://arxiv.org/abs/1712.08430}{{\ttfamily 1712.08430}}].

\bibitem{Dorsch:2017nza}
G.C.~Dorsch, S.J.~Huber, K.~Mimasu and J.M.~No, \emph{{The Higgs Vacuum Uplifted: Revisiting the Electroweak Phase Transition with a Second Higgs Doublet}}, \href{https://doi.org/10.1007/JHEP12(2017)086}{\emph{JHEP} {\bfseries 12} (2017) 086} [\href{https://arxiv.org/abs/1705.09186}{{\ttfamily 1705.09186}}].

\bibitem{Andersen:2017ika}
J.O.~Andersen, T.~Gorda, A.~Helset, L.~Niemi, T.V.I.~Tenkanen, A.~Tranberg et~al., \emph{{Nonperturbative Analysis of the Electroweak Phase Transition in the Two Higgs Doublet Model}}, \href{https://doi.org/10.1103/PhysRevLett.121.191802}{\emph{Phys. Rev. Lett.} {\bfseries 121} (2018) 191802} [\href{https://arxiv.org/abs/1711.09849}{{\ttfamily 1711.09849}}].

\bibitem{Kainulainen:2019kyp}
K.~Kainulainen, V.~Keus, L.~Niemi, K.~Rummukainen, T.V.I.~Tenkanen and V.~Vaskonen, \emph{{On the validity of perturbative studies of the electroweak phase transition in the Two Higgs Doublet model}}, \href{https://doi.org/10.1007/JHEP06(2019)075}{\emph{JHEP} {\bfseries 06} (2019) 075} [\href{https://arxiv.org/abs/1904.01329}{{\ttfamily 1904.01329}}].

\bibitem{Su:2020pjw}
W.~Su, A.G.~Williams and M.~Zhang, \emph{{Strong first order electroweak phase transition in 2HDM confronting future Z \& Higgs factories}}, \href{https://doi.org/10.1007/JHEP04(2021)219}{\emph{JHEP} {\bfseries 04} (2021) 219} [\href{https://arxiv.org/abs/2011.04540}{{\ttfamily 2011.04540}}].

\bibitem{Davoudiasl:2021syn}
H.~Davoudiasl, I.M.~Lewis and M.~Sullivan, \emph{{Multi-TeV signals of baryogenesis in a Higgs troika model}}, \href{https://doi.org/10.1103/PhysRevD.104.015024}{\emph{Phys. Rev. D} {\bfseries 104} (2021) 015024} [\href{https://arxiv.org/abs/2103.12089}{{\ttfamily 2103.12089}}].

\bibitem{Aoki:2021oez}
M.~Aoki, T.~Komatsu and H.~Shibuya, \emph{{Possibility of a multi-step electroweak phase transition in the two-Higgs doublet models}}, \href{https://doi.org/10.1093/ptep/ptac068}{\emph{PTEP} {\bfseries 2022} (2022) 063B05} [\href{https://arxiv.org/abs/2106.03439}{{\ttfamily 2106.03439}}].

\bibitem{Goncalves:2021egx}
D.~Gon\c{c}alves, A.~Kaladharan and Y.~Wu, \emph{{Electroweak phase transition in the 2HDM: Collider and gravitational wave complementarity}}, \href{https://doi.org/10.1103/PhysRevD.105.095041}{\emph{Phys. Rev. D} {\bfseries 105} (2022) 095041} [\href{https://arxiv.org/abs/2108.05356}{{\ttfamily 2108.05356}}].

\bibitem{Goncalves:2023svb}
D.~Gon\c{c}alves, A.~Kaladharan and Y.~Wu, \emph{{Gravitational waves, bubble profile, and baryon asymmetry in the complex 2HDM}}, \href{https://doi.org/10.1103/PhysRevD.108.075010}{\emph{Phys. Rev. D} {\bfseries 108} (2023) 075010} [\href{https://arxiv.org/abs/2307.03224}{{\ttfamily 2307.03224}}].

\bibitem{Coleman:1973jx}
S.R.~Coleman and E.J.~Weinberg, \emph{{Radiative Corrections as the Origin of Spontaneous Symmetry Breaking}}, \href{https://doi.org/10.1103/PhysRevD.7.1888}{\emph{Phys. Rev. D} {\bfseries 7} (1973) 1888}.

\bibitem{Branco:2011iw}
G.C.~Branco, P.M.~Ferreira, L.~Lavoura, M.N.~Rebelo, M.~Sher and J.P.~Silva, \emph{{Theory and phenomenology of two-Higgs-doublet models}}, \href{https://doi.org/10.1016/j.physrep.2012.02.002}{\emph{Phys. Rept.} {\bfseries 516} (2012) 1} [\href{https://arxiv.org/abs/1106.0034}{{\ttfamily 1106.0034}}].

\bibitem{Aoki:2009ha}
M.~Aoki, S.~Kanemura, K.~Tsumura and K.~Yagyu, \emph{{Models of Yukawa interaction in the two Higgs doublet model, and their collider phenomenology}}, \href{https://doi.org/10.1103/PhysRevD.80.015017}{\emph{Phys. Rev. D} {\bfseries 80} (2009) 015017} [\href{https://arxiv.org/abs/0902.4665}{{\ttfamily 0902.4665}}].

\bibitem{Grinstein:2015rtl}
B.~Grinstein, C.W.~Murphy and P.~Uttayarat, \emph{{One-loop corrections to the perturbative unitarity bounds in the CP-conserving two-Higgs doublet model with a softly broken $ {\mathrm{\mathbb{Z}}}_2 $ symmetry}}, \href{https://doi.org/10.1007/JHEP06(2016)070}{\emph{JHEP} {\bfseries 06} (2016) 070} [\href{https://arxiv.org/abs/1512.04567}{{\ttfamily 1512.04567}}].

\bibitem{Akeroyd:2000wc}
A.G.~Akeroyd, A.~Arhrib and E.-M.~Naimi, \emph{{Note on tree level unitarity in the general two Higgs doublet model}}, \href{https://doi.org/10.1016/S0370-2693(00)00962-X}{\emph{Phys. Lett. B} {\bfseries 490} (2000) 119} [\href{https://arxiv.org/abs/hep-ph/0006035}{{\ttfamily hep-ph/0006035}}].

\bibitem{Ginzburg:2005dt}
I.F.~Ginzburg and I.P.~Ivanov, \emph{{Tree-level unitarity constraints in the most general 2HDM}}, \href{https://doi.org/10.1103/PhysRevD.72.115010}{\emph{Phys. Rev. D} {\bfseries 72} (2005) 115010} [\href{https://arxiv.org/abs/hep-ph/0508020}{{\ttfamily hep-ph/0508020}}].

\bibitem{Gerard:2007kn}
J.M.~Gerard and M.~Herquet, \emph{{A Twisted custodial symmetry in the two-Higgs-doublet model}}, \href{https://doi.org/10.1103/PhysRevLett.98.251802}{\emph{Phys. Rev. Lett.} {\bfseries 98} (2007) 251802} [\href{https://arxiv.org/abs/hep-ph/0703051}{{\ttfamily hep-ph/0703051}}].

\bibitem{Haber:2010bw}
H.E.~Haber and D.~O'Neil, \emph{{Basis-independent methods for the two-Higgs-doublet model III: The CP-conserving limit, custodial symmetry, and the oblique parameters S, T, U}}, \href{https://doi.org/10.1103/PhysRevD.83.055017}{\emph{Phys. Rev. D} {\bfseries 83} (2011) 055017} [\href{https://arxiv.org/abs/1011.6188}{{\ttfamily 1011.6188}}].

\bibitem{Belle:2016ufb}
{\scshape Belle} collaboration, \emph{{Measurement of the inclusive $B\to X_{s+d} \gamma$ branching fraction, photon energy spectrum and HQE parameters}},  in \emph{{38th International Conference on High Energy Physics}}, 8, 2016 [\href{https://arxiv.org/abs/1608.02344}{{\ttfamily 1608.02344}}].

\bibitem{Misiak:2017bgg}
M.~Misiak and M.~Steinhauser, \emph{{Weak radiative decays of the B meson and bounds on $M_{H^\pm }$ in the Two-Higgs-Doublet Model}}, \href{https://doi.org/10.1140/epjc/s10052-017-4776-y}{\emph{Eur. Phys. J. C} {\bfseries 77} (2017) 201} [\href{https://arxiv.org/abs/1702.04571}{{\ttfamily 1702.04571}}].

\bibitem{Bagnaschi:2018ofa}
E.~Bagnaschi et~al., \emph{{MSSM Higgs Boson Searches at the LHC: Benchmark Scenarios for Run 2 and Beyond}}, \href{https://doi.org/10.1140/epjc/s10052-019-7114-8}{\emph{Eur. Phys. J. C} {\bfseries 79} (2019) 617} [\href{https://arxiv.org/abs/1808.07542}{{\ttfamily 1808.07542}}].

\bibitem{ATLAS:2021upq}
{\scshape ATLAS} collaboration, \emph{{Search for charged Higgs bosons decaying into a top quark and a bottom quark at $ \sqrt{\mathrm{s}} $ = 13 TeV with the ATLAS detector}}, \href{https://doi.org/10.1007/JHEP06(2021)145}{\emph{JHEP} {\bfseries 06} (2021) 145} [\href{https://arxiv.org/abs/2102.10076}{{\ttfamily 2102.10076}}].

\bibitem{CMS:2018rmh}
{\scshape CMS} collaboration, \emph{{Search for additional neutral MSSM Higgs bosons in the $\tau\tau$ final state in proton-proton collisions at $\sqrt{s}=$ 13 TeV}}, \href{https://doi.org/10.1007/JHEP09(2018)007}{\emph{JHEP} {\bfseries 09} (2018) 007} [\href{https://arxiv.org/abs/1803.06553}{{\ttfamily 1803.06553}}].

\bibitem{CMS:2019bnu}
{\scshape CMS} collaboration, \emph{{Search for a heavy Higgs boson decaying to a pair of W bosons in proton-proton collisions at $\sqrt{s} =$ 13 TeV}}, \href{https://doi.org/10.1007/JHEP03(2020)034}{\emph{JHEP} {\bfseries 03} (2020) 034} [\href{https://arxiv.org/abs/1912.01594}{{\ttfamily 1912.01594}}].

\bibitem{ATLAS:2020zms}
{\scshape ATLAS} collaboration, \emph{{Search for heavy Higgs bosons decaying into two tau leptons with the ATLAS detector using $pp$ collisions at $\sqrt{s}=13$ TeV}}, \href{https://doi.org/10.1103/PhysRevLett.125.051801}{\emph{Phys. Rev. Lett.} {\bfseries 125} (2020) 051801} [\href{https://arxiv.org/abs/2002.12223}{{\ttfamily 2002.12223}}].

\bibitem{Posch:2010hx}
P.~Posch, \emph{{Enhancement of h ---\ensuremath{>} gamma gamma in the Two Higgs Doublet Model Type I}}, \href{https://doi.org/10.1016/j.physletb.2011.01.003}{\emph{Phys. Lett. B} {\bfseries 696} (2011) 447} [\href{https://arxiv.org/abs/1001.1759}{{\ttfamily 1001.1759}}].

\bibitem{Djouadi:2005gj}
A.~Djouadi, \emph{{The Anatomy of electro-weak symmetry breaking. II. The Higgs bosons in the minimal supersymmetric model}}, \href{https://doi.org/10.1016/j.physrep.2007.10.005}{\emph{Phys. Rept.} {\bfseries 459} (2008) 1} [\href{https://arxiv.org/abs/hep-ph/0503173}{{\ttfamily hep-ph/0503173}}].

\bibitem{ATLAS:2022tnm}
{\scshape ATLAS} collaboration, \emph{{Measurement of the properties of Higgs boson production at $\sqrt{s} = 13$ TeV in the $H\to\gamma\gamma$ channel using $139$ fb$^{-1}$ of $pp$ collision data with the ATLAS experiment}}, \href{https://doi.org/10.1007/JHEP07(2023)088}{\emph{JHEP} {\bfseries 07} (2023) 088} [\href{https://arxiv.org/abs/2207.00348}{{\ttfamily 2207.00348}}].

\bibitem{CMS:2021kom}
{\scshape CMS} collaboration, \emph{{Measurements of Higgs boson production cross sections and couplings in the diphoton decay channel at $ \sqrt{\mathrm{s}} $ = 13 TeV}}, \href{https://doi.org/10.1007/JHEP07(2021)027}{\emph{JHEP} {\bfseries 07} (2021) 027} [\href{https://arxiv.org/abs/2103.06956}{{\ttfamily 2103.06956}}].

\bibitem{Mathematica:2022}
{Wolfram Research, Inc.}, ``{Mathematica, Version 13.1}.'' \url{https://www.wolfram.com/mathematica/}, 2022.

\bibitem{Wainwright:2011kj}
C.L.~Wainwright, \emph{{CosmoTransitions: Computing Cosmological Phase Transition Temperatures and Bubble Profiles with Multiple Fields}}, \href{https://doi.org/10.1016/j.cpc.2012.04.004}{\emph{Comput. Phys. Commun.} {\bfseries 183} (2012) 2006} [\href{https://arxiv.org/abs/1109.4189}{{\ttfamily 1109.4189}}].

\bibitem{Kirzhnits:1976ts}
D.A.~Kirzhnits and A.D.~Linde, \emph{{Symmetry Behavior in Gauge Theories}}, \href{https://doi.org/10.1016/0003-4916(76)90279-7}{\emph{Annals Phys.} {\bfseries 101} (1976) 195}.

\bibitem{Mohapatra:1979qt}
R.N.~Mohapatra and G.~Senjanovic, \emph{{Soft CP Violation at High Temperature}}, \href{https://doi.org/10.1103/PhysRevLett.42.1651}{\emph{Phys. Rev. Lett.} {\bfseries 42} (1979) 1651}.

\bibitem{Salomonson:1984px}
P.~Salomonson and B.-S.K.~Skagerstam, \emph{{High Temperature Phases in an O($N$) X O($N$) Symmetric Four $\epsilon$ Dimensional Vector Model}}, \href{https://doi.org/10.1016/0370-2693(85)91039-1}{\emph{Phys. Lett. B} {\bfseries 155} (1985) 100}.

\bibitem{Bimonte:1995sc}
G.~Bimonte and G.~Lozano, \emph{{On Symmetry nonrestoration at high temperature}}, \href{https://doi.org/10.1016/0370-2693(95)01395-4}{\emph{Phys. Lett. B} {\bfseries 366} (1996) 248} [\href{https://arxiv.org/abs/hep-th/9507079}{{\ttfamily hep-th/9507079}}].

\bibitem{Bimonte:1995xs}
G.~Bimonte and G.~Lozano, \emph{{Can symmetry nonrestoration solve the monopole problem?}}, \href{https://doi.org/10.1016/0550-3213(95)00626-5}{\emph{Nucl. Phys. B} {\bfseries 460} (1996) 155} [\href{https://arxiv.org/abs/hep-th/9509060}{{\ttfamily hep-th/9509060}}].

\bibitem{Dvali:1996zr}
G.R.~Dvali, A.~Melfo and G.~Senjanovic, \emph{{Nonrestoration of spontaneously broken P and CP at high temperature}}, \href{https://doi.org/10.1103/PhysRevD.54.7857}{\emph{Phys. Rev. D} {\bfseries 54} (1996) 7857} [\href{https://arxiv.org/abs/hep-ph/9601376}{{\ttfamily hep-ph/9601376}}].

\bibitem{Pietroni:1996zj}
M.~Pietroni, N.~Rius and N.~Tetradis, \emph{{Inverse symmetry breaking and the exact renormalization group}}, \href{https://doi.org/10.1016/S0370-2693(97)00150-0}{\emph{Phys. Lett. B} {\bfseries 397} (1997) 119} [\href{https://arxiv.org/abs/hep-ph/9612205}{{\ttfamily hep-ph/9612205}}].

\bibitem{Patel:2013zla}
H.H.~Patel, M.J.~Ramsey-Musolf and M.B.~Wise, \emph{{Color Breaking in the Early Universe}}, \href{https://doi.org/10.1103/PhysRevD.88.015003}{\emph{Phys. Rev. D} {\bfseries 88} (2013) 015003} [\href{https://arxiv.org/abs/1303.1140}{{\ttfamily 1303.1140}}].

\bibitem{Kilic:2015joa}
C.~Kilic and S.~Swaminathan, \emph{{Can A Pseudo-Nambu-Goldstone Higgs Lead To Symmetry Non-Restoration?}}, \href{https://doi.org/10.1007/JHEP01(2016)002}{\emph{JHEP} {\bfseries 01} (2016) 002} [\href{https://arxiv.org/abs/1508.05121}{{\ttfamily 1508.05121}}].

\bibitem{Ramsey-Musolf:2017tgh}
M.J.~Ramsey-Musolf, P.~Winslow and G.~White, \emph{{Color Breaking Baryogenesis}}, \href{https://doi.org/10.1103/PhysRevD.97.123509}{\emph{Phys. Rev. D} {\bfseries 97} (2018) 123509} [\href{https://arxiv.org/abs/1708.07511}{{\ttfamily 1708.07511}}].

\bibitem{Glioti:2018roy}
A.~Glioti, R.~Rattazzi and L.~Vecchi, \emph{{Electroweak Baryogenesis above the Electroweak Scale}}, \href{https://doi.org/10.1007/JHEP04(2019)027}{\emph{JHEP} {\bfseries 04} (2019) 027} [\href{https://arxiv.org/abs/1811.11740}{{\ttfamily 1811.11740}}].

\bibitem{Carena:2019une}
M.~Carena, Z.~Liu and Y.~Wang, \emph{{Electroweak phase transition with spontaneous Z$_{2}$-breaking}}, \href{https://doi.org/10.1007/JHEP08(2020)107}{\emph{JHEP} {\bfseries 08} (2020) 107} [\href{https://arxiv.org/abs/1911.10206}{{\ttfamily 1911.10206}}].

\bibitem{Kozaczuk:2019pet}
J.~Kozaczuk, M.J.~Ramsey-Musolf and J.~Shelton, \emph{{Exotic Higgs boson decays and the electroweak phase transition}}, \href{https://doi.org/10.1103/PhysRevD.101.115035}{\emph{Phys. Rev. D} {\bfseries 101} (2020) 115035} [\href{https://arxiv.org/abs/1911.10210}{{\ttfamily 1911.10210}}].

\bibitem{Matsedonskyi:2020mlz}
O.~Matsedonskyi and G.~Servant, \emph{{High-Temperature Electroweak Symmetry Non-Restoration from New Fermions and Implications for Baryogenesis}}, \href{https://doi.org/10.1007/JHEP09(2020)012}{\emph{JHEP} {\bfseries 09} (2020) 012} [\href{https://arxiv.org/abs/2002.05174}{{\ttfamily 2002.05174}}].

\bibitem{Carena:2021onl}
M.~Carena, C.~Krause, Z.~Liu and Y.~Wang, \emph{{New approach to electroweak symmetry nonrestoration}}, \href{https://doi.org/10.1103/PhysRevD.104.055016}{\emph{Phys. Rev. D} {\bfseries 104} (2021) 055016} [\href{https://arxiv.org/abs/2104.00638}{{\ttfamily 2104.00638}}].

\bibitem{Carena:2022qpf}
M.~Carena, Y.-Y.~Li, T.~Ou and Y.~Wang, \emph{{Anatomy of the electroweak phase transition for dark sector induced baryogenesis}}, \href{https://doi.org/10.1007/JHEP02(2023)139}{\emph{JHEP} {\bfseries 02} (2023) 139} [\href{https://arxiv.org/abs/2210.14352}{{\ttfamily 2210.14352}}].

\bibitem{Carena:2022yvx}
M.~Carena, J.~Kozaczuk, Z.~Liu, T.~Ou, M.J.~Ramsey-Musolf, J.~Shelton et~al., \emph{{Probing the Electroweak Phase Transition with Exotic Higgs Decays}}, \href{https://doi.org/10.31526/lhep.2023.432}{\emph{LHEP} {\bfseries 2023} (2023) 432} [\href{https://arxiv.org/abs/2203.08206}{{\ttfamily 2203.08206}}].

\bibitem{LIGOScientific:2016aoc}
{\scshape LIGO Scientific, Virgo} collaboration, \emph{{Observation of Gravitational Waves from a Binary Black Hole Merger}}, \href{https://doi.org/10.1103/PhysRevLett.116.061102}{\emph{Phys. Rev. Lett.} {\bfseries 116} (2016) 061102} [\href{https://arxiv.org/abs/1602.03837}{{\ttfamily 1602.03837}}].

\bibitem{NANOGrav:2023gor}
{\scshape NANOGrav} collaboration, \emph{{The NANOGrav 15 yr Data Set: Evidence for a Gravitational-wave Background}}, \href{https://doi.org/10.3847/2041-8213/acdac6}{\emph{Astrophys. J. Lett.} {\bfseries 951} (2023) L8} [\href{https://arxiv.org/abs/2306.16213}{{\ttfamily 2306.16213}}].

\bibitem{EPTA:2023fyk}
{\scshape EPTA, InPTA:} collaboration, \emph{{The second data release from the European Pulsar Timing Array - III. Search for gravitational wave signals}}, \href{https://doi.org/10.1051/0004-6361/202346844}{\emph{Astron. Astrophys.} {\bfseries 678} (2023) A50} [\href{https://arxiv.org/abs/2306.16214}{{\ttfamily 2306.16214}}].

\bibitem{Caprini:2015zlo}
C.~Caprini et~al., \emph{{Science with the space-based interferometer eLISA. II: Gravitational waves from cosmological phase transitions}}, \href{https://doi.org/10.1088/1475-7516/2016/04/001}{\emph{JCAP} {\bfseries 04} (2016) 001} [\href{https://arxiv.org/abs/1512.06239}{{\ttfamily 1512.06239}}].

\bibitem{Cai:2017cbj}
R.-G.~Cai, Z.~Cao, Z.-K.~Guo, S.-J.~Wang and T.~Yang, \emph{{The Gravitational-Wave Physics}}, \href{https://doi.org/10.1093/nsr/nwx029}{\emph{Natl. Sci. Rev.} {\bfseries 4} (2017) 687} [\href{https://arxiv.org/abs/1703.00187}{{\ttfamily 1703.00187}}].

\bibitem{Caprini:2018mtu}
C.~Caprini and D.G.~Figueroa, \emph{{Cosmological Backgrounds of Gravitational Waves}}, \href{https://doi.org/10.1088/1361-6382/aac608}{\emph{Class. Quant. Grav.} {\bfseries 35} (2018) 163001} [\href{https://arxiv.org/abs/1801.04268}{{\ttfamily 1801.04268}}].

\bibitem{Romano:2016dpx}
J.D.~Romano and N.J.~Cornish, \emph{{Detection methods for stochastic gravitational-wave backgrounds: a unified treatment}}, \href{https://doi.org/10.1007/s41114-017-0004-1}{\emph{Living Rev. Rel.} {\bfseries 20} (2017) 2} [\href{https://arxiv.org/abs/1608.06889}{{\ttfamily 1608.06889}}].

\bibitem{Christensen:2018iqi}
N.~Christensen, \emph{{Stochastic Gravitational Wave Backgrounds}}, \href{https://doi.org/10.1088/1361-6633/aae6b5}{\emph{Rept. Prog. Phys.} {\bfseries 82} (2019) 016903} [\href{https://arxiv.org/abs/1811.08797}{{\ttfamily 1811.08797}}].

\bibitem{Athron:2023xlk}
P.~Athron, C.~Bal\'azs, A.~Fowlie, L.~Morris and L.~Wu, \emph{{Cosmological phase transitions: from perturbative particle physics to gravitational waves}},  \href{https://arxiv.org/abs/2305.02357}{{\ttfamily 2305.02357}}.

\bibitem{Guo:2020grp}
H.-K.~Guo, K.~Sinha, D.~Vagie and G.~White, \emph{{Phase Transitions in an Expanding Universe: Stochastic Gravitational Waves in Standard and Non-Standard Histories}}, \href{https://doi.org/10.1088/1475-7516/2021/01/001}{\emph{JCAP} {\bfseries 01} (2021) 001} [\href{https://arxiv.org/abs/2007.08537}{{\ttfamily 2007.08537}}].

\bibitem{Balazs:2023kuk}
C.~Bal\'azs, Y.~Xiao, J.M.~Yang and Y.~Zhang, \emph{{New vacuum stability limit from cosmological history}}, \href{https://doi.org/10.1016/j.nuclphysb.2024.116533}{\emph{Nucl. Phys. B} {\bfseries 1002} (2024) 116533} [\href{https://arxiv.org/abs/2301.09283}{{\ttfamily 2301.09283}}].

\bibitem{Vaskonen:2016yiu}
V.~Vaskonen, \emph{{Electroweak baryogenesis and gravitational waves from a real scalar singlet}}, \href{https://doi.org/10.1103/PhysRevD.95.123515}{\emph{Phys. Rev. D} {\bfseries 95} (2017) 123515} [\href{https://arxiv.org/abs/1611.02073}{{\ttfamily 1611.02073}}].

\bibitem{Alves:2018jsw}
A.~Alves, T.~Ghosh, H.-K.~Guo, K.~Sinha and D.~Vagie, \emph{{Collider and Gravitational Wave Complementarity in Exploring the Singlet Extension of the Standard Model}}, \href{https://doi.org/10.1007/JHEP04(2019)052}{\emph{JHEP} {\bfseries 04} (2019) 052} [\href{https://arxiv.org/abs/1812.09333}{{\ttfamily 1812.09333}}].

\bibitem{Kosowsky:1991ua}
A.~Kosowsky, M.S.~Turner and R.~Watkins, \emph{{Gravitational radiation from colliding vacuum bubbles}}, \href{https://doi.org/10.1103/PhysRevD.45.4514}{\emph{Phys. Rev. D} {\bfseries 45} (1992) 4514}.

\bibitem{Kosowsky:1992vn}
A.~Kosowsky and M.S.~Turner, \emph{{Gravitational radiation from colliding vacuum bubbles: envelope approximation to many bubble collisions}}, \href{https://doi.org/10.1103/PhysRevD.47.4372}{\emph{Phys. Rev. D} {\bfseries 47} (1993) 4372} [\href{https://arxiv.org/abs/astro-ph/9211004}{{\ttfamily astro-ph/9211004}}].

\bibitem{Jinno:2016vai}
R.~Jinno and M.~Takimoto, \emph{{Gravitational waves from bubble collisions: An analytic derivation}}, \href{https://doi.org/10.1103/PhysRevD.95.024009}{\emph{Phys. Rev. D} {\bfseries 95} (2017) 024009} [\href{https://arxiv.org/abs/1605.01403}{{\ttfamily 1605.01403}}].

\bibitem{Hindmarsh:2015qta}
M.~Hindmarsh, S.J.~Huber, K.~Rummukainen and D.J.~Weir, \emph{{Numerical simulations of acoustically generated gravitational waves at a first order phase transition}}, \href{https://doi.org/10.1103/PhysRevD.92.123009}{\emph{Phys. Rev. D} {\bfseries 92} (2015) 123009} [\href{https://arxiv.org/abs/1504.03291}{{\ttfamily 1504.03291}}].

\bibitem{Hindmarsh:2017gnf}
M.~Hindmarsh, S.J.~Huber, K.~Rummukainen and D.J.~Weir, \emph{{Shape of the acoustic gravitational wave power spectrum from a first order phase transition}}, \href{https://doi.org/10.1103/PhysRevD.96.103520}{\emph{Phys. Rev. D} {\bfseries 96} (2017) 103520} [\href{https://arxiv.org/abs/1704.05871}{{\ttfamily 1704.05871}}].

\bibitem{Hindmarsh:2013xza}
M.~Hindmarsh, S.J.~Huber, K.~Rummukainen and D.J.~Weir, \emph{{Gravitational waves from the sound of a first order phase transition}}, \href{https://doi.org/10.1103/PhysRevLett.112.041301}{\emph{Phys. Rev. Lett.} {\bfseries 112} (2014) 041301} [\href{https://arxiv.org/abs/1304.2433}{{\ttfamily 1304.2433}}].

\bibitem{Giblin:2013kea}
J.T.~Giblin, Jr. and J.B.~Mertens, \emph{{Vacuum Bubbles in the Presence of a Relativistic Fluid}}, \href{https://doi.org/10.1007/JHEP12(2013)042}{\emph{JHEP} {\bfseries 12} (2013) 042} [\href{https://arxiv.org/abs/1310.2948}{{\ttfamily 1310.2948}}].

\bibitem{Hindmarsh:2016lnk}
M.~Hindmarsh, \emph{{Sound shell model for acoustic gravitational wave production at a first-order phase transition in the early Universe}}, \href{https://doi.org/10.1103/PhysRevLett.120.071301}{\emph{Phys. Rev. Lett.} {\bfseries 120} (2018) 071301} [\href{https://arxiv.org/abs/1608.04735}{{\ttfamily 1608.04735}}].

\bibitem{Hindmarsh:2019phv}
M.~Hindmarsh and M.~Hijazi, \emph{{Gravitational waves from first order cosmological phase transitions in the Sound Shell Model}}, \href{https://doi.org/10.1088/1475-7516/2019/12/062}{\emph{JCAP} {\bfseries 12} (2019) 062} [\href{https://arxiv.org/abs/1909.10040}{{\ttfamily 1909.10040}}].

\bibitem{Jinno:2019jhi}
R.~Jinno, H.~Seong, M.~Takimoto and C.M.~Um, \emph{{Gravitational waves from first-order phase transitions: Ultra-supercooled transitions and the fate of relativistic shocks}}, \href{https://doi.org/10.1088/1475-7516/2019/10/033}{\emph{JCAP} {\bfseries 10} (2019) 033} [\href{https://arxiv.org/abs/1905.00899}{{\ttfamily 1905.00899}}].

\bibitem{Konstandin:2017sat}
T.~Konstandin, \emph{{Gravitational radiation from a bulk flow model}}, \href{https://doi.org/10.1088/1475-7516/2018/03/047}{\emph{JCAP} {\bfseries 03} (2018) 047} [\href{https://arxiv.org/abs/1712.06869}{{\ttfamily 1712.06869}}].

\bibitem{Cai:2023guc}
R.-G.~Cai, S.-J.~Wang and Z.-Y.~Yuwen, \emph{{Hydrodynamic sound shell model}}, \href{https://doi.org/10.1103/PhysRevD.108.L021502}{\emph{Phys. Rev. D} {\bfseries 108} (2023) L021502} [\href{https://arxiv.org/abs/2305.00074}{{\ttfamily 2305.00074}}].

\bibitem{Caprini:2006jb}
C.~Caprini and R.~Durrer, \emph{{Gravitational waves from stochastic relativistic sources: Primordial turbulence and magnetic fields}}, \href{https://doi.org/10.1103/PhysRevD.74.063521}{\emph{Phys. Rev. D} {\bfseries 74} (2006) 063521} [\href{https://arxiv.org/abs/astro-ph/0603476}{{\ttfamily astro-ph/0603476}}].

\bibitem{Kahniashvili:2008pf}
T.~Kahniashvili, A.~Kosowsky, G.~Gogoberidze and Y.~Maravin, \emph{{Detectability of Gravitational Waves from Phase Transitions}}, \href{https://doi.org/10.1103/PhysRevD.78.043003}{\emph{Phys. Rev. D} {\bfseries 78} (2008) 043003} [\href{https://arxiv.org/abs/0806.0293}{{\ttfamily 0806.0293}}].

\bibitem{Kahniashvili:2008pe}
T.~Kahniashvili, L.~Campanelli, G.~Gogoberidze, Y.~Maravin and B.~Ratra, \emph{{Gravitational Radiation from Primordial Helical Inverse Cascade MHD Turbulence}}, \href{https://doi.org/10.1103/PhysRevD.78.123006}{\emph{Phys. Rev. D} {\bfseries 78} (2008) 123006} [\href{https://arxiv.org/abs/0809.1899}{{\ttfamily 0809.1899}}].

\bibitem{Kahniashvili:2009mf}
T.~Kahniashvili, L.~Kisslinger and T.~Stevens, \emph{{Gravitational Radiation Generated by Magnetic Fields in Cosmological Phase Transitions}}, \href{https://doi.org/10.1103/PhysRevD.81.023004}{\emph{Phys. Rev. D} {\bfseries 81} (2010) 023004} [\href{https://arxiv.org/abs/0905.0643}{{\ttfamily 0905.0643}}].

\bibitem{Caprini:2009yp}
C.~Caprini, R.~Durrer and G.~Servant, \emph{{The stochastic gravitational wave background from turbulence and magnetic fields generated by a first-order phase transition}}, \href{https://doi.org/10.1088/1475-7516/2009/12/024}{\emph{JCAP} {\bfseries 12} (2009) 024} [\href{https://arxiv.org/abs/0909.0622}{{\ttfamily 0909.0622}}].

\bibitem{Kisslinger:2015hua}
L.~Kisslinger and T.~Kahniashvili, \emph{{Polarized Gravitational Waves from Cosmological Phase Transitions}}, \href{https://doi.org/10.1103/PhysRevD.92.043006}{\emph{Phys. Rev. D} {\bfseries 92} (2015) 043006} [\href{https://arxiv.org/abs/1505.03680}{{\ttfamily 1505.03680}}].

\bibitem{Yang:2021uid}
J.~Yang and L.~Bian, \emph{{Magnetic field generation from bubble collisions during first-order phase transition}}, \href{https://doi.org/10.1103/PhysRevD.106.023510}{\emph{Phys. Rev. D} {\bfseries 106} (2022) 023510} [\href{https://arxiv.org/abs/2102.01398}{{\ttfamily 2102.01398}}].

\bibitem{Di:2020kbw}
Y.~Di, J.~Wang, R.~Zhou, L.~Bian, R.-G.~Cai and J.~Liu, \emph{{Magnetic Field and Gravitational Waves from the First-Order Phase Transition}}, \href{https://doi.org/10.1103/PhysRevLett.126.251102}{\emph{Phys. Rev. Lett.} {\bfseries 126} (2021) 251102} [\href{https://arxiv.org/abs/2012.15625}{{\ttfamily 2012.15625}}].

\bibitem{Dolgov:2002ra}
A.D.~Dolgov, D.~Grasso and A.~Nicolis, \emph{{Relic backgrounds of gravitational waves from cosmic turbulence}}, \href{https://doi.org/10.1103/PhysRevD.66.103505}{\emph{Phys. Rev. D} {\bfseries 66} (2002) 103505} [\href{https://arxiv.org/abs/astro-ph/0206461}{{\ttfamily astro-ph/0206461}}].

\bibitem{Athron:2022jyi}
P.~Athron, C.~Balazs, A.~Fowlie, L.~Morris, G.~White and Y.~Zhang, \emph{{How arbitrary are perturbative calculations of the electroweak phase transition?}}, \href{https://doi.org/10.1007/JHEP01(2023)050}{\emph{JHEP} {\bfseries 01} (2023) 050} [\href{https://arxiv.org/abs/2208.01319}{{\ttfamily 2208.01319}}].

\bibitem{Gould:2021oba}
O.~Gould and T.V.I.~Tenkanen, \emph{{On the perturbative expansion at high temperature and implications for cosmological phase transitions}}, \href{https://doi.org/10.1007/JHEP06(2021)069}{\emph{JHEP} {\bfseries 06} (2021) 069} [\href{https://arxiv.org/abs/2104.04399}{{\ttfamily 2104.04399}}].

\bibitem{Andreassen:2016cvx}
A.~Andreassen, D.~Farhi, W.~Frost and M.D.~Schwartz, \emph{{Precision decay rate calculations in quantum field theory}}, \href{https://doi.org/10.1103/PhysRevD.95.085011}{\emph{Phys. Rev. D} {\bfseries 95} (2017) 085011} [\href{https://arxiv.org/abs/1604.06090}{{\ttfamily 1604.06090}}].

\bibitem{Dunne:2005rt}
G.V.~Dunne and H.~Min, \emph{{Beyond the thin-wall approximation: Precise numerical computation of prefactors in false vacuum decay}}, \href{https://doi.org/10.1103/PhysRevD.72.125004}{\emph{Phys. Rev. D} {\bfseries 72} (2005) 125004} [\href{https://arxiv.org/abs/hep-th/0511156}{{\ttfamily hep-th/0511156}}].

\bibitem{Ivanov:2022osf}
A.~Ivanov, M.~Matteini, M.~Nemev\v{s}ek and L.~Ubaldi, \emph{{Analytic thin wall false vacuum decay rate}}, \href{https://doi.org/10.1007/JHEP03(2022)209}{\emph{JHEP} {\bfseries 03} (2022) 209} [\href{https://arxiv.org/abs/2202.04498}{{\ttfamily 2202.04498}}].

\bibitem{Athron:2023rfq}
P.~Athron, L.~Morris and Z.~Xu, \emph{{How robust are gravitational wave predictions from cosmological phase transitions?}}, \href{https://doi.org/10.1088/1475-7516/2024/05/075}{\emph{JCAP} {\bfseries 05} (2024) 075} [\href{https://arxiv.org/abs/2309.05474}{{\ttfamily 2309.05474}}].

\bibitem{Ekstedt:2021kyx}
A.~Ekstedt, \emph{{Higher-order corrections to the bubble-nucleation rate at finite temperature}}, \href{https://doi.org/10.1140/epjc/s10052-022-10130-5}{\emph{Eur. Phys. J. C} {\bfseries 82} (2022) 173} [\href{https://arxiv.org/abs/2104.11804}{{\ttfamily 2104.11804}}].

\bibitem{Athron:2022mmm}
P.~Athron, C.~Bal\'azs and L.~Morris, \emph{{Supercool subtleties of cosmological phase transitions}}, \href{https://doi.org/10.1088/1475-7516/2023/03/006}{\emph{JCAP} {\bfseries 03} (2023) 006} [\href{https://arxiv.org/abs/2212.07559}{{\ttfamily 2212.07559}}].

\bibitem{Dorsch:2014qja}
G.C.~Dorsch, S.J.~Huber, K.~Mimasu and J.M.~No, \emph{{Echoes of the Electroweak Phase Transition: Discovering a second Higgs doublet through $A_0 \rightarrow ZH_0$}}, \href{https://doi.org/10.1103/PhysRevLett.113.211802}{\emph{Phys. Rev. Lett.} {\bfseries 113} (2014) 211802} [\href{https://arxiv.org/abs/1405.5537}{{\ttfamily 1405.5537}}].

\bibitem{CMS:2016xnc}
{\scshape CMS} collaboration, \emph{{Search for neutral resonances decaying into a Z boson and a pair of b jets or $\tau$ leptons}}, \href{https://doi.org/10.1016/j.physletb.2016.05.087}{\emph{Phys. Lett. B} {\bfseries 759} (2016) 369} [\href{https://arxiv.org/abs/1603.02991}{{\ttfamily 1603.02991}}].

\bibitem{ATLAS:2018oht}
{\scshape ATLAS} collaboration, \emph{{Search for a heavy Higgs boson decaying into a $Z$ boson and another heavy Higgs boson in the $\ell\ell bb$ final state in $pp$ collisions at $\sqrt{s}=13$ TeV with the ATLAS detector}}, \href{https://doi.org/10.1016/j.physletb.2018.07.006}{\emph{Phys. Lett. B} {\bfseries 783} (2018) 392} [\href{https://arxiv.org/abs/1804.01126}{{\ttfamily 1804.01126}}].

\bibitem{CMS:2019ogx}
{\scshape CMS} collaboration, \emph{{Search for new neutral Higgs bosons through the H$\to$ ZA $\to \ell^{+}\ell^{-} \mathrm{b\bar{b}}$ process in pp collisions at $\sqrt{s} =$ 13 TeV}}, \href{https://doi.org/10.1007/JHEP03(2020)055}{\emph{JHEP} {\bfseries 03} (2020) 055} [\href{https://arxiv.org/abs/1911.03781}{{\ttfamily 1911.03781}}].

\bibitem{Noble:2007kk}
A.~Noble and M.~Perelstein, \emph{{Higgs self-coupling as a probe of electroweak phase transition}}, \href{https://doi.org/10.1103/PhysRevD.78.063518}{\emph{Phys. Rev. D} {\bfseries 78} (2008) 063518} [\href{https://arxiv.org/abs/0711.3018}{{\ttfamily 0711.3018}}].

\bibitem{Huang:2015tdv}
P.~Huang, A.~Joglekar, B.~Li and C.E.M.~Wagner, \emph{{Probing the Electroweak Phase Transition at the LHC}}, \href{https://doi.org/10.1103/PhysRevD.93.055049}{\emph{Phys. Rev. D} {\bfseries 93} (2016) 055049} [\href{https://arxiv.org/abs/1512.00068}{{\ttfamily 1512.00068}}].

\bibitem{ATLAS:2022jtk}
{\scshape ATLAS} collaboration, \emph{{Constraints on the Higgs boson self-coupling from single- and double-Higgs production with the ATLAS detector using pp collisions at s=13 TeV}}, \href{https://doi.org/10.1016/j.physletb.2023.137745}{\emph{Phys. Lett. B} {\bfseries 843} (2023) 137745} [\href{https://arxiv.org/abs/2211.01216}{{\ttfamily 2211.01216}}].

\bibitem{CMS:2022dwd}
{\scshape CMS} collaboration, \emph{{A portrait of the Higgs boson by the CMS experiment ten years after the discovery.}}, \href{https://doi.org/10.1038/s41586-022-04892-x}{\emph{Nature} {\bfseries 607} (2022) 60} [\href{https://arxiv.org/abs/2207.00043}{{\ttfamily 2207.00043}}].

\bibitem{Goncalves:2018qas}
D.~Gon\c{c}alves, T.~Han, F.~Kling, T.~Plehn and M.~Takeuchi, \emph{{Higgs boson pair production at future hadron colliders: From kinematics to dynamics}}, \href{https://doi.org/10.1103/PhysRevD.97.113004}{\emph{Phys. Rev. D} {\bfseries 97} (2018) 113004} [\href{https://arxiv.org/abs/1802.04319}{{\ttfamily 1802.04319}}].

\bibitem{Kling:2016lay}
F.~Kling, T.~Plehn and P.~Schichtel, \emph{{Maximizing the significance in Higgs boson pair analyses}}, \href{https://doi.org/10.1103/PhysRevD.95.035026}{\emph{Phys. Rev. D} {\bfseries 95} (2017) 035026} [\href{https://arxiv.org/abs/1607.07441}{{\ttfamily 1607.07441}}].

\bibitem{Cepeda:2019klc}
M.~Cepeda et~al., \emph{{Report from Working Group 2}: {Higgs Physics at the HL-LHC and HE-LHC}}, \href{https://doi.org/10.23731/CYRM-2019-007.221}{\emph{CERN Yellow Rep. Monogr.} {\bfseries 7} (2019) 221} [\href{https://arxiv.org/abs/1902.00134}{{\ttfamily 1902.00134}}].

\bibitem{Mlynarikova:2023bvx}
{\scshape ATLAS, CMS} collaboration, \emph{{Higgs Physics at HL-LHC}},  in \emph{{30th International Workshop on Deep-Inelastic Scattering and Related Subjects}}, 7, 2023 [\href{https://arxiv.org/abs/2307.07772}{{\ttfamily 2307.07772}}].

\bibitem{Grojean:2006bp}
C.~Grojean and G.~Servant, \emph{{Gravitational Waves from Phase Transitions at the Electroweak Scale and Beyond}}, \href{https://doi.org/10.1103/PhysRevD.75.043507}{\emph{Phys. Rev. D} {\bfseries 75} (2007) 043507} [\href{https://arxiv.org/abs/hep-ph/0607107}{{\ttfamily hep-ph/0607107}}].

\bibitem{Roshan:2024qnv}
R.~Roshan and G.~White, \emph{{Using gravitational waves to see the first second of the Universe}}, \href{https://doi.org/10.1103/RevModPhys.97.015001}{\emph{Rev. Mod. Phys.} {\bfseries 97} (2025) 015001} [\href{https://arxiv.org/abs/2401.04388}{{\ttfamily 2401.04388}}].

\bibitem{Martin:2014bca}
S.P.~Martin, \emph{{Taming the Goldstone contributions to the effective potential}}, \href{https://doi.org/10.1103/PhysRevD.90.016013}{\emph{Phys. Rev. D} {\bfseries 90} (2014) 016013} [\href{https://arxiv.org/abs/1406.2355}{{\ttfamily 1406.2355}}].

\bibitem{Elias-Miro:2014pca}
J.~Elias-Miro, J.R.~Espinosa and T.~Konstandin, \emph{{Taming Infrared Divergences in the Effective Potential}}, \href{https://doi.org/10.1007/JHEP08(2014)034}{\emph{JHEP} {\bfseries 08} (2014) 034} [\href{https://arxiv.org/abs/1406.2652}{{\ttfamily 1406.2652}}].

\bibitem{Blinov:2015vma}
N.~Blinov, S.~Profumo and T.~Stefaniak, \emph{{The Electroweak Phase Transition in the Inert Doublet Model}}, \href{https://doi.org/10.1088/1475-7516/2015/07/028}{\emph{JCAP} {\bfseries 07} (2015) 028} [\href{https://arxiv.org/abs/1504.05949}{{\ttfamily 1504.05949}}].

\bibitem{Cutting:2019zws}
D.~Cutting, M.~Hindmarsh and D.J.~Weir, \emph{{Vorticity, kinetic energy, and suppressed gravitational wave production in strong first order phase transitions}}, \href{https://doi.org/10.1103/PhysRevLett.125.021302}{\emph{Phys. Rev. Lett.} {\bfseries 125} (2020) 021302} [\href{https://arxiv.org/abs/1906.00480}{{\ttfamily 1906.00480}}].

\bibitem{Turner:1992tz}
M.S.~Turner, E.J.~Weinberg and L.M.~Widrow, \emph{{Bubble nucleation in first order inflation and other cosmological phase transitions}}, \href{https://doi.org/10.1103/PhysRevD.46.2384}{\emph{Phys. Rev. D} {\bfseries 46} (1992) 2384}.

\bibitem{Linde:1981zj}
A.D.~Linde, \emph{{Decay of the False Vacuum at Finite Temperature}}, \href{https://doi.org/10.1016/0550-3213(83)90072-X}{\emph{Nucl. Phys. B} {\bfseries 216} (1983) 421}.

\bibitem{Apreda:2001us}
R.~Apreda, M.~Maggiore, A.~Nicolis and A.~Riotto, \emph{{Gravitational waves from electroweak phase transitions}}, \href{https://doi.org/10.1016/S0550-3213(02)00264-X}{\emph{Nucl. Phys. B} {\bfseries 631} (2002) 342} [\href{https://arxiv.org/abs/gr-qc/0107033}{{\ttfamily gr-qc/0107033}}].

\bibitem{Espinosa:2010hh}
J.R.~Espinosa, T.~Konstandin, J.M.~No and G.~Servant, \emph{{Energy Budget of Cosmological First-order Phase Transitions}}, \href{https://doi.org/10.1088/1475-7516/2010/06/028}{\emph{JCAP} {\bfseries 06} (2010) 028} [\href{https://arxiv.org/abs/1004.4187}{{\ttfamily 1004.4187}}].

\bibitem{Hindmarsh:2020hop}
M.B.~Hindmarsh, M.~L\"uben, J.~Lumma and M.~Pauly, \emph{{Phase transitions in the early universe}}, \href{https://doi.org/10.21468/SciPostPhysLectNotes.24}{\emph{SciPost Phys. Lect. Notes} {\bfseries 24} (2021) 1} [\href{https://arxiv.org/abs/2008.09136}{{\ttfamily 2008.09136}}].

\bibitem{Pen:2015qta}
U.-L.~Pen and N.~Turok, \emph{{Shocks in the Early Universe}}, \href{https://doi.org/10.1103/PhysRevLett.117.131301}{\emph{Phys. Rev. Lett.} {\bfseries 117} (2016) 131301} [\href{https://arxiv.org/abs/1510.02985}{{\ttfamily 1510.02985}}].

\bibitem{Weir:2017wfa}
D.J.~Weir, \emph{{Gravitational waves from a first order electroweak phase transition: a brief review}}, \href{https://doi.org/10.1098/rsta.2017.0126}{\emph{Phil. Trans. Roy. Soc. Lond. A} {\bfseries 376} (2018) 20170126} [\href{https://arxiv.org/abs/1705.01783}{{\ttfamily 1705.01783}}].

\bibitem{No:2011fi}
J.M.~No, \emph{{Large Gravitational Wave Background Signals in Electroweak Baryogenesis Scenarios}}, \href{https://doi.org/10.1103/PhysRevD.84.124025}{\emph{Phys. Rev. D} {\bfseries 84} (2011) 124025} [\href{https://arxiv.org/abs/1103.2159}{{\ttfamily 1103.2159}}].

\bibitem{Laurent:2022jrs}
B.~Laurent and J.M.~Cline, \emph{{First principles determination of bubble wall velocity}}, \href{https://doi.org/10.1103/PhysRevD.106.023501}{\emph{Phys. Rev. D} {\bfseries 106} (2022) 023501} [\href{https://arxiv.org/abs/2204.13120}{{\ttfamily 2204.13120}}].

\bibitem{Ekstedt:2024fyq}
A.~Ekstedt, O.~Gould, J.~Hirvonen, B.~Laurent, L.~Niemi, P.~Schicho et~al., \emph{{How fast does the WallGo? A package for computing wall velocities in first-order phase transitions}}, \href{https://doi.org/10.1007/JHEP04(2025)101}{\emph{JHEP} {\bfseries 04} (2025) 101} [\href{https://arxiv.org/abs/2411.04970}{{\ttfamily 2411.04970}}].

\bibitem{Carena:2025flp}
M.~Carena, A.~Ireland, T.~Ou and I.R.~Wang, \emph{{The Discriminant Power of Bubble Wall Velocities: Gravitational Waves and Electroweak Baryogenesis}},  \href{https://arxiv.org/abs/2504.17841}{{\ttfamily 2504.17841}}.

\end{thebibliography}\endgroup
\bibliographystyle{JHEP}

\end{document}